\documentclass[a4paper,11pt]{article}
\usepackage[margin=1in]{geometry}

\usepackage[latin2]{inputenc}
\usepackage[english]{babel}
\usepackage[all=normal,paragraphs=tight]{savetrees}
\usepackage{caption}
\usepackage{subcaption}
\usepackage{float}

\usepackage{enumerate}
\newcommand{\useless}[1]{}

\newcommand{\lcd}{\textsc{$p$-$\lambda$-ECS}\xspace}
\newcommand{\wlcd}{\textsc{$p$-Weighted  $\lambda$-ECS}\xspace}

\newcommand{\medg}{\textsc{Minimum Equivalent Digraph}\xspace}

\newcommand{\scss}{\textsc{Strongly Connected Steiner Subgraph}\xspace}

\newcommand{\C}{{\cal C}}
\newcommand{\D}{{\cal D}}
\newcommand{\Z}{{\cal Z}}

\newcommand{\del}{{\sf del}}
\newcommand{\undel}{{\sf undel}}

\newcommand{\cut}[1]{{(#1, \overline{{#1}})}\xspace}
\newcommand{\co}[1]{\overline{#1}\xspace}

\newcommand{\myparagraph}[1]{\smallskip\paragraph{\textbf{\sffamily #1} \ }}

\newcommand{\Yes}{{\sc Yes}}
\newcommand{\No}{{\sc No}}

\newcommand{\longversion}[1]{}
\newcommand{\bigoh}{{\cal O}}
\newcommand{\cO}{{\cal O}}

\newcommand{\paradefn}[4]{
 \vspace{3mm}
\noindent\fbox{
  \begin{minipage}{.95\textwidth}
  \begin{tabular*}{\textwidth}{@{\extracolsep{\fill}}lr} \textsc{#1}  & {\bf{Parameter:}} #3 \\ \end{tabular*}
  {\bf{Input:}} #2  \\
  {\bf{Question:}} #4
  \end{minipage}
  }
  \vspace{2mm}
}

\usepackage[usenames,svgnames]{xcolor}
\usepackage{graphicx}
\usepackage{tikz}
\usetikzlibrary{decorations.shapes}

\usepackage{wrapfig}
\usepackage{titlesec}
\usepackage{titling}

\usepackage{xspace}

\usepackage[ruled,noline,linesnumbered]{algorithm2e}
\usepackage{multicol}
\usepackage{framed}
\usepackage{boxedminipage}
\usepackage{stmaryrd}

\usepackage{amssymb,amsthm}
\usepackage[fleqn]{amsmath}
\usepackage{pdfpages}
\usepackage{lastpage}

\definecolor{wine-stain}{rgb}{0.5,0,0}
\usepackage{hyperref}
\hypersetup{
    pdftex,
    pdfpagelabels,
    bookmarks,
    hyperindex,
    hyperfigures,
    linktocpage,
    colorlinks,
    allcolors=wine-stain,
    linktoc=all
}

\newcommand{\FPT}{\textsf{\textup{FPT}}\xspace}

\newcommand{\NPC}{\textrm{\textup{\textsf{NP}-complete}}\xspace}

\newcommand{\NP}{{\ensuremath{\sf{NP}}}}
\newcommand{\classP}{{\ensuremath{\sf{P}}}}

\newcommand{\NO}{\textsc{No}}

\newcommand{\OO}{{\mathcal O}}

\newcommand{\R}{{\mathcal R}}

\usepackage{todonotes}
\usepackage{microtype}
\usepackage{amsthm}
\usepackage{amssymb}
\usepackage{thm-restate}

\newtheorem{theorem}{Theorem}[section]
\newtheorem{lemma}{Lemma}[section]

\newtheorem{claim}{Claim}

\newtheorem{definition}{Definition}[section]
\newtheorem{observation}{Observation}[section]
\newtheorem{proposition}{Proposition}[section]

\makeatletter
\@addtoreset{claim}{lemma}
\@addtoreset{claim}{theorem}
\@addtoreset{claim}{observation}
\@addtoreset{claim}{proposition}
\@addtoreset{claim}{corollary}
\@addtoreset{redrule}{section}
\@addtoreset{subclaim}{claim}
\makeatother

\newcommand{\altdelta}{\partial}

\title{On finding highly connected spanning subgraphs
\thanks{ P. Misra is partially supported by the European Research Council (ERC) grant ``Rigorous Theory of Preprocessing'', reference 267959. M. S. Ramanujan is supported by Austrian Science Funds (FWF), project P26696. 
S. Saurabh is supported by PARAPPROX, ERC starting grant no. 306992.
}}
\author{
 Manu Basavaraju\thanks{
    Department of Computer Science and Engineering, NITK Surathkal, India, \texttt{manub@nitk.ac.in}.
  } \and 
Pranabendu Misra\thanks{
    Institute of Mathematical Sciences, HBNI, India,   \texttt{pranabendu@imsc.res.in}.
  }\and 
M. S. Ramanujan\thanks{
   Algorithms and Complexity Group, TU Wien, Vienna, \texttt{ramanujan@ac.tuwien.ac.at}.
  }\and 
  Saket Saurabh\thanks{
    Institute of Mathematical Sciences, HBNI, India, \texttt{saket@imsc.res.in}, and
    Department of Informatics, University of Bergen, Norway, \texttt{Saket.Saurabh@ii.uib.no}.
  }
}
\date{}

\begin{document}
\begin{titlepage}
\def\thepage{}
\thispagestyle{empty}
\maketitle
\vspace{-1cm}
\begin{abstract}
In the {\sc Survivable Network Design Problem} (SNDP), the input is an edge-weighted (di)graph $G$ and an integer $r_{uv}$ for every pair of vertices $u,v\in V(G)$. The objective is to construct a subgraph $H$ of minimum weight which contains $r_{uv}$ edge-disjoint (or node-disjoint) $u$-$v$ paths. This is a fundamental problem in combinatorial optimization that captures numerous well-studied problems in graph theory and graph algorithms.
%

An important restriction of this problem is the case when the connectivity demands are equal for \emph{every} pair of vertices in the graph. 
In this paper, we consider the the edge-connectivity version of this problem
which is called the {\sc $\lambda$-Edge Connected Subgraph} ({\sc $\lambda$-ECS}) problem. 
In this problem, we are given a $\lambda$-edge connected (di)graph $G$ with a non-negative weight function $w$ on the edges and an integet $k$, and the objective is to find a minimum weight spanning subgraph $H$ that is also $\lambda$-edge connected, and has at upto $k$ fewer edges than $G$. In other words, we are asked to compute a maximum weight subset of edges, of cardinality upto $k$, which may be safely deleted from $G$. Motivated by this question, we investigate the connectivity properties of $\lambda$-edge connected (di)graphs and  obtain algorithmically significant  structural results. One of our  central structural results can be roughly stated  as follows.   
\begin{center}
	
\begin{quote}
{\em In polynomial time, one can either find a set of $k$ edges which can be deleted from the given (di)graph without violating the connectivity constraints, or correctly conclude that  the (di)graph contains only $\OO(\lambda k^3)$ `interesting' edges.} 

\end{quote}
\end{center}

%
\noindent
We demonstrate the importance of our structural results  by presenting an algorithm running in time $2^{\cO(k \log k)} |V(G)|^{\cO(1)}$ for   {\sc $\lambda$-ECS}, thus proving its fixed-parameter tractability.  
We follow up on this result and obtain the {\em first  polynomial compression} for {\sc $\lambda$-ECS} on unweighted graphs. As a consequence, we also obtain the first fixed parameter tractable algorithm, and a polynomial kernel for a  parameterized version of the classic {\sc Mininum Equivalent Graph} problem.
We believe that our structural results are of independent interest and will play a crucial role in  the design  of algorithms for connectivity-constrained problems in general and the SNDP problem in particular.


\end{abstract}


\end{titlepage}

\newpage 


\section{Introduction}
Network design problems, and the \textsc{Survivable Network Design Problem}(SNDP) in particular, are some of the most fundamental research topics in combinatorial optimization, algorithm design and graph theory, because of their wide spread applications.
It involves designing a cost effective communication network that remains operational despite a number of equipment failures.
Such failures may be caused by any number of things such as a hardware or software faults, a broken link between two network components, human error an so on. This class of problems are modeled as graphs, with the nodes representing the network components (such as computers, routers, etc.), edges representing the communication links between the components and the associated costs of the vertices and edges. Then the network design problem becomes, the problem of finding a subgraph satisfying certain connectivity constraints, or the problem of augmenting a the graph to achieve certain connectivity requirements,
at a minimum cost.
 
The most general variant of these problems, is called the {\sc Survivable Network Design Problem} (SNDP). Here, the input is an edge-weighted graph $G$ and an integer $r_{uv}$ for every pair of vertices $u,v\in V(G)$. The objective is to construct a subgraph $H$ of minimum weight which contains $r_{uv}$ edge-disjoint (or node-disjoint) $u$-$v$ paths for every pair of vertices $u,v$. Depending on the type of demands or weights allowed, it generalizes numerous network design problems, and consequently there is a long line of research into the design of polynomial-time exact algorithms as well as approximation algorithms for these problems. Let us note that almost all such problems turn out to be {\NP}-hard.
A highlight of this line of research is the $2$-approximation algorithm of Jain \cite{jain2001factor} for the {\em edge-connectivity} version of SNDP. This work introduced the iterative rounding technique which has subsequently become a essential part of the approximation algorithms toolkit. Kortsarz et al. \cite{KortsarzKL04} were the first to prove a lower bound for the {\em node-connectivity} of SNDP and showed that this problem cannot be approximated within a factor of $2^{\log^{1-\epsilon}n}$ for any $\epsilon>0$. Subsequently, Chakraborty et al. \cite{ChakrabortyCK08} improved this lower bound to  $p^\epsilon$ where $p$, the maximum connectivity demand exceeds $p_0$ with $p_0$ and $\epsilon$ being fixed constants. More recently, Chuzhoy and Khanna \cite{ChuzhoyK12} gave an $\bigoh(p^3 \log n)$-factor approximation algorithm for this problem, where $p$ is again the maximum of the connectivity demands. There is also a significant amount of literature on the \emph{directed} versions of SNDP. Here, there is an integer $r_{uv}$ for every {\em ordered pair} $(u,v)\in V(G)\times V(G)$.
We direct the reader to \cite{kortsarz2010approximating,khuller1996approximation} for surveys on this  topic.

An important and well-studied restriction of SNDP is the version where the demands are \emph{uniform} for every pair of vertices in the graph. That is, for some $\lambda$, $r_{uv}=\lambda$ for every $u,v\in V(G)$. This restriction is termed $\lambda$-{\sc SNDP with Uniform Demands}, and when the demands are on the edge-connectivity of the graph, it is called $\lambda$-{\sc Edge Connected Subgraph} ($\lambda$-ECS).
This problem generalizes many other well studied problems such as {\sc Hamiltonian Cycle}, {\sc Minimum Strongly Connected Spanning Subgraph(MSCSS)}, {\sc 2-Edge Connected Spanning Subgraph} etc. 
It was shown by Khuller and Vishkin \cite{khuller1994biconnectivity} that this problem admits a $2$-approximation algorithm.
We again direct the reader to the surverys \cite{kortsarz2010approximating,khuller1996approximation} for more details.
%

In this paper we investigate the edge connectivity properties of (di)graphs, motivated by the following question that is derived from $\lambda$-ECS.
\begin{quote}
Let $G$ be a $\lambda$-edge connected (di)graph and let $w$ be a non-negative weight function $w$ on the edges. Find a maximum weight subset of edges, $F$, of cardinality upto $k$, such that $G - S$ is also $\lambda$-connected.
\end{quote}
We obtain new structural results on $\lambda$-connected (di)graphs, which could be of independent interest.
The following is a brief description of our results.
Consider a $\lambda$-edge connected (di)graph, and call an edge deletable if it can removed without decreasing the connectivity of the graph. 
Then the following statement holds for any directed graph, and for undirected graphs when $\lambda$ is an even number.
In polynomial time, either we can find a set of $k$ edges which can be removed from the graph without decreasing it's connectivity, or we conclude that the graph contains $\Omega(\lambda k^2)$ deletable edges. 
For odd values of $\lambda$ in undirected graphs, this statement is obviously false, e.g. consider a cycle and $\lambda = 1$. In this case, we show the following. In polynomial time, either we can find a set of $k$ edges which can be removed from the graph without decreasing it's connectivity, or we conclude that all but $\Omega(\lambda k^3)$ of the edges in the graph are irrelevant. Here a set of edges in the graph is irrelevant if there is some deletion set of cardniality $k$ that is disjoint from it. More formally,
\begin{theorem}\label{thm:struct}
    Let $G$ be a (di)graph such that it is $\lambda$-connected, and $k$ be any integer. Then there is a polynomial time algorithm, that either computes a subset $F$ of edges, of cardinality $k$, such that $G-F$ is $\lambda$-edge connected, or finds a deletable edge $e$ that is irrelevant, or concludes that total number of deletable edges that are not irrelevant is bounded by $7 \lambda k^3$. 
    
    Furthermore, in digraphs and in undirected graphs with an even value of $\lambda$, no edges are marked as irrelevant, and we can bound the total number of deletable edges to $\lambda k^2$ and $2 \lambda k^2$, respectively.
\end{theorem}
We postpone the discussion of our methods and techniques to prove the above theorem to section~\ref{sec:overview}, and move on to the algorithmic applications of our result.
Our results directly lead to a fixed parameter tractable algorithm for $\lambda$-ECS, when parameterized by the size of the deletion set.
Let us state this result more formally.
In \emph{parameterized complexity}, we consider instances of the form $(x,k)$, where $x$ is a  problem instance, and $k$ is a positive integer called the {\em parameter} which reflects some structural property of the instance $x$. The notion of tractability in parameterized complexity is called {\em fixed parameter tractability} (\FPT). This entails solvability of $(x, k)$ in time $\tau(k)\cdot |x|^{\bigoh(1)} $, where $\tau$ is an arbitrary function, by taking advantage of structural properties that are ensured by the parameter.
We refer to textbooks \cite{cygan2015parameterized,downey2012parameterized} for an introduction to parameterized complexity.
Typically, the most natural parameterization when studying an \NPC problem is the \emph{size} of the solution.
In case of $\lambda$-ECS, that would be the number of edges in $H$. 
However, observe that $H$ is a spanning subgraph of $G$ that is $\lambda$-edge connected and therefore every vertex in $H$ has degree at least $\lambda$, implying that $H$ has at least $\frac{\lambda n}{2}$ edges. Hence, if we consider the minimum number of edges in a $\lambda$-connected subgraph of $G$ as a parameter, denoted by $\ell$, then either $\frac{\lambda n}{2}>\ell$, in which case there is no such subgraph, or $n\leq \frac{2\ell}{\lambda}$ in which case we can just go over all edge subsets of $G$ of size at most $\ell$, resulting in a trivial {\FPT} algorithm.  Then, perhaps the next question would be whether there is a subgraph $H$ on at most $\frac{\lambda n}{2}+\ell$ edges, where $\ell$ is the parameter \footnote{Such parameterizations are called \emph{above / below guarantee} parameterization; we refer to \cite{GutinY12,mahajan2009parameterizing} for an introduction to this topic.}. 
However, in this case there cannot even be an algorithm of the with a running time of $\bigoh(n^{g(\ell)})$ unless {\classP} = {\NP}, for any function $g$. The reason is simply that, any $2$-edge connected graph $G$ has a Hamiltonian Cycle if and only if it has a $2$-edge connected spanning subgraph with exactly $\frac{2\cdot n}{2}+0 = n$ edges, 
and therefore such an algorithm will solve the {\sc Hamiltonian Cycle} problem in polynomial time.
Hence, a more meaningful parameterization of $\lambda$-ECS is in terms of the `dual' parameter, which is the number of edges of $G$ that are \emph{not} present in a minimum $\lambda$-connected spanning subgraph $H$.

\paradefn{\lcd}
		   {A graph or digraph $G$ which is $\lambda$-connected and an integer $k$}{$k$}
         {Is there a set $F\subseteq E(G)$ of size at least $k$
         such that $H=G - F$ is also $\lambda$-connected?}

\noindent
\vspace{-2 pt}
\begin{restatable}{theorem}{lcd-thm}\label{thm:lcd-thm}
\lcd can be solved in time $2^{\cO(k \log k)}n^{\cO(1)}$.
\end{restatable}

\noindent
Our result extends to the \emph{weighted} version of this problem, which is defined as follows.

\paradefn{\wlcd}
		   {A graph or digraph $G$ which is $\lambda$-connected, $w:E(G)\to {\mathbb R}_{\geq 0}$, a target weight $\alpha\in {\mathbb R}$ and an integer $k$}{$k$}
         {Is there a set $F\subseteq E(G)$ of size at most $k$
         such that $H=G - F$ is also $\lambda$-connected and  $w(F)\geq \alpha$?}\\
\begin{restatable}{theorem}{wlcd-thm}\label{thm:wlcd-thm}
    \wlcd can be solved in time $2^{\cO(k \log k)}n^{\cO(1)}$.
\end{restatable}
\noindent
While this algorithm doesn't directly follow from Theorem~\ref{thm:struct}, it builds upon the structural properties of the input graph provided by it.
We would like to emphasize the fact that the exponent of $n$ in the polynomial component of the running time is in fact independent of $\lambda$. 
Hence, \wlcd is solvable in polynomial time for $\lambda = \log^{\OO(1)} n$ and $k = \OO(\frac{\log n}{\log \log n})$.

We also obtain a polynomial compression of \lcd, i.e. a smaller instance of a related problem that is equivalent to the input instance. Formally, a parameterized problem $\Pi \subseteq \Sigma^*\times \mathbb{N}$ is  said to admit a {\em polynomial kernel}, if there is a polynomial time algorithm which given an instance $(x,k)\in \Pi$, returns an instance $(x',k')\in \Pi$ such that, $(x,k)\in \Pi$ if and only if  $(x',k')\in \Pi$ and $|x'|, k' \leq k^{\cO(1)}$. 
A {\em polynomial compression} is a relaxation of polynomial kernelization where 
the output may be an instance of a different parameterized problem. 
\begin{theorem}
\label{thm:poly-compression-full}
	For any $\delta > 0$, there exists a randomized compression for 
	\lcd\ of size $ \cO(k^{18} \lambda^6 
	(\log k \lambda + \log(1/\delta))$, such that the error probability is upper bounded by $1-\delta$. 
\end{theorem}

This compression routine could be a good starting point for   streaming and dynamic graph algorithms for connectivity based problems. 
Finally, an immediate corollary of our fixed-parameter tractability result for {\lcd} is the first fixed-parameter algorithm for a parameterized version of the classic \medg problem. In this problem, the goal is to find a minimum spanning subgraph which is ``equivalent'' 
to the input graph.
Two graphs $G$ and $H$ are said to be {\em equivalent} if for any two vertices $u,v$, the vertex $v$ is reachable from $u$ in $G$, if and only if $v$ is reachable from $u$ in $H$.
This problem is easily seen to be \NPC, by a reduction from the {\sc Hamiltonian Cycle} problem~\cite{garey1979computers}. The natural  \emph{parameterized} version of this problem asks, given a graph $G$ and integer $k$, whether there is a subgraph $H$ on at most $m-k$ edges which is equivalent to $G$. It is well known that \medg can be reduced to an input $G'$ which is strongly connected (that is, there is directed path between every pair of vertices in $G'$).  The following proposition due to  Moyles and Thompson~\cite{moyles1969algorithm},  see also \cite[Sections 2.3]{bang2008digraphs}, 
reduces the problem of finding a minimum equivalent sub-digraph of an arbitrary $G$ to a strong digraph. 

\begin{proposition}\label{prop:medredtostrong}
Let $G$ be a digraph  on $n$ vertices  with strongly connected components $C_1,\ldots,C_r$.  Given a minimum equivalent subdigraph $C_i'$ for each $C_i$, $i\in [r]$, one can obtain a minimum equivalent subdigraph $G'$ of $G$ containing each of $C'_i$ in $\cO(n^\omega)$ time. Here, $\omega$ is the exponent of the fastest known matrix multiplication algorithm and   $\omega$ is currently upper bounded  by $2.376$.
\end{proposition}

\noindent 
Proposition~\ref{prop:medredtostrong} allows us to reduce an instance of \medg on a general digraph to instances
where the graph is strongly connected, in polynomial time. 
We now solve \medg by executing the algorithm of Theorem \ref{thm:lcd-thm} with $\lambda=1$ for each strongly connected component of the input digraph. \\

\myparagraph{Related work.} Network design problems are very well studied in the framework of approximation algorithms, and we direct the reader to the surveys \cite{kortsarz2010approximating,khuller1996approximation} for more details.
However, not much is known about the parameterized complexity of these problems,
and we state the few known results. Based on the fact that any strongly-connected graph has an equivalent subdigraph containing at most $2n-2$ arcs, Bang-Jensen and Yeo \cite{bang2008minimum} study the parameterization of $1$-ECS below $2n-2$ (instead of $m$, the total number of edges), and obtain an algorithm that runs in time $2^{\bigoh(k \log k)}n^{\bigoh(1)}$ that decides whether a given strongly connected digraph has an equivalent digraph with at most $2n-2-k$ edges. 
However, note that this parameterization is of limited use in the cases where $m \leq 2n -2 - k$ or when the graph is weighted.
Marx and V\'{e}gh studied the problem of augmenting the edge connectivity of an undirected graph from $\lambda - 1$ to $\lambda$ ~\cite{marx2015fixed}, via a 
minimum cost set of upto $k$ new links, and obtain a {\FPT} algorithm and polynomial kernel for it.
Basavaraju et.al. ~\cite{basavaraju2014} improve the running time this algorithm and,
extend these results to a different variant of this problem. 
Exact exponential algorithms for these problems have also been studied.
The first exact algorithm for \textsc{Minimum Equivalent Graph(MEG)} and MSCSS, running in time $2^{\OO(m)}$ time,  was given in by Moyles and Thompson~\cite{moyles1969algorithm} in 1969, where $m$ is the number of edges in the graph.
Very recently, Fomin et.al. \cite{fomin2016efficient} gave the first single-exponential algorithm for MEG and MSCSS, 
i.e. with a running time of $2^{\OO(n)}$.
\textsc{Hamiltonian Cycle}, which is a special case of MEG, has 
a classic algorithm, running in time $\OO(2^n)$, known from 1960s\cite{held1962dynamic, bellman1962dynamic}.
It was recently improved to $\OO(1.657^n)$ for undirected graphs~\cite{bjorklund2014determinant}, and to $\OO(1.888^n)$ for bipartite digraphs~\cite{cygan2013fast}. 
A survey of these results may be found in Chapter 12 of the textbook of Bang-Jensen and Gutin \cite{bang2008digraphs}.

\paragraph{Organization of the paper.} 
We first give an overview of our results  and a sketch of our methods and techniques in Section \ref{sec:overview}. 
We recall some relevant terminology and graph-theoretic results in Section \ref{sec:prelim}, and subsequently we prove certain results based on min-cuts in (di)graphs which are used throughout the paper. 
We then present the full descriptions of our algorithm on directed and undirected graphs. These can be found in Section \ref{sec:directed} and Section \ref{sec:undirected}  respectively. 
Section \ref{sec:undirected} is composed of two parts depending on the parity of $\lambda$. Subsequently, we build upon the results from earlier sections to prove Theorem \ref{thm:wlcd-thm} (Section \ref{sec:weighted}). 
Finally, we arrive at the design of a randomized polynomial compression (Theorem \ref{thm:poly-compression-full}). This is presented in Section \ref{sec:compression}, where we first handle the case when $G$ is a digraph, and then argue that similar arguments work extend to the case when $G$ is an undirected graph. We then conclude with some open problems in Section~\ref{sec:conclusion}.
\section{An overview of our results.} 
\label{sec:overview}
This section presents a brief overview and the main ideas presented in this paper. We refer the reader to Section~\ref{sec:prelim} for the definitions of many of the notation and terms that are used here.
A crucial notion we will use repeatedly is that of deletable edges.
An edge in a graph is \emph{deletable}, if removing it does not violate the required connectivity constraints, and otherwise it is \emph{undeletable}. 
We denote by $\del(G)$ the set of deletable edges in $G$, and by $\undel(G)$ the set of undeletable edges in $G$. 
It is clear that any subset of edges $F$ (called a \emph{deletion set}) of cardinality $k$, such that $G-F$ is $\lambda$-connected, is always a subset of the collection of deletable edges.
We show the following structural result, relating the number of deletable edges to the cardinality of a deletion set.
\begin{quote}
$(\star)$ If a graph contains $\Omega(\lambda k^2)$ deletable edges then there is a set of $k$ edges which can be removed from the graph
without violating the connectivity constraints.
\end{quote}
At a first glance, this is obviously false.
For example, set $\lambda = 1$ and consider an arbitrarily long cycle.
Then every edge is a deletable edge but no more than one edge may be deleted without disconnecting the graph.
Note that, this example can be generalized to any \emph{odd} value of $\lambda$.
However, we show that the statement does indeed hold for digraphs (for any value of $\lambda$),
and for undirected graphs whenever $\lambda$ is even. Hence, we prove the following lemma for undirected graphs.
%
\begin{restatable}{lemma}{lambdaundirmain}
    \label{lemma:lambda-undir:upper bound}
    Let $G$ be an undirected graph and $k$ be an integer, such that $G$ is $\lambda$-connected where $\lambda$ is an even integer. Then in polynomial time, we can either find a set $F$ of cardinality $k$ such that $G-F$ is $\lambda$-connected, or conclude that $G$ has at most $2\lambda k^2$ deletable edges in total.
\end{restatable}
%
\noindent
For digraphs, the parameters in the lemma may be slightly improved to obtain the following statement.
%
%
\begin{restatable}{lemma}{lambdadirmain}\label{lemma:lambda-dir:upper bound}
    Let $G$ be a digraph and $k$ be an integer such that, $G$ is $\lambda$-connected for some integer $\lambda$.  
    Then in polynomial time, we can either find a set $F$ of cardinality $k$ such that $G-F$ is $\lambda$-connected, or conclude that $G$ has at most $\lambda k^2$ deletable edges in total.
\end{restatable}

Our proof of these statements is built upon a close examination of a greedily constructed maximal set deletion set $F$.
We may assume that the graph has more than $\OO(\lambda k^2)$ deletable edges to begin with, as otherwise the claims are trivially true.
Now, if the greedy deletion set has $k$ or more edges, then we are done.
Otherwise, we delete the edges of the greedy deletion set in an arbitrary but fixed sequence and examine its effect on the other deletable edges of the graph.
Each time an edge in the greedy deletion set is deleted 
several other deletable edges in the graph may become undeletable in the remaining graph.
Since at the end of this deletion sequence, all the remaining edges are undeletable, there must be a step where $\Omega(\lambda k)$ edges turn undeletable after having been deletable prior to this step. We show that
we can extract another deletion set of $k$ edges from this collection of edges as required to prove our claim. The process of extracting a deletion set of cardinality $k$ from this collection of edges is as follows. We show that there is a subset of $\Theta(k)$ edges in this collection such that there are no $\lambda$-cuts in the current graph which separate the endpoints of more than one edge in this subset. We then show that there is a way to pick $k$ edges from this subset such that these edges form a deletion set. 
We essentially show that, should our algorithm fail to find a desired deletion set even though the number of deletable edges exceeds the stated bounds, 
then the input graph must itself violates the required connectivity properties. While our algorithms are quite simple, the analysis is fairly techincal building upon the \emph{submodularity of cuts}. Interestingly, the analysis is much simpler in the case of digraphs when compared to the case of undirected graphs. 


The case of undirected graphs and an odd value of $\lambda$ is much more involved. In this case, we show that the statement ($\star$) is essentially true
if we restrict our deletion set to a well chosen subset of the deletable edges,
which can be computed in polynomial time.
Additionally we must increase the bound on the number of deletable edges to $\Omega(\lambda k^3)$. As can be observed in the above example of a cycle, it is possible to identify certain deletable edges as being disjoint from \emph{some} deletion set of cardinality $k$ in the given graph. We call an edge satisfying this property, an \emph{irrelevant} edge.
We give a polynomial time procedure that identifies certain edges as irrelevant
in the given graph. We use this procedure to iteratively grow the set of irrelevant edges, always ensuring that if there is a deletion set of $k$ edges then there is one that is disjoint from this set. Finally, by excluding these irrelevant edges from the set of deletable edges, we show that the proposed statement holds true.
\begin{restatable}{lemma}{mainodd}
    \label{lem:main_odd}
    Let $\lambda\in {\mathbb N}$ be odd. Let $G$ be an undirected graph such that $G$ is $\lambda$ connected, $k$ be  an integer and let $\R$ a subset of edges of $G$. Then there is a polynomial time algorithm that, either    
    either computes a subset of edges  $F \subseteq E(G) \setminus R$ of cardinality $k$ such that $G-F$ is $\lambda$-connected, or
    finds an edge $e$ in $E(G)\setminus \R$ such that the given graph has a deletion set of cardinality $k$ that is disjoint from $\R$ if and only if it has such a set disjoint from $\R\cup \{e\}$, or concludes that there are at most $\lambda(6k^3+9k^2+k)$ deletable edges in $E(G) \setminus \R$.
\end{restatable}
\noindent
The above algorithm, and the corresponding analysis, is much more involved.
It starts off with the approach of the earlier algorithms, but requires a deeper examination of the structure of the graph.
Recall the example of a cycle for $\lambda=1$.
We build upon the intuition provided by this example to show the following
structural result.
If a particular deletion set of cardinality $k$ that is proposed by algorithm, is actually incorrect,
then the graph can be decomposed into a ``cycle-like'' structure,
which then allows us to identify and mark a new deletable edge as irrelevant.
More precisely, we obtain a partition of the vertex set of the graph
such that, the sets in the partition can be arranged in a cycle
with each subset being ``adjacent'' only to two neighboring subsets.
It is clear that combining the above three lemmas gives a proof of Theorem~\ref{thm:struct}.

The above results directly imply \textsf{FPT} algorithm for $\lcd$ in any unweighted (di)graph. This is because in polynomial time, we can either compute a solution or conclude that the set of deletable edges (which are not irrelevant) is bounded by a polynomial in $k$. This implies a branching algorithm for \lcd with the claimed running time.
The above results also form the starting point of  our \emph{polynomial compression} for the {\lcd} problem. This is because, we have proved that unless the number of deletable edges in the instance is bounded, we can always compute a solution in polynomial time.
Hence, we may assume that the instance has $\cO(\lambda k^3)$ deletable edges and
we use the results of Assadi et.al.~\cite{AssadiKLT15}
to give a randomized polynomial compression for such instances.
Assadi et.al.~\cite{AssadiKLT15} give a dynamic sketching scheme 
for finding min-cuts between a fixed pair of vertices in a
dynamic graph, where the dynamic edge set is restricted 
to the edges between a fixed subset of vertices.
We obtained the claimed compression by treating the deletable edges of the graph as the afore-mentioned set of dynamic edges
and using certain structural properties of a solution.

Finally, we turn to \wlcd.
First, note that our results can be used to solve a more general version of {\lcd}. In this generalization, there is an additional requirement that the solution \emph{must} be contained in a given subset $W$ of the edges of the graph.
To be precise we give a polynomial time algorithm that, given a set $W$ 
containing $7 \lambda k^3$ deletable edges of the graph ($\lambda k^2$ deletable edges for digraphs),
finds a deletion set of cardinality $k$(if one exists) that is additionally a subset of $W$. While the set $W$ is not explicitly mentioned in the statements
of our lemmas and theorems, we always assume that 
the set of deletable edges is restricted to be a subset of $W$.
This fact comes in handy for designing an \FPT algorithm for \wlcd,
where we must find a deletion set of maximum total weight which contains upto $k$ edges.
We use the following simple observation
which leads us to the algorithm for weighted instances.
\begin{quote}
Let $W$ be the set of the `heaviest' $7\lambda k^3$ deletable edges,
(only $\lambda k^2$ edges for digraphs).
Then there is a polynomial time algorithm that either correctly concludes that there is a solution (of the required kind) which intersects this set $W$
(and this is the only possibility for digraphs),
or computes an edge $e \in W$ which can be safely added to the set of irrelevant edges.
\end{quote}
For digraphs and undirected graphs with an even value of $\lambda$, this result follows easily from the arguments in the unweighted case
as there are no irrelevant edges to deal with.
For undirected graphs with an odd value of $\lambda$,
we have to be more careful while marking an edge as irrelevant,
lest it affect the weight of the required solutions.
For this, we use a  modification of our scheme for finding irrelevant
edges in the unweighted odd $\lambda$ case. 
Finally let us note that, as a consequence of our algorithms, \wlcd is solvable in polynomial time for $\lambda = (\log n)^{\OO(1)}$ and $k = \OO(\frac{\log n}{\log \log n})$.
Let us continue on to an overview of our algorithms and analysis.



\subsection{Directed Graphs}
\label{sec:short-directed}
Let us sketch the results and methods for directed graph as presented in Section~\ref{sec:directed}.
For any edge $e=(u,v)$, we denote by $\D(e)$ the set of deletable edges of $G$ which are undeletable in $G-e$. That is, those edges for which the edge $e$ is `critical'. 
We will deal with a \emph{fixed} deletable  edge $e^* = (u^*, v^*)$ in $G$ such that $\D(e^*)$ has at least $k \lambda$ edges. 
The main lemma we require for our algorithm is the following.
\begin{lemma}
    \label{lemma:short-lambda-dir:alternate solution}
    Let $G$ be a digraph and $\lambda\in {\mathbb N}$ such that $G$ is a $\lambda$-connected digraph. If there is a deletable edge $e^*\in E(G)$ such that $|\D(e^*)| \geq k\lambda$ then there is a set $\Z \subseteq \D(e^*)$ of $k$ edges such that $G- \Z$ is $\lambda$-connected.
\end{lemma}
We denote by $G^*$ the graph  $G - e^*$. Since $e^*$ is by definition, deletable in $G$, it follows that $G^*$ is a $\lambda$-connected digraph. Furthermore, for the fixed edge $e^*$, we denote by $\Z(e^*)$ a subset $\{e_1,\dots, e_k\}$ of $\D(e^*)$ which has the property that for any $\lambda$-cut $\cut{X}$ in $G^*$ that separates the pair $\{u^*,v^*\}$, the intersection of the edges of this cut with $\Z(e^*)$ is at most 1. We note that the fact that such a set exists is non-trivial and requires a proof.
For every $j\in [k]$, we let $e_j=(u_j,v_j)\in \Z(e^*)$. Finally, for every $i \in [k]$, we denote by $\Z_i$ the set $\{e_1, e_2, \ldots, e_i\} \subseteq \Z(e^*)$ and by $G^*_i$ the subgraph $G^*-Z_i$. Note that $\Z_{k} = \Z(e^*)$.

 In order to prove Lemma \ref{lemma:short-lambda-dir:alternate solution}, we prove that the digraph $G - \Z(e^*)$ is $\lambda$-connected. Since $|\Z(e^*)|=k$ by definition, Lemma \ref{lemma:short-lambda-dir:alternate solution} follows. 
 Hence, it remains to prove that $G - \Z(e^*)$ is $\lambda$-connected.
%

\begin{definition}

A cut $\cut{X}$ in $G^*_i$ {\rm (}for any $i\in [k]${\rm )} is called a cut of {\bf Type 1} if it separates the ordered pair $\{ u^*, v^* \}$ and a cut of {\bf Type 2} otherwise. We call $\cut{X}$ a {\bf violating cut} if $\cut{X}$ is a cut of Type 1 and $\delta_{G^*_i}(X) \leq\lambda - 2$ or $\cut{X}$ is a cut of Type 2 and $\delta_{G^*_i}(X) \leq\lambda - 1$.
\end{definition}




We now prove a lemma that shows that for any $i\in [k]$ and in particular, for $i=k$, the digraph $G_i^*$ excludes violating cuts. For this, we first exclude the possibility of violating cuts of Type 1 and then use the structure guaranteed by this conclusion to argue the exclusion of violating cuts of Type 2 (Lemma \ref{lemma:short-lambda-dir:alternate solution}). Using Lemma \ref{lemma:short-lambda-dir:G_i cut properties_type1},  we obtain Lemma \ref{lemma:short-lambda-dir:alternate solution} which provides a way to compute a deletion set from $\D(e^*)$. We will then use this lemma to prove Lemma~\ref{lemma:lambda-dir:upper bound}.

\begin{lemma}\label{lemma:short-lambda-dir:G_i cut properties_type1}\label{lemma:short-lambda-dir:G_i cut properties_type2}
For every $i\in [k]$, the digraph $G^*_i$
has no violating cuts.
%
\end{lemma}

\paragraph{Proof of Lemma~\ref{lemma:short-lambda-dir:alternate solution}.}
We define the set $\cal Z$ in the statement of the lemma to be the set ${\cal Z}(e^*)={\cal Z}_k$. In order to prove that $\cal Z$ satisfies the required properties, we need to argue that $G'=G-{\cal Z}$ remains $\lambda$-connected. If this were not the case then there is a cut $\cut{X}$ in $G'$ such that $\delta_{G'}(X)\leq \lambda-1$. We now consider the following cases. In the first case, $X$ is crossed by the edge $(u^*,v^*)$. In this case, it follows that $X$ is a cut of Type 1 in $G_k^*$ and furthermore, $\delta_{G_k^*}(X)=\delta_{G'}(X)-1\leq \lambda-2$. But this implies the presence of a violating cut of Type 1 in $G_k^*$, a contradiction to Lemma \ref{lemma:short-lambda-dir:G_i cut properties_type1}. In the second case, $X$ is not crossed by the edge $(u^*,v^*)$. In this case, it follows that $X$ is a cut of Type 2 in $G_k^*$ and $\delta_{G_k^*}(X)=\delta_{G'}(X)\leq \lambda-1$. But this implies the presence of a violating cut of Type 2 in $G_k^*$, which is again a contradiction to Lemma \ref{lemma:short-lambda-dir:G_i cut properties_type2}. Hence, we conclude that $\cal Z$ indeed satisfies the required properties. This completes the proof of the lemma. 
\qed



\paragraph{Proof of Lemma~\ref{lemma:lambda-dir:upper bound}.}
    Let $F = \{ f_1, f_2, \ldots, f_p\}$ be an arbitrary maximal set of edges such that $G - F$ is $\lambda$-connected.
    If $|F| = p \geq k$, then we already have the required deletion set.
    Therefore, we may assume that $p\leq k-1$.
Now, consider the graphs $G_0,\dots, G_p$ with $G_0=G$ and $G_i$ defined as  $G_i = G - \{f_1, \ldots f_i\}$ for all $i\in [p]$.
    Note that $G_{i+1} = G_i - f_{i+1}$ and $G_p = G - F$.
    Observe that each $G_i$ is $\lambda$-connected, by the definition of $F$.
    Let $\D_i$ be the set of deletable edges in $G_i$ which are undeletable in $G_{i+1}$.
    Observe that $\D_i$ is the set of edges that turn undeletable when $f_i$ is deleted.

  Now consider any deletable edge of $G$.
    It is either contained in $F$, 
    or there is some $i\in \{0,\dots, p-1\}$ such that it is deletable in $G_i$ but undeletable in $G_{i+1}$.
    In other words, the set $F \cup \D_1 \cup \D_2 \ldots \cup \D_p$ covers all the deletable edges of $G$. Since $p \leq k-1$ and the number of deletable edges in $G$ is at least $k^2\lambda$, it follows that for some $i\in [p]$, the set $\D_i$ has size at least $k\cdot \lambda$.
%
%
%
         Let $\Z_i$ be the set of at least $k$ edges corresponding to $\D_i$ guaranteed by Lemma~\ref{lemma:short-lambda-dir:alternate solution}. We know that $G_i - \Z_i$ is $\lambda$-connected.         
    Since $G_i$ is a subgraph of $G$ on the same set of vertices, it follows that $G - \Z_i$ is also $\lambda$-connected, which gives us a required deletion set.
%
%
\qed

%
%
%
%

\subsection{Undirected Graphs}
\label{sec:short-undirected}

Now we outline the results for $\lambda$-connected undirected graphs, as presented in Section~\ref{sec:undirected}. As mentioned earlier, we need to handle even-connectivity and odd-connectivity separately. When $\lambda$ is even, we closely follow the strategy used for digraphs, albeit with a more involved analysis, and we refer the reader to Section~\ref{sec:undirected} for details.


When $\lambda$ is an odd number, it is possible that the number of deletable edges is unbounded in $k$ in spite of the presence of a deletion set of size $k$. Indeed, recall the following example. Let $G$ be a cycle on $n$ vertices, $\lambda=1$ and $k=2$. Clearly, every edge in $G$ is deletable, but there is no deletion set of cardinality $2$.
 In order to overcome this obstacle, we design a subroutine that either find a required deletion set, or detects an edge which is disjoint from \emph{some} deletion set of cardinality $k$ in the graph. Before we formally state the corresponding lemma, we additionally define a subset of irrelevant edges
 and a deletion set is now defined to be a subset $F$ of $E(G)\setminus \R$ of size $k$ such that $G-F$ is $\lambda$-connected. Finally, we note that the set $\R$ contains all the undeletable edges of $G$. 

\mainodd*

We can then iteratively execute the algorithm of this lemma to either find a required deletion set or grow the set of irrelevant edges.
From now onward, we represent the input to our algorithm as $(G,k,\R)$,
and assume that $\lambda$ is an odd integer.
%
%
%
%
%
%
%
%
%
We begin by proving  the following lemma which says that if the graph admits a ``cycle-like'' decomposition, then certain deletable edges may be safely added to the set $\R$ without affecting the existence of a deletion set. 
\begin{lemma}\label{lemma:short-lambda-undir:odd irrelevent edge}
    Let $(G,k,\R)$ be an input, where $G$ is $\lambda$-connected,
    and let $X_1, X_2, \ldots X_{2k+2}$ be a partition of $V(G)$ into non-empty subsets  
    such that the following properties hold in the graph $G$.
    \begin{enumerate}
\setlength{\itemsep}{-2pt}
        \item $\delta_G(X_1, X_2) = \delta_G(X_2,X_3) \ldots = \delta_G(X_{2k+2}, X_1) = \frac{\lambda + 1}{2}$.
        
        \item Every edge of the graph either has both endpoints in some $X_i$ for $i\in [2k+2]$,
        or contained in one of the edge sets mentioned above.
        
        \item There are deletable edges $e_1, e_2, \ldots, e_{2k+2}$ in $E(G) \setminus \R$ such that
        $e_i \in \altdelta(X_i, X_{i+1})$ for $i \in [2k+2]$.
        (Here $X_{2k+3}$ denotes the set $X_1$.)
        
    \end{enumerate}
    Then $(G,k,\R)$ has a deletion set of cardinality $k$ if and only if $(G,k,\R \cup \{ e_1 \}$ has a deletion set of cardinality $k$.

\end{lemma}
Next, we set up some notation which will be used in subsequent lemmas. 
Let $S^*$ denote a fixed subset of $E(G)\setminus \R$ of at most $k-1$ edges such that the graph $G_{S^*}=G - S^*$ is $\lambda$-connected. We let $e^*\notin \R$ denote a deletable edge in $G_{S^*}$ such that $\D(e^*)=(\del(G_{S^*})\cap \undel(G_{S^*}- \{e^*\}))\setminus \R$ has at least $\eta \lambda$ edges where $\eta = 3k(2k+3) + 1$.
We denote by $G^*$ the graph $G_{S^*}-\{e^*\}$. 
Let $\Z(e^*)=\{e_1,\dots, e_\eta\}$ be a collection of edges in $\D(e^*)$ as before in the case of directed graphs. Furthermore, let $\C(e^*)=\{C_1,\dots, C_{\eta}\}$ be a collection of $\eta$ $\lambda$-cuts in $G^*$ such that, for each $e_i \in \Z(e^*)$ there is a unique cut $C_i \in {\C(e^*)}$ which separates the endpoints of $e_i$ and, for every $i\in [\eta-1]$, $C_i\subset C_{i+1}$. Again, the existence of such a collection requires a proof, which may be found in Section~\ref{sec:undirected}.
Furthermore, we may assume that both these collections are known to us. We remark that \emph{computing} these collections was not particularly important in the case of digraphs or the case of even $\lambda$ in undirected graphs. This is because the main structural lemmas we proved were only required to be \emph{existential}. 
However, in the odd case, it is crucial that we are able to \emph{compute} these collections when given the graph $G_{S^*}$ and the edge $e^*$. 
For every $i\in [\eta]$, we let $(u_i,v_i)$ denote the endpoints of the edge $e_i$.

Let $\widehat{\Z}= \{e_{(2k+3)i + 1} \in \Z(e^*) \mid 0 \leq i \leq 3k \}$ and observe that $|\widehat{\Z}| = 3k + 1$.
Let $\widehat{\C}$ be the subcollection of $\C(e^*)$ corresponding to $\widehat{\Z}$. 
Let $\C$ be defined as the set  $\{C_i \in \widehat{\C} \mid (C_{i} \setminus C_{i-(2k+3)}) \cap V(S^*) = \emptyset\}$ 
where $V(S^*)$ denotes the set of endpoints of edges in $S^*$. 
Since $|S^*|\leq k-1$ at most $2(k-1)$ cuts of $\widehat{\C}$ are excluded from $\C$ and hence, $|\C| \geq k$.
Let $\Z$ be the subcollection of $\widehat{\Z}$ corresponding to $\C$.
For any $i\in [\eta]$ such that $e_i \in \Z$, we define $\Z_i = \{e_j \in \Z \,|\, j \leq i \}$ and $G^*_i = G^* - \Z_i$. 
From now onwards, whenever we talk about the set $Z_i$ and graph $G_i$, we assume that the corresponding edge $e_i\in \Z$ and hence these are well-defined.

\begin{definition}
	Let $i\in [\eta]$ such that $e_i \in \Z$.
	A cut $\cut{X}$ in $G^*_i$ {\rm (}for any $i\in [k]${\rm )} is called a cut of {\bf Type 1} if it separates the  pair $\{ u^*, v^* \}$ and a cut of {\bf Type 2} otherwise. We call $\cut{X}$ a {\bf violating cut} if $\cut{X}$ is a cut of Type 1 and $\delta_{G^*_i}(X) \leq\lambda - 2$ or $\cut{X}$ is a cut of Type 2 and $\delta_{G^*_i}(X) \leq\lambda - 1$.
\end{definition}

%
%
\noindent
As before, we have the following lemma for handling Type 1 cuts.\begin{lemma}\label{lemma:short-lambda-undir:odd type 1 violating cut}
    For any $i\in [\eta]$ such that $e_i\in \Z$, the graph $G_i^*$ has no violating cuts of Type 1.
\end{lemma}

To handle the violating cuts of Type 2, we define a violating triple $(X,i,j)$ and we prove several structural lemmas based on this definition.
\begin{definition}
    Let $i\in [\eta]$ such that $e_i \in \Z$. Let $\cut{X}$ be a violating cut of Type 2 in $G_i^*$ such that $u^*,v^*\notin X$, $e_i$ crosses $\cut{X}$ and $X$ is inclusion-wise minimal. Let $j<i$ be such that $e_j\in \Z$, $e_j$ crosses the cut $\cut{X}$ in $G^*$ and there is no $r$ such that $r$ satisfies these properties and $j<r<i$. Then we call the tuple $(X,i,j)$ a {\bf violating triple}.
\end{definition}

Observe that for any violating triple $(X,i,j)$, it holds that $j\leq i-(2k+3)$ and hence, 
there are cuts $C_j \subset C_{i-(2k+2)} \subset C_{i-(2k+1)} \ldots \subset C_{i-1}\subset C_{i}$ 
such that they are all $\lambda$-cuts in $G^*$ and all but $C_{j}$ and $C_{i}$ are $\lambda$-cuts in $G_i^*$ as well.
For the sake of convinience, let us rename these cuts as follows.
Let $C_j \subset C_{2k+2} \subset C_{2k+1} \ldots \subset C_1 \subset C_i$ denote the sets 
$C_j \subset C_{i-(2k+2)} \subset C_{i-(2k+2)} \ldots \subset C_{i-1} \subset C_i$ respectively, 
and let $\C_{ij}$ denote this ordered collection.
Additionally, we may refer to the cuts $C_0$ and $C_{2k+3}$,
which denote the cuts $C_i$ and $C_j$ respectively. The following lemma ties the existence of violating cuts of Type 2 to the existence of violating triples.

\begin{lemma}\label{lem:short-violating_triple_exists} Let $i\in [\eta]$ such that $e_i \in \Z$ and let $\cut{X}$ be a violating cut of Type 2 in $G_i^*$ such that $G_{i-1}^*$ has no such violating cut, $u^*,v^*\notin X$ and $X$ is inclusion-wise minimal. Then, there is a $j<i$ such that $(X,i,j)$ is a violating triple. Furthermore given $G,i,X$, we can compute $j$ in polynomial time. Finally,  the following properties hold with regards to the triple $(X,i,j)$.
       {\bf(1)} $\delta_{G^*}(X)\geq \lambda+1$,  
         {\bf(2)} $X\subseteq C_i \setminus C_j$, 
         {\bf(3)} $e_i$ and $e_j$ are the only edges of $\Z$ which cross the cut $\cut{X}$ in $G^*$, 
         {\bf(4)} $\delta_{G^*_i}(X)=\lambda-1$.
\end{lemma}

Let $C_a$ and $C_b$ be two consecutive cuts in $\C_{ij}$ such that $b = a+1$ and observe that $C_j = C_j \subset C_b \subset C_a \subset C_i = C_i$.
Let $X_1 = X \cap (C_i \setminus C_a)$, $X_2 = X \cap (C_a \setminus C_b)$ and $X_3 = C_b \setminus C_j$. We will show that these sets are non-empty and more interestingly, $X_2$ is in fact \emph{all} of $C_a \setminus C_b$.


%

\begin{lemma}\label{lem:short-define_X_intersection}
    Let $i\in [\eta]$ such that $e_i \in \Z$ and let $(X,i,j)$ be a violating triple. Let $X_1\uplus X_2\uplus X_3$ be the partition of $X$ as defined above.
    The sets $X_1,X_2,X_3$ are all non-empty and furthermore, $X_2=C_a \setminus C_b$.
\end{lemma}

Moving forward, when dealing with a violating triple $(X,i,j)$, we continue to use the notation defined earlier. That is, the sets $X_1,X_2, X_3$ are defined to be the intersections of $X$ with the sets $C_i\setminus C_a$, $C_a\setminus C_b$ and $C_b$ respectively with $X_2=C_a\setminus C_b$. 
 Furthermore, we may assume 
that $\delta_{G^*_i}(X_1)=\delta_{G^*_i}(X_3)=\delta_{G^*_i}(X_1\cup X_2)=\delta_{G^*_i}(X_3\cup X_2)=\lambda$. 
This is justified by the proof of the above lemma.
inally, $\delta_{G^*_i}(X_2)=\lambda+1$.

Recall that our main objective in the rest of the section is to show that the sets $X_1,X_2,X_3$ satisfy the premises of Lemma \ref{lemma:short-lambda-undir:odd irrelevent edge}. 
For this, we begin by showing that these sets satisfy similar properties with respect to the graph $G^*$ instead of the graph $G$ (which is what is required for Lemma \ref{lemma:short-lambda-undir:odd irrelevent edge}). Following this, we show how to `lift' the required properties to the graph $G$ (Lemma \ref{lemma:short-lambda-undir:odd irrelevent edge_premise1_lift1}), which will allow us to satisfy the premises of  Lemma \ref{lemma:short-lambda-undir:odd irrelevent edge}.


\begin{lemma}\label{lemma:short-lambda-undir:odd irrelevent edge_premise1}
    Let $i\in [\eta]$ such that $e_i \in \Z$ and let $(X,i,j)$ be a violating triple. 
    Let $X_1\uplus X_2\uplus X_3$ be the partition of $X$ as defined above. Let $W = V(G) \setminus X$. 
    Then, $\delta_{G_i^*}(W, X_1) = \delta_{G_i^*}(X_3, W)= \frac{\lambda - 1}{2}$, $\delta_{G_i^*}(X_1,X_2) = \delta_{G_i^*}(X_2,X_3)  = \frac{\lambda + 1}{2}$. Furthermore, $\delta_{G_i^*}(X_2,W)=\delta_{G_i^*}(X_1,X_3)=0$.
\end{lemma}

\begin{lemma} \label{lemma:short-lambda-undir:odd irrelevent edge_premise1_lift1}
    Let $i\in [\eta]$ such that $e_i \in \Z$ and let $(X,i,j)$ be a violating triple. 
    Let $X_1\uplus X_2\uplus X_3$ be the partition of $X$ as defined above. Let $W=V(G)\setminus X$. 
    Then, $\delta_{G}(W, X_1) = \delta_{G}(X_1,X_2) = \delta_{G}(X_2,X_3)  = \frac{\lambda + 1}{2}$. Furthermore, $\delta_{G}(X_2,W)=\delta_{G}(X_1,X_3)=0$.
\end{lemma}

\myparagraph{Proof of Lemma \ref{lem:main_odd}.}  Let $F = \{ f_1, f_2, \ldots, f_p\}$ be an arbitrary maximal set of edges disjoint from $\R$ such that $G - F$ is $\lambda$-connected.
If $|F| = p \geq k$, then we already have the required deletion set in $G$.
Therefore, we may assume that $p\leq k-1$.

Now, consider the graphs $G_0,\dots, G_p$ with $G_0=G$ and $G_i$ defined as  $G_i = G - \{f_1, \ldots f_i\}$ for all $i\in [p]$.
Note that $G_{i+1} = G_i - f_{i+1}$ and $G_p = G - F$.
Observe that each $G_i$ is $\lambda$-connected by the definition of $F$.
Let $\D_i$ be the set of deletable edges in $G_i$ which are undeletable in $G_{i+1}$.
Observe that $\D_i=\D(f_i)$ in the graph $G_i$.

Now consider any deletable edge of $G$.
It is either contained in $F$, 
or there is some $r\in \{0,\dots, p-1\}$ such that it is deletable in $G_i$ but undeletable in $G_{r+1}$.
In other words, the set $F \cup \D_1 \cup \D_2 \ldots \cup \D_p$ covers all the deletable edges of $G$. Since $p \leq k-1$ and the number of deletable edges in $G$ is greater than $\eta\lambda$, it follows that for some $r\in [p]$, the set $\D_r$ has size more than $\eta\lambda$. We fix one such $r\in [p]$ and if $r>1$, then we define $S^*=\{f_1,\dots, f_{r-1}\}$ and $S^*=\emptyset$ otherwise. We define $e^*=e_r$.

We then construct the sets $\Z(e^*), \C(e^*), \widehat \Z, \widehat \C$.
Then we construct the  cut-collection $\C$ and the corresponding edge set $\Z$
by using the set $S^*$. Recall that $\Z$ contains at least $k$ edges and is by definition disjoint from $\R$. Consider the graph $G^*=G-S^*\cup \{e^*\}=G-\{f_1,\dots, f_r\}$ and note that $G^*$ is $\lambda$-connected.

We check whether $G^*-\Z$ is $\lambda$-connected. If so, then we are done since $\Z$ is a deletion set for the graph. Otherwise, we know that $G^*-\Z$ contains a violating cut. Lemma \ref{lemma:short-lambda-undir:odd type 1 violating cut} implies that such a violating cut cannot be of Type 1. Hence, we compute in polynomial time (using Lemma \ref{lem:short-violating_triple_exists}) a violating triple $(X,i,j)$ in the graph $G_i^*$ for some $i\in [\eta]$. We now invoke Lemma \ref{lemma:short-lambda-undir:odd irrelevent edge} with the resulting decomposition  to compute an irrelevant edge $e\in E(G)\setminus \R$ in polynomial time and return it. 
 This completes the proof of the lemma. \qed

\vspace{1em}
\noindent
We now proceed to give a full description of our results.
\section{Preliminaries}
\label{sec:prelim}

For a finite set $V$, $2^V$ denotes the collection of all subsets of $V$.

\paragraph{Graphs and Digraphs.}
A {\em graph} $G$ consists of a set of $n$ {\em vertices} $V(G) = \{v_1, v_2, \ldots, v_n\}$ and a set of $m$ {\em undirected edges} $E(G) \subseteq V(G) \times V(G)$.
For any $v_i, v_j \in V(G)$, $(v_i,v_j)$ and $(v_j, v_i)$ denote the same edge,
and $v_i$ and $v_j$ are called {\em neighbours}.
The {\em degree} of a vertex $v$ is the size of the $N(v)$,
which denotes the set of all neighbours of $v$.
Similarly, a {\em digraph} $H$ consists of a set of $n$ {\em vertices} $V(H) = \{v_1, v_2, \ldots, v_n\}$ and a set of $m$ {\em directed edges} $E(H) \subseteq V(H) \times V(H)$.
For an edge $e = (v_i, v_j) \in E(H)$ we say that the edge is directed from $v_i$ to $v_j$, and $v_i$ and $v_j$ are called the {\em tail} and the {\em head} of $e$ respectively.
For a vertex $v$, an edge $e$ is called an {\em out-edge} of $v$ if $v$ is the tail of $e$, 
and it is called an {\em in-edge} of $v$ if $v$ is the head of $e$.  A vertex $u$ of $H$ is an {\em in-neighbor} ({\em out-neighbor}) of a vertex $v$ if $(u,v)\in E(H)$ ($(v,u)\in E(H)$, respectively). The {\em in-degree} $d^-(v)$ ({\em out-degree}
$d^+(v)$) of a vertex $v$ is the number of its in-neighbors
(out-neighbors). We denote the set of in-neighbors and out-neighbors of a vertex $v$  by $N^{-}(v)$  and $N^{+}(v)$ correspondingly.  A digraph $H$ is
{\em strong} if for every pair $x,y$ of vertices there
are directed paths from $x$ to $y$ and from $y$ to $x.$ A maximal
strongly connected subdigraph of $H$ is called a {\em strong
component}. A {\em walk} $W$ in $H$ consists of a sequence of edges $ \{e_1, e_2, \ldots, e_\ell\} \subseteq E(H)$, such that for two consecutive edges $e_i$ and $e_{i+1}$ in the walk, the head of $e_i$ is the same as the tail of $e_{i+1}$.
We say that a walk $W$ visits a vertex $v$ if $W$ contains an edge incident on $v$.
A walk is called a closed walk if the tail of $e_1$ and the head of $e_\ell$ are the same vertex.
Observe that if $W$ is not a closed walk, then the tail of $e_1$ and the head of $e_\ell$ have
exactly one out-arc and one in-arc incident on them, respectively.
The tail of $e_1$ is called the start vertex of $W$, and the head of $e_\ell$ is called the end vertex of $W$.
All other vertices visited by $W$ are called internal vertices, and for any internal vertex $v$
there is at least one in-arc of $v$ and at least one out arc of $v$ which is present in $W$.
A path $P$ in $D$ is a walk which visits any vertex at most once, i.e.
there are at most two arcs in $P$ which are incident on any vertex visited by $W$.
Observe that any edge occurs at most once in a path $P$ and this induces an ordering of these edges.
We say that $P$ visits these edges in that order.
Similarly, for the collection of vertices which are present in $P$, 
$P$ induces ordering of these vertices and we say that $P$ visits them in that order.
Let $P$ be a path which visita a vertex $u$ and then visits a vertex $v$.
We write $P[u,v]$ to denote the sub-path of $P$ which starts from $u$ and ends at $v$.
For two path $P$ and $Q$ such that the end vertex of $P$ is same as the start vertex of $Q$,
we write $P + Q$ to denote the walk from the start vertex of $P$ to the end vertex of $Q$.

Let $G$ be a graph or a digraph.
For a collection of edges $F \subseteq E(G)$, we use $G -F$ to denote
the subgraph obtained from $G$ by removing the edges in $F$ from $E(G)$.
If $F$ contains only a single edge $e$, then we simply write $G - e$.
For an introduction to graph theory and directed graphs we refer to the textbooks of Diestel~\cite{diestel2012graph} and Bang-Jensen and Gutin~\cite{bang2008digraphs}. Let $G$ be a digraph. A  {\em subdivision} of an 
edge $e=(u,v)$ of $G$  yields a new digraph, $G'$, containing one new vertex $w$, and with an edge set 
replacing $e$ by two new edges, $(u,w)$ and $(w,v)$. That is, $V(G')=V(G)\cup \{w\}$ and $E(G')=(E(G)\setminus \{(u,v)\}) \cup \{(u,w),(w,v)\}$. 

\paragraph{Cuts in a graph.}
For a subset $X$ of a set $V$, $\co{X}$ denotes the set $V \setminus X$.
Let $X$ and $Y$ be subsets of a set $V$.
We say that $X$ and $Y$ {\em cross} in $V$ if all of $X \cap Y, \co{X} \cap \co{Y}, X \setminus Y, Y \setminus X$ are non-empty.
Otherwise we say that $X$ and $Y$ are {\em uncrossed}.
Observe that if $X$ and $Y$ cross in $V$, then $X \cup Y$ is a proper subset of $V$.

Let $G$ be a graph or a digraph.
A {\em cut} $(X,Y)$ is a ordered partiton of $V(G)$.
Therefore, for any subset $X$ of $V(G)$ we have a cut $\cut{X}$.
We also use the term ``the cut $X$'' to denote $\cut{X}$.
In undirected graphs $(X,\co{X})$ and $(\co{X},X)$ denote the same cut.
We say that $\cut{X}$ {\em separates a pair of vertices} $\{u,v\}$ if exactly one of these vertices is in $X$.
We say that an edge $e=(u,v)$ {\em crosses} $\cut{X}$, if the cut separates $\{u, v\}$.
In directed graphs we distinguish between the cuts $(X,\co{X})$ and $(\co{X},X)$.
We say that $\cut{X}$ {\em separates an ordered pair of vertices} $\{u,v\}$ only if $u \in X, v \in \co{X}$.
We say that an edge $e=(u,v)$ crosses $\cut{X}$, if the cut seperates the ordered pair $\{u, v\}$.
For a subgraph $H$ of $G$ and a cut $\cut{X}$, we define $\altdelta_{H}(X)$ 
as the set of edges in $H$ which cross this cut.
%
We use $\delta_{H}(X)$ to denote the \emph{number} of edges in $H$ 
which cross this cut, that is $|\altdelta_H(X)|$.
%
We also say that an edge {\em $e$ is part of the cut $\cut{X}$} if $e$ crosses $\cut{X}$.
For a number $\lambda$ and a graph or a digraph $H$, we say that $\cut{X}$ is a $\lambda$-cut in $H$
if $\delta_{H}(X) = \lambda$.
Often, when the graph is clear from the context, we shall skip the subscript and write $\delta(X)$.
We say that two cuts $\cut{X}$ and $\cut{Y}$ cross, if the sets $X$ and $Y$ cross in $V$.
Otherwise these cuts are uncrossed.

The key tool in our arguments is the submodularity of graph cuts.
A function $f:2^V \rightarrow \mathbb{R}$, where $V$ is a finite set,
is called \emph{submodular} if for any $X, Y \subseteq V$ the following
holds.
$$f(X \cap Y) + f(X \cup Y) \leq f(X) + f(Y)$$

It is well known that graph cuts are submodular~\cite{frank2011connections, nagamochi2008algorithmic}.
\begin{proposition}\label{lemma:submod}
	Let $G$ be a (di)graph.
	Let $\cut{X}$ and $\cut{Y}$ be two cuts in $G$.
	Then $\delta(X \cap Y) + \delta(X \cup Y) \leq \delta(X) + \delta(Y)$.
	Furthermore, if $e \in \delta(X \cap Y) \cup \delta(X \cup Y)$, then $e \in \delta(X) \cup \delta(Y)$.
\end{proposition}
And using the submodularity of cuts we can obtain the following well known result. It implies that the $\lambda$-cuts in a $\lambda$-connected graph where $\lambda$ is \emph{odd}, form a laminar family.
\begin{proposition}\label{prop:uncrossing}
	Let $\lambda\in {\mathbb N}$ be odd and let $G$ be a $\lambda$-connected graph. 
	Let $\cut{X}$ and $\cut{Y}$ be two $\lambda$-cuts in $G$ such that $X \cup Y \neq V(G)$.
	Then, $\cut{X}$ and $\cut{Y}$ do not cross and we have that, either $X \subseteq Y$ or $Y \subseteq X$.
\end{proposition}

\subsection{Structural properties of $\lambda$-connected graphs}
       In this part, we begin by recalling some elementary structural results regarding connectivity in graphs and digraphs. Following that, we state and prove the properties that will be required in the description as well as proof of correctness of our algorithms.

\begin{definition}
	A connected undirected graph $G$ is {\bf $\lambda$ edge-connected} if deleting any set of $\lambda - 1$ or fewer edges
	leaves the resulting graph connected. Equivalently, an undirected graph $G$ is $\lambda$ edge-connected 
	if there are $\lambda$ 	edge-disjoint paths between every pair of vertices in $G$.

	A strongly connected digraph $G$ is {\bf $\lambda$ edge-connected} if deleting any set of $\lambda - 1$ or fewer edges
	leaves the resulting graph strongly connected. Equivalently, a digraph $G$ is $\lambda$ edge-connected if for any pair of 
	vertices $u$ and $v$ in $G$, there are $\lambda$ edge-disjoint paths from $u$ to $v$.
\end{definition}

Since we are interested in only the edge-connectivity of graphs, we will refer to $\lambda$ edge-connected graphs/digraphs
 simply as $\lambda$-connected graphs/digraphs. 
An immediate consequence of the definition of $\lambda$-connectivity in undirected graphs is that every vertex 
in $G$ has degree at least $\lambda$. 
And for digraphs, every vertex must have both in-degree and out-degree at least $\lambda$.
We now formally define a notion of  \emph{deletable} and \emph{undeletable} edges in a given (di)graph $G$.

\begin{definition}
	Let $G$ be a (di)graph and $\lambda\in {\mathbb N}$ such that $G$ is $\lambda$-connected.
	Then, an edge $e \in E(G)$ is called \textbf{deletable} if $G - e$ is a $\lambda$-connected (di)graph.
	Otherwise $e$ is an \textbf{undeletable} edge in $G$.
	We denote by $\del(G)$ the set of deletable edges in $G$, and by $\undel(G)$ the set of undeletable edges in $G$.
\end{definition}

Observe that any deletion set in the graph is a subset of the deletable edges.
For the weighted version of the problem, we will often focus on a subset $W$ of the edges in the graph,
and we will only be interested in those deletion sets in the graph that are subsets of $W$.
In such cases, we define the set $\del(G)$ as those deletable edges of the graph $G$ which are also present in $W$,
and say that \emph{$\del(G)$ is restricted to $W$}.
This modification also carries over to all the subsequent results and definitions,
and we implicitly assume that $\del(G)$ is restricted to $W$. 

\begin{definition}\label{def:special_edges}
	Let $G$ be a (di)graph and $\lambda\in {\mathbb N}$ such that $G$ is $\lambda$-connected, and let $e^*=(u^*,v^*)\in E(G)$ be a deletable edge. 
	We denote by $\D(e^*)$ the set $\del(G)\cap \undel(G - e^*)$.
\end{definition}

\begin{observation}\label{obs:newly_undeletable_cross}
Let $G$ be a (di)graph and $\lambda\in {\mathbb N}$ such that $G$ is $\lambda$-connected.  If $e$ is a deletable edge in $G$ then it does not cross \emph{any} $\lambda$-cut in $G$,
and if $e$ is undeletable then it \emph{must} cross some $\lambda$-cut in $G$.
\end{observation}


%
%

\begin{lemma}\label{lemma:lambda:cuts with D(e*) edges}
	Let $G$ be a (di)graph and $\lambda\in {\mathbb N}$ such that $G$ is $\lambda$-connected.
	Let $e^*=(u^*, v^*)$ be a deletable edge in $G$, and let $G^* = G - e^*$.
	Let $\cut{X}$ be $\lambda$-cut in $G^* $ crossed by $e=(u,v) \in \D(e^*)$.
	Then, $\cut{X}$ is also crossed by the edge $e^*$ in $G$.
\end{lemma}
\begin{proof}	 
	We only argue the case when  $G$ is a digraph.
	The arguments for the case when $G$ is an undirected graph are similar.
	Since $e$ is deletable in $G$, it cannot be the case that $\cut{X}$ is a $\lambda$-cut in $G$. Since $\cut{X}$ is a $\lambda$ cut in $G^*=G-e^*$, it must be the case that $e^*$ also crossed $\cut{X}$ in $G$. This completes the proof of the lemma.
%
%
%
%
%
\end{proof}

From the set of edges $\D(e^*)$, we will compute a set of edges $\Z(e^*)$ with 
some very useful properties.

\begin{lemma} \label{lemma:lambda:special edges}\label{lem:lambda_special_edges}
Let $G$ be a (di)graph and $\lambda\in {\mathbb N}$ such that $G$ is $\lambda$-connected. Let $e^*=(u^*,v^*)$ be a deletable edge in $G$ such that $|\D(e^*)| \geq \ell \lambda$ for some $\ell\in {\mathbb N}$ and let $G^*=G-e^*$. Then there is a set $\Z'(e^*)\subseteq \D(e^*)$  such that,
\begin{itemize}
\setlength{\itemsep}{-2pt}
	\item $|\Z'(e^*)| = \ell$ and
	\item $| \delta_{G^*}(X) \cap \Z'(e^*) | \leq 1$	for any $\lambda$-cut $(X,\overline{X})$ in $G^*$ which separates the (ordered) pair $\{u^*,v^*\}$.
\end{itemize}

Further, there is an algorithm that, given $G,e^*$ and $\ell$, runs in time $\bigoh(\lambda(m+n))$ and computes the set $\Z'(e^*)$, where $m$ and $n$ are the number of edges and vertices in $G$ respectively.
\end{lemma}
\begin{proof} 
	We only prove the lemma for the case when $G$ is a digraph.
	The arguments for the case when $G$ is an undirected graph are similar.
	Since, $G^*$ is a $\lambda$-connected digraph, there are $\lambda$ edge disjoint paths 
	from $u^*$ to $v^*$ in $G^*$.
	Let ${\cal P} = \{ P_1, P_2, \ldots, P_\lambda \}$ be such a collection of paths. Note that $\cal P$ can be computed in time $\bigoh(\lambda(m+n))$ 
    via several well known algorithms such as the
    Ford-Fulkerson algorithm or the Edmonds-Karp algorithm (for details, see e.g., \cite{nagamochi2008algorithmic}).
	Now, by Observation \ref{obs:newly_undeletable_cross},  every edge $e \in \D(e^*)$ crosses a $\lambda$-cut which separates the ordered pair $\{u^*, v^*\}$,
	and therefore $e$ is contained in exactly one of these paths.
	Let $P_i$ be the path such that $E(P_i) \cap \D(e^*)$ is maximized, and let $Z_i = E(P_i) \cap \D(e^*)$.
	Observe that $|Z_i| \geq \frac{|\D(e^*)|}{\lambda} \geq \ell$ and we order the edges of $\Z'(e^*)$ as per the order they occur in the 
	path from $u^*$ to $v^*$.
	
	We define $\Z'(e^*)$ to be first $\ell$ edges of the ordered set $Z_i$.
	It remains to prove that $\Z'(e^*)$ satisfies the claimed properties. 
	By definition, it holds that $|\Z'(e^*)| = \ell$. 
	Now, let $\cut{X}$ be a $\lambda$-cut separating the pair $\{u^*, v^*\}$.
	Suppose that $|\altdelta_{G^*}(X)\cap \Z'(e^*)|\geq 2$. Recall that $\Z'(e^*)$ is a subset of the edges in the path $P_i\in {\cal P}$. Since $\cal P$ contains exactly $\lambda$ paths, it must be the case that $\altdelta_{G^*}(X)$ is disjoint from at least one of these paths. But this contradicts our assumption that $\cut{X}$ separates the pair $\{u^*,v^*\}$.
%
%
%
This completes the proof of this lemma.
	\end{proof}
	
The following lemma gives us a crucial subroutine that
is used in computing a deletion set in the graph in the directed case

\begin{lemma}\label{lemma:lambda:special cuts 1}\label{lem:lambda_special_cuts_1}
There is an algorithm that, given $G$, $\lambda$, $\ell$, $e^*=(u^*,v^*)$ and $\Z'(e^*)$  as in the statement of Lemma \ref{lemma:lambda:special edges}, runs
in polynomial time
and computes an ordered collection of edges $\Z(e^*) \subseteq \D(e^*)$, and
an ordered family of $\lambda$-cuts in $G^*=G-e^*$, $\C(e^*) = \{ C_1, C_2, \ldots, C_{\ell}\}$ 
with each cut separating the (ordered) pair $\{u^*, v^*\}$ such that the following statements hold.
	\begin{enumerate}
\setlength{\itemsep}{-2pt}
    	\item For every $i \in [\ell]$, $\altdelta_{G^*}(C_i) \cap \Z(e^*) = \{ e_i \}$ and $e_i \notin \delta_{G^*}(C_j)$ for $i \neq j$.
        And for any $\lambda$-cut $\cut{Y}$ which separates the ordered pair $u^*, v^*$, $|\Z(e^*) \cap \altdelta_{G^*}(Y)| \leq 1$.

    	\item For every $i \in [\ell - 1]$, $C_{i} \subset C_{i+1}$. 
    		   In particular for every $j < i$, both the endpoints of the edge $e_j$ lie in $C_i$.
%
    \end{enumerate}
\end{lemma}
\begin{proof}
	We assume that $G$ is a digraph. The proof for the case where $G$ is an undirected graph is similar.
	Let $\Z(e^*) = \Z'(e^*)$ and it will remain unchanged throughout.
    Observe that this satisfies the second part of the first property,
    which is guaranteed by Lemma~\ref{lemma:lambda:special edges}.
	Since $e_1, e_2, \ldots e_\ell \in \D(e^*)$, by definition, 
    there are subsets of vertices, $C_1, C_2, \ldots, C_\ell$, which define $\lambda$-cuts in $G^*$ such that,
	they separate the ordered pair $\{u^*, v^*\}$ and $e_i \in \altdelta_{G^*}(C_i)$ for every $i \in [\ell]$.
	Further, by Lemma~\ref{lemma:lambda:special edges} all these cuts are distinct.
	Observe that this collection of cuts satisfies the first property required by this lemma.
	Also note that, for every $i \in [\ell]$, $u^* \in C_i$ and $v^* \in \co{C_i}$.
 	Hence $\forall i,j \in [\ell]$, $C_i \cap C_j$ and $C_i \cup C_j$ are non-empty proper subsets of $V(G^*)$.	We now prove the following claim which allows us to uncross a pair of crossing cuts while preserving certain properties of the original cuts.
	

	\begin{claim}\label{clm:pre_iterative_uncrossing}
		Let $e_i=(u_i,v_i),e_j=(u_j,v_j)\in \Z(e^*)$ and $C_i,C_j$ be distinct $\lambda$-cuts in $G^*$ which separate $\{u_i,v_i\}$ and $\{u_j,v_j\}$ respectively.
		Then exist cuts $C_i'$ and $C_j'$ such that,
		\begin{itemize}
\setlength{\itemsep}{-2pt}
			\item $C_j' \subset C_i'$,
			\item amongst the edges $e_i$ and $e_j$, $C_j'$ separates exactly one of the the two edges,
			      while $C_i'$ separates only the other edge.
			\item $C_j' \subseteq C_j$, and $C_i' \supseteq C_i$.
		\end{itemize}
	\end{claim}
	\begin{proof}
		If $C_i$ and $C_j$ do not cross, then either $C_j \subset C_i$ in which case we are done by setting $C_j'=C_j$ and $C_i'=C_i$, 
		or $C_i \subset C_j$, in which case we are done by setting $C_i' = C_j$ and $C_j' = C_i$.
		It easy to see that they satisfy the required properties.
		
		Otherwise, the two cuts cross and by the submodularity of cuts (Proposition \ref{lemma:submod}) we have
		$\delta_{G^*}(C_i \cap C_j) + \delta_{G^*}(C_i \cup C_j) \leq \delta_{G^*}(C_i) + \delta_{G^*}(C_j)$.
		Furthermore, any edge crossing one of the two new cuts $C_i\cap C_j$ and $C_i\cup C_j$ must also cross one of the two original cuts $C_i$ and $C_j$,
		i.e. $\altdelta_{G^*}(C_i \cap C_j) \cup \altdelta_{G^*}(C_i \cup C_j) \subseteq \altdelta_{G^*}(C_j) \cup \altdelta_{G^*}(C_j)$.
		Let $C_j' = C_i \cap C_j$ and $C_i' = C_i \cup C_j$ and note that $C_j' \subset C_i'$, $C_j'\subseteq C_j$, and $C_i'\supseteq C_i$.
		Since, $G^*$ is a $\lambda$-connected graph and $C_i$ and $C_j$ formed $\lambda$-cuts in $G^*$ which
		separate the ordered pair $\{ u^*, v^* \}$,
		it must be the case that $C_i'$ and $C_j'$ are also $\lambda$-cuts in $G^*$ 
		and they also separate the ordered pair $\{ u^*, v^* \}$.
		Therefore, by Lemma~\ref{lemma:lambda:special edges}, one of $e_i$ and $e_j$
		crosses only $C_i'$ and the other crosses only $C_j'$.
		This concludes the proof of this claim.
	\end{proof}
	
	Now we argue that starting from $C_1$, we can iteratively use this claim to uncross each $C_j$ from all $C_i$ for $i > j$.

	\begin{claim}\label{clm:iterative_uncrossing}
		Let $\C=\{C_1, C_2, \ldots, C_\ell\}$ be a collection of cuts such that for every $i\in [\ell]$, 
		there is a unique edge  $e_i\in\Z(e)$ which crosses the cut $C_i$. Furthermore, suppose that for some $j\in [\ell]$, we have $\forall p<j$, $\forall q>p$, the set $C_p\subset C_q$ (see Figure \ref{fig:partial_laminar}).
        
%

%
%
		Then, there is a  collection of cuts $\C'=\{C_1', C_2', \ldots, C_\ell'\}$ such that for every $i\in [\ell]$, there is a unique edge $e_i\in \Z(e)$ which crosses the cut $C'_i$ 
        and $\forall p \leq  j$, $\forall q>p$, the set  $C_p \subset C_q$.
         Furthermore, given $\C$ the collection $\C'$ can be computed in polynomial time.
	\end{claim}

	\begin{proof}	
	Consider the cut $C_j$. If $C_j\subset C_q$ for every $q>j$, then we are done by setting $C'_i=C_i$ for every $i\in[\ell]$. Therefore, we may assume that $C_j$ crosses $C_q$ for some $q>j$.   We now invoke Claim \ref{clm:pre_iterative_uncrossing} on the pair $C_j$ and $C_q$ to obtain $C_j'$ and $C_q'$ with the stated properties. We now redefine the collection $\C$ as $\C=\{C_1,\dots, C_{j-1},C_j \leftarrow C'_j,C_{j+1},\dots,C_{q-1},C_q \leftarrow C'_q,\dots, C_\ell\}$. Observe that due to Claim \ref{clm:pre_iterative_uncrossing}, $\C$ still satisfies all the properties mentioned in the premise of the lemma. Furthermore, the size of the set $C_j$ has now strictly decreased. Hence, after a finite number of iterations of this step, we will reach a collection where $C_j\subset C_q$ for every $q>j$, completing the proof of existence of the collection $\C'$. It is clear from the description of this iterative process that given $\C$, the collection $\C'$ can be computed in polynomial time. 
    This completes the proof of the claim.
%
%
%
%
%
%
%
%
%
%
%
%
%
%
	\end{proof}
	
	We now return to the proof of the lemma and argue that the collection $C_1, C_2, \ldots, C_\ell$ mentioned in the statement of the lemma can be computed 
    as follows.
	\begin{figure}[t]
	
	\begin{center}

  \includegraphics{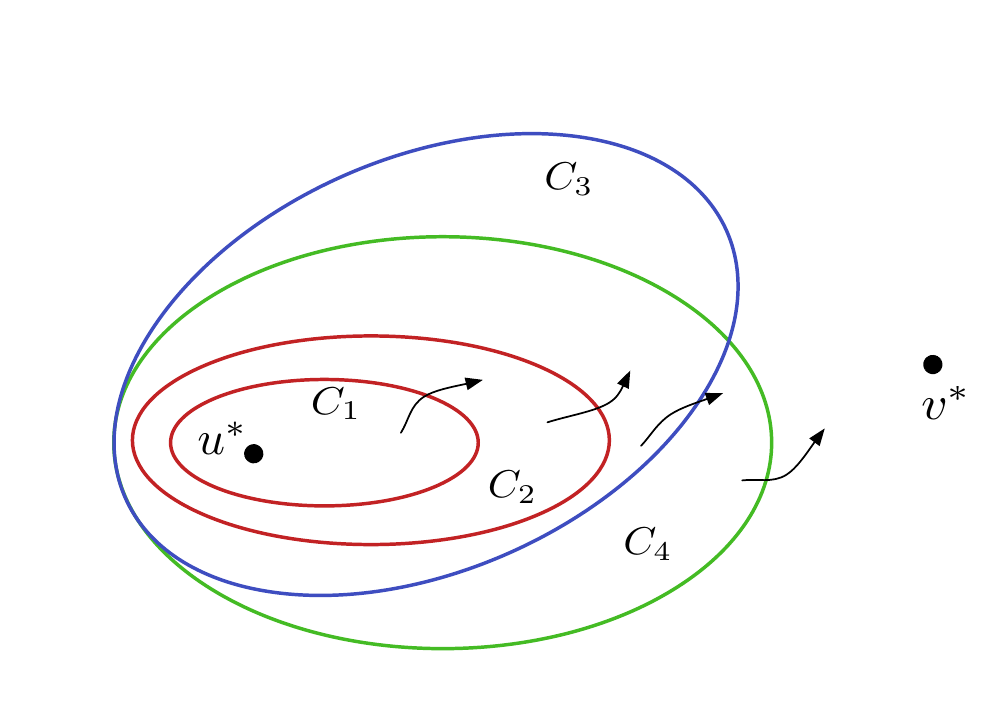}
  	\end{center}
  \caption{An illustration of the sets in the collection $\C$ defined in the statement of Claim \ref{clm:iterative_uncrossing}.}
  \label{fig:partial_laminar}
\end{figure}
	For each $i\in [\ell]$ and $e_i=(u_i, v_i)$, we find an arbitrary cut $C_i$ between the sets $\{u^*, u_i\}$ and $\{v^*, v_i\}$.	We now start from the initial collection  $\C=\{C_1, C_2, \ldots, C_\ell\}$
	and repeatedly invoke Claim \ref{clm:iterative_uncrossing} until we obtain a collection of cuts which satisfies 
	the second property of this lemma. We define this collection of cuts to be the set $\C(e^*)$.
	Each invocation of the above claim, reduces by one, the number of pairs of cuts which 
	violate the second property, while preserving the first property.
	Therefore after $\ell(\ell -1)$ executions of the algorithm of Claim \ref{clm:iterative_uncrossing}, we will obtain a collection of cuts which satisfies the second 
	property.
\end{proof}

\section{Directed Graphs}
\label{sec:directed}
In this section, we provide the details of our results on digraphs. Recall that the main lemma we require for our algorithm is Lemma~\ref{lemma:lambda-dir:upper bound}, which we restate for the sake of completeness.

\lambdadirmain*

\myparagraph{Setting up notation.} 
For the remainder of this section, we will deal with a \emph{fixed} deletable  edge $e^* = (u^*, v^*)$ in $G$ such that $\D(e^*)$ has at least $k \lambda$ edges. 
We also denote by $G^*$ the graph  $G - e^*$. Since $e^*$ is by definition, deletable in $G$, it follows that $G^*$ is a $\lambda$-connected digraph. Furthermore, for the fixed edge $e^*$, we denote by $\Z(e^*)$ the subset $\{e_1,\dots, e_k\}$ of $\D(e^*)$ guaranteed by Lemma \ref{lemma:lambda:special edges} and by $\C(e^*)$, the collection of cuts guaranteed by Lemma \ref{lemma:lambda:special cuts 1}. 
For every $j\in [k]$, we let $e_j=(u_j,v_j)\in \Z(e^*)$. Finally, for every $i \in [k]$, we denote by $\Z_i$ the set $\{e_1, e_2, \ldots, e_i\} \subseteq \Z(e^*)$ and by $G^*_i$ the subgraph $G^*-Z_i$. Note that $\Z_{k} = \Z(e^*)$.
We will prove that the digraph $G - \Z(e^*)$ is $\lambda$-connected.
%
But before we proceed to the formal proofs, we need a  final definition.

\begin{definition}

A cut $\cut{X}$ in $G^*_i$ {\rm (}for any $i\in [k]${\rm )} is called a cut of {\bf Type 1} if it separates the ordered pair $\{ u^*, v^* \}$ and a cut of {\bf Type 2} otherwise. We call $\cut{X}$ a {\bf violating cut} if $\cut{X}$ is a cut of Type 1 and $\delta_{G^*_i}(X) \leq\lambda - 2$ or $\cut{X}$ is a cut of Type 2 and $\delta_{G^*_i}(X) \leq\lambda - 1$.
\end{definition}




We now prove two lemmas that show that for any $i\in [k]$ and in particular, for $i=k$, the digraph $G_i^*$ excludes violating cuts. For this, we first exclude violating cuts of Type 1 and then use the structure guaranteed by this lemma to argue the exclusion of violating cuts of Type 2.

\begin{lemma}\label{lemma:lambda-dir:G_i cut properties_type1}
For every $i\in [k]$, the digraph $G^*_i$
has no violating cuts of Type 1.
\footnote{A shorter proof of this lemma can be obtained by using the characterization that, any $\lambda-1$ connected directed graph has a collection of $\lambda - 1$ arc disjoint spanning out-trees rooted at $u^*$. We would like to thank an anonymous reviewer for pointing out this fact.}
%
\end{lemma}

\begin{proof}
	Suppose that for some $i\in [k]$, the digraph $G^*_i$ has a violating cut of Type 1 and let $i$ be the \emph{least} integer for which this happens. Let $\cut{X}$ be a violating cut of Type 1 in $G^*_i$ such that $X$ is a set of minimum size. 
	
	We first observe that the cut $\cut{X}$ separates the ordered pair $\{u_i,v_i\}$ as well. Indeed, if this were not the case then $\cut{X}$ is also a violating cut of Type 1 in the graph $G_i^*+e_i$ which is precisely the graph $G_{i-1}^*$. However, this contradicts our choice of $i$ as the least integer in $[k]$ such that $G_i^*$ contains a violating cut of Type 1. Furthermore, recall that $G^*$ is $\lambda$-connected. Hence, it follows that $\delta_{G^*}(X)\geq \lambda$.

	We next observe that $i>1$. Suppose to the contrary that $i=1$. Recall that $e^*$ is deletable in $G$. This implies that $\delta_{G^*}(X)\geq \lambda$. Furthermore $G_1^*=G^*-e_1$, implying that $\delta_{G^*}(X)\geq \lambda-1$, a contradiction to $X$ being a violating cut of Type 1 in $G_1^*$. Hence we conclude that $i>1$. In fact, for the same reason, it must be the case that for \emph{some} $j\in [i-1]$, the cut $\cut{X}$ separates the ordered pair $\{u_j,v_j\}$. Moving forward, we choose $j$ to be the \emph{largest} integer less than $i$ such that the cut $\cut{X}$ separates the ordered pair $\{u_j,v_j\}$.
	
	  \begin{figure}[t]
	  \begin{center}
  \includegraphics{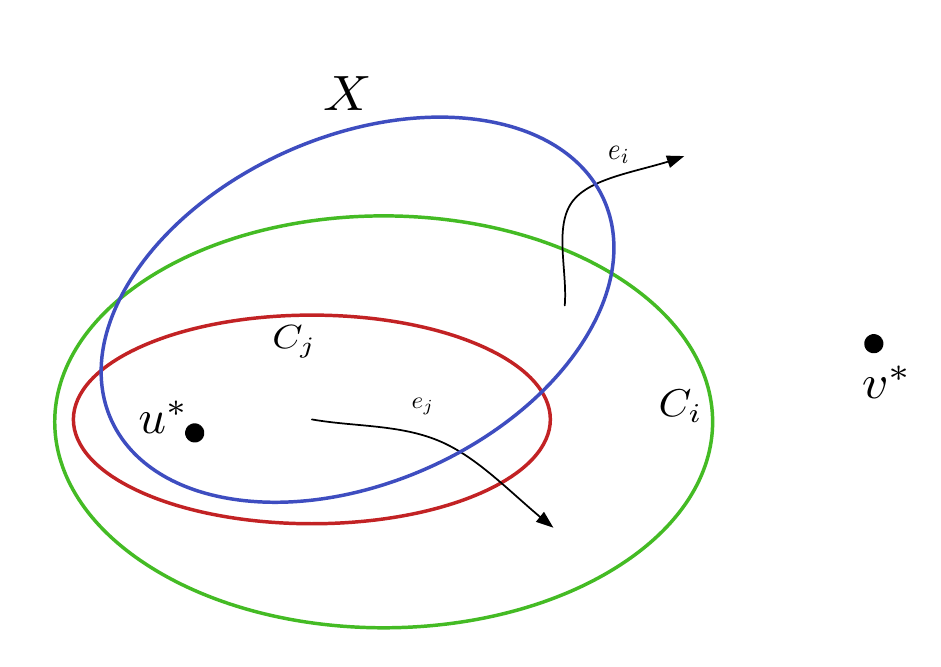}
  	  	\end{center}
  \caption{An illustration of the sets $X,C_i,C_j$ and the edges $e_i$ and $e_j$ in the proof of Lemma \ref{lemma:lambda-dir:G_i cut properties_type1}.}
  \label{fig:dir_type1}
\end{figure}

	  Recall that  Lemma \ref{lem:lambda_special_cuts_1} ensures that there are cuts $C_i,C_j\in \C(e^*)$ such that $e_i$ crosses $\cut{C_i}$, $e_j$ crosses $\cut{C_j}$, $e_i$ \emph{does not} cross $\cut{C_j}$ and $e_j$ does not cross $\cut{C_i}$. Furthermore, $C_j\subset C_i$ (see Figure \ref{fig:dir_type1}).

	  We will now inspect the sets $X,C_i,C_j$ and make a few observations regarding their `interaction'. Observe that $X\cap C_j$ contains $u^*$ and the complement of $X\cup C_j$ contains $v^*$. Hence, $X\cap C_j$ and $X\cup C_j$ are both cuts separating the ordered pair $\{u^*,v^*\}$. Furthermore, observe that $\delta_{G^*_i}(C_j)=\lambda-1$. This is because $C_j$ is a $\lambda$-cut in $G^*$ and $e_j$ is the \emph{only} edge in $\Z_i$ which crosses this cut. Invoking the submodularity of the cuts $C_j$ and $X$, we infer that 
	  
	  $$\delta_{G^*_i}(X\cap C_j)+\delta_{G^*_i}(X\cup C_j)\leq \delta_{G^*_i}(X)+\delta_{G^*_i}(C_j)\leq 2\lambda-3$$
	  
	  Hence, it must be the case that $\delta_{G^*_i}(X\cup C_j)\leq \lambda-2$ or $\delta_{G^*_i}(X\cap C_j)\leq \lambda-2$. We consider each case separately. 
	  
	  Consider the former case. Note that by Lemma \ref{lem:lambda_special_cuts_1}, for any $r<j$, it must be the case that $u_r,v_r\in C_j$, implying that $u_r,v_r\in C_j\cup X$. Hence, the edge $e_r$ cannot cross the cut $\cut{C_j\cup X}$. Similarly, it cannot be the case that for some $r>j$, the edge $e_r$ crosses the cut $\cut{C_j\cup X}$ because this would then imply that $e_r$ crosses the cut $X$ ($e_r$ cannot cross $C_j$ since $r\neq j$), contradicting the choice of $j$ as the highest possible index less than $i$ for which such an edge exists. Therefore, we conclude that $e_j$ and $e_i$  are the only two \emph{potential} edges crossing the cut $\cut{C_j\cup X}$ in $G_i^*$. Since $\delta_{G^*}(C_j\cup X)\geq \lambda$ and $\delta_{G_i^*}(C_j\cup X)\leq \lambda-2$, it must be the case that both $e_j$ \emph{and} $e_i$ cross the cut $\cut{C_j\cup X}$ in $G_i^*$. But this contradicts Lemma \ref{lemma:lambda:special edges}, which states that at most one of the edges in $\Z(e^*)$ can cross any $\lambda$-cut of Type 1 in $G^*$. As a result, we conclude that $\delta_{G^*_i}(X\cup C_j)\geq \lambda-1$,  implying that $\delta_{G^*_i}(X\cap C_j)\leq \lambda-2$. In the second case, we have the following two subcases.
	  
	 \begin{description}
	 	\item[Subcase 1:] $X\cap C_j\subset X$. In this subcase, we observe that $\cut{X\cap C_j}$ is also a violating cut of Type 1 in $G_i^*$, contradicting our choice of $X$ as the minimum possible such set. Indeed, $X\cap C_j$ separates the ordered pair $\{u^*,v^*\}$ and hence is a cut of Type 1. In the case we are in, we already know that $\delta_{G_i^*}(X\cap C_j)\leq \lambda-2$, implying that $\cut{X\cap C_j}$ is also a violating cut of Type 1 in $G_i^*$, completing the argument for this subcase.

	 	\item[Subcase 2:] $X\subset C_j$. In this case, we have demonstrated the presence of a set $X$ which contains $u_i$ and is disjoint from $v_i$, as well as a set $C_j$ which also does not contain $v_i$. Hence, it must be the case that $e_i$ (along with $e_j$) crosses $C_j$, a contradiction to the definition of the family $\C(e^*)$. This completes the argument for the second subcase.
	 \end{description}
	  
	  Having obtained a contradiction in each case, we conclude that the digraph $G_i^*$ has no violating cuts of Type 1. This completes the proof of the lemma.
	\end{proof}

Given Lemma \ref{lemma:lambda-dir:G_i cut properties_type1}, we now argue that $G_i^*$ also excludes violating cuts of Type 2.

\begin{lemma}\label{lemma:lambda-dir:G_i cut properties_type2}
	For every $i\in [k]$, the digraph $G_i^*$ has no violating cuts of Type 2.
\end{lemma}

\begin{proof}
Suppose that for some $i\in [k]$, the digraph $G^*_i$ has a violating cut of Type 2 and let $i$ be the \emph{least} integer for which this happens. Again, we choose $\cut{X}$ such that $|X|$ is minimized. Note that due to the asymmetry of cuts in digraphs, there are three possible cases. Either $u^*,v^*\in X$ or $u^*,v^*\in \bar X$ or $u^*\in \bar X$, $v^*\in X$. Precisely, if $u^*\in X$, it must be the case that $v^*\in X$. 
	
	Observe that the cut $\cut{X}$ separates the ordered pair $\{u_i,v_i\}$. If this were not true, then $\cut{X}$ is also a violating cut of Type 2 in the graph $G_i^*+e_i=G_{i-1}$, contradicting our choice of $i$.
	Furthermore, since $e_i$ is a deletable edge in $G$ and it crosses the cut $\cut{X}$, it follows that $\delta_{G}(X)\geq \lambda+1$. Since the edge $e^*$ does not cross this cut, it must be the case that $\delta_{G^*}(X)\geq \lambda+1$ as well.
	
	Invoking the same arguments as in the proof of Lemma \ref{lemma:lambda-dir:G_i cut properties_type1} we conclude that $i>1$ and that there is a $j\in [k]$ such that $j<i$ and $\cut{X}$ also separates the ordered pair $\{u_j,v_j\}$. We choose $j$ to be the \emph{largest} integer less than $i$ such that the cut $\cut{X}$ separates the ordered pair $\{u_j,v_j\}$.
	Recall that  Lemma \ref{lem:lambda_special_cuts_1} ensures that there are cuts $C_i,C_j\in \C(e^*)$ such that $e_i$ crosses $\cut{C_i}$, $e_j$ crosses $\cut{C_j}$, $e_i$ \emph{does not} cross $\cut{C_j}$ and $e_j$ does not cross $\cut{C_i}$. Furthermore, $C_j\subset C_i$. 
	We now argue that $v^*\notin X$.

	\begin{claim} 
		$v^*\notin X$.
	\end{claim}
	
	\begin{proof}
Suppose to the contrary that $v^*\in X$. Invoking the submodularity of the cuts $C_i$ and $X$, we infer that 
	$$\delta_{G^*_i}(X\cap C_i)+\delta_{G^*_i}(X\cup C_i)\leq \delta_{G^*_i}(X)+\delta_{G^*_i}(C_i)\leq 2\lambda-2$$
	Suppose that $\delta_{G^*_i}(X\cap C_i)\leq \lambda-2$.
	We now consider the following two subcases.
	
	\begin{description}
		\item[Subcase 1:] $u^*\in X$. in this subcase, we argue that $X\cap C_i$ is a violating cut of Type 1 in $G_i^*$, which contradicts Lemma \ref{lemma:lambda-dir:G_i cut properties_type1}. First of all, observe that $u^*,v^*\in X$, $u^*\in C_i$ and $v^*\notin C_i$. Hence, the cut $X\cap C_i$ indeed separates the ordered pair $\{u^*,v^*\}$ and hence is a cut of Type 1 in $G_i^*$. It remains to argue that $X\cap C_i$ is a violating cut in $G_i^*$. But this follows from our assumption that $\delta_{G^*_i}(X\cap C_i)\leq \lambda-2$.

		\item[Subcase 2:] $u^*\notin X$. In this subcase, we argue that $X\cap C_i$ is also a violating cut of Type 2 in $G_i^*$ and $X\cap C_i\subset X$, contradicting our choice of $X$ as a minimal such set. First of all, observe that $v^*\in X\setminus C_i$, implying that $X\cap C_i\subset X$. Furthermore, since $u^*\notin X$ and $v^*\notin C_i$, it follows that $u^*,v^*\notin X\cap C_i$ and hence $X\cap C_i$ is a cut of Type 2 in $G_i^*$. It remains to argue that $X\cap C_i$ is a violating cut in $G_i^*$. But this follows from our assumption that $\delta_{G^*_i}(X\cap C_i)\leq \lambda-2$.

	\end{description}

	Having reached a contradiction in either subcase, we conclude that $\delta_{G^*_i}(X\cap C_i)\geq \lambda-1$, which in turn implies that $\delta_{G^*_i}(X\cup C_i)\leq \lambda-1$. Furthermore, $u^*,v^*\in X\cup C_i$. However, since $G^*_i=(G-e^*)-{\cal Z}_i$ and $e_i$ is the \emph{only} edge of ${\cal Z}_i\cup \{e^*\}$ that can cross $C_i$, it follows that $\delta_{G^*_i}(X\cup C_i)=\delta_{G}(X\cup C_i)-1$. Since $\delta_{G^*_i}(X\cap C_i)\geq \lambda-1$, it follows that $\delta_{G}(X\cup C_i)\leq \lambda$. Since $e_i$ crosses this cut in $G$, it follows that $e_i$ is undeletable in $G$, contradicting our assumption that $e_i\in {\cal Z}(e^*)$. 	Hence we conclude that
	 $v^*\notin X$, completing the proof of the claim.
	 \end{proof}

	 Since $\cut{X}$ is a cut of Type 2, it must be the case that $u^*\notin X$ as well. 
	Invoking the submodularity of the cuts $C_j$ and $X$, we infer that 
	 
	 $$\delta_{G^*_i}(X\cap C_j)+\delta_{G^*_i}(X\cup C_j)\leq \delta_{G^*_i}(X)+\delta_{G^*_i}(C_j)\leq 2\lambda-2$$

	 We again begin with the case when $\delta_{G^*_i}(X\cap C_j)\leq \lambda-1$. In this case, we argue that $X\cap C_j$ is a violating cut of Type 2 in $G^*_i$ and $X\cap C_j\subset X$, contradicting our choice of $X$. First of all, we argue that $X \cap C_j$ is a non-empty proper subset of $X$. This is because $u_j,u_i\in X$, $v_j,v_i\in \bar X$ and by Lemma \ref{lem:lambda_special_cuts_1}, $u_i\in \bar C_j$. As a result, $u_j\in X\cap C_j$ and $u_i\in X\cap \bar C_j$, implying that $X\cap C_j$ is a non-empty proper subset of $X$. Furthermore, since $u^*\notin X$ and $v^*\notin C_j$, it follows that $u^*,v^*\notin X\cap C_j$ and hence $X\cap C_j$ is a cut of Type 2 in $G_i^*$. It remains to argue that $X\cap C_j$ is a violating cut in $G_i^*$. But this follows from our assumption that $\delta_{G^*_i}(X\cap C_j)\leq \lambda-1$.

	Finally, we consider the case when $\delta_{G^*_i}(X\cap C_j)> \lambda-1$. In this case, we know that $\delta_{G^*_i}(X\cup C_j)\leq  \lambda-2$. Since $u^*\in C_j$ and $v^*\notin X\cup C_j$, it follows that $X\cup C_j$ separates the ordered pair $\{u^*,v^*\}$ in $G_i^*$. That is, $X\cup C_j$ is a cut of Type 1 in $G_i^*$. However, since we are in the case when $\delta_{G^*_i}(X\cup C_j)\leq  \lambda-2$, we conclude that $X\cup C_j$ is in fact a \emph{violating} cut of Type 1 in $G_i^*$, contradicting Lemma \ref{lemma:lambda-dir:G_i cut properties_type1}. 
%
%
%
%
%
%
%
%
%
	This completes the proof of the lemma.
\end{proof}

Having proved Lemma \ref{lemma:lambda-dir:G_i cut properties_type1} and Lemma \ref{lemma:lambda-dir:G_i cut properties_type2}, we have the following lemma
for computing a deletion set from $\D(e^*)$.
\begin{lemma}
\label{lemma:lambda-dir:alternate solution}
    Let $G$ be a digraph and $\lambda\in {\mathbb N}$ such that $G$ is a $\lambda$-connected digraph. If there is a deletable edge $e^*\in E(G)$ such that $|\D(e^*)| \geq k\lambda$ then there is a set $\Z \subseteq \D(e^*)$ of $k$ edges such that $G- \Z$ is $\lambda$-connected.
\end{lemma}
\begin{proof}
We define the set $\cal Z$ in the statement of the lemma to be the set ${\cal Z}(e^*)={\cal Z}_k$. In order to prove that $\cal Z$ satisfies the required properties, we need to argue that $G'=G-{\cal Z}$ remains $\lambda$-connected. If this were not the case then there is a cut $\cut{X}$ in $G'$ such that $\delta_{G'}(X)\leq \lambda-1$. We now consider the following cases. In the first case, $X$ is crossed by the edge $(u^*,v^*)$. In this case, it follows that $X$ is a cut of Type 1 in $G_k^*$ and furthermore, $\delta_{G_k^*}(X)=\delta_{G'}(X)-1\leq \lambda-2$. But this implies the presence of a violating cut of Type 1 in $G_k^*$, a contradiction to Lemma \ref{lemma:lambda-dir:G_i cut properties_type1}. In the second case, $X$ is not cross by the edge $(u^*,v^*)$. In this case, it follows that $X$ is a cut of Type 2 in $G_k^*$ and $\delta_{G_k^*}(X)=\delta_{G'}(X)\leq \lambda-1$. But this implies the presence of a violating cut of Type 2 in $G_k^*$, a contradiction to Lemma \ref{lemma:lambda-dir:G_i cut properties_type2}. Hence, we conclude that $\cal Z$ indeed satisfies the required properties. This completes the proof of the lemma. 
\end{proof}


We now prove Lemma~\ref{lemma:lambda-dir:upper bound},
using Lemma~\ref{lemma:lambda-dir:alternate solution}

\paragraph{Proof of Lemma~\ref{lemma:lambda-dir:upper bound}.}
    Let $F = \{ f_1, f_2, \ldots, f_p\}$ be an arbitrary maximal set of edges such that $G - F$ is $\lambda$-connected.
    If $|F| = p \geq k$, then we already have a required deletion set.
    Therefore, we may assume that $p\leq k-1$.
Now, consider the graphs $G_0,\dots, G_p$ with $G_0=G$ and $G_i$ defined as  $G_i = G - \{f_1, \ldots f_i\}$ for all $i\in [p]$.
    Note that $G_{i+1} = G_i - f_{i+1}$ and $G_p = G - F$.
    Observe that each $G_i$ is $\lambda$-connected, by the definition of $F$.
    Let $\D_i$ be the set of deletable edges in $G_i$ which are undeletable in $G_{i+1}$.
    Observe that $\D_i=\D(f_i)$ (see Definition \ref{def:special_edges}) in the graph $G_i$.

  Now consider any deletable edge of $G$.
    It is either contained in $F$, 
    or there is some $i\in \{0,\dots, p-1\}$ such that it is deletable in $G_i$ but undeletable in $G_{i+1}$.
    In other words, the set $F \cup \D_1 \cup \D_2 \ldots \cup \D_p$ covers all the deletable edges of $G$. Since $p \leq k-1$ and the number of deletable edges in $G$ is at least $k^2\lambda$, it follows that for some $i\in [p]$, the set $\D_i$ has size at least $k\cdot \lambda$.
    

         Let $\Z_i$ be the set of at least $k$ edges corresponding to $\D_i$ guaranteed by Lemma~\ref{lemma:lambda-dir:alternate solution}. We know that $G_i - \Z_i$ is $\lambda$-connected.         
    Since $G_i$ is a subgraph of $G$ on the same set of vertices, it follows that $G - \Z_i$ is also $\lambda$-connected, which again gives us a deletion set of cardinality $k$.
%
%
\qed

A straightforward consequence of the above lemma is an {\FPT} algorithm for {\lcd} on digraphs.

\begin{lemma}\label{thm:lambda-dir:main}
	\lcd in directed graphs can be solved  in time $2^{\cO(k \log k)} + n^{\cO(1)}$.
\end{lemma}

This completes the section on digraphs and in the rest of the paper, we will deal exclusively with undirected graphs.

\section{Undirected Graphs}
\label{sec:undirected}

In this section, we present our results for undirected graphs.
As it often happens when dealing with the connectivity of graphs, the \emph{parity} of the size of the min-cuts plays a crucial role in the design of our algorithms. As a result, we need to handle even-connectivity and odd-connectivity separately. The first subsection contains the details of our results when $\lambda$ is even. At a high level, we follow the strategy used for digraphs. However, the case when $\lambda$ is odd is much more involved and is discussed in the next subsection.
\subsection{Even Connectivity}
%
%

We begin by restating the main lemma of this subsection.

\lambdaundirmain*

\myparagraph{Setting up the notation.} For the rest of this subsection, we fix a deletable edge $e^* = (u^*, v^*)$ in $G$ such that $\D(e^*)$ contains at least $2k \lambda$ edges.
Let $G^* = G - e^*$. Since $e^*$ is deletable in $G$, it follows that  $G^*$ is a $\lambda$-connected graph.

Then, using Lemma~\ref{lemma:lambda:special edges} and Lemma~\ref{lemma:lambda:special cuts 1}
we can obtain $\Z(e^*)$ and $\C(e^*)$.
Let $\Z$ be the set $\{ e_{2i -1} \in \Z(e^*) \; | \; i \in [k] \}$, which is a subset of $\Z(e^*)$. 
We will show that $G - \Z$ is $\lambda$-connected.
Let $G_i$ be the graph $G^* - \{e_1, e_3, \ldots, e_i \}$, for $i \in \{1,2, \ldots ,2k-1 \}$.
For any odd number $i < 2k$, let $\Z_i$ be the subset $\{e_1, e_3, \ldots, e_i\}$ of $\Z$. For each $j\in [2k]$, we let $u_j$ and $v_j$ be the endpoints of the edge $e_j$.


We now recall the definition of cuts of Type 1 and Type 2 but in the setting of undirected graphs.

\begin{definition}
A cut $\cut{X}$ in $G^*_i$ {\rm (}for any $i\in [k]${\rm )} is called a cut of {\bf Type 1} if it separates the  pair $\{ u^*, v^* \}$ and a cut of {\bf Type 2} otherwise. We call $\cut{X}$ a {\bf violating cut} if $\cut{X}$ is a cut of Type 1 and $\delta_{G^*_i}(X) \leq\lambda - 2$ or $\cut{X}$ is a cut of Type 2 and $\delta_{G^*_i}(X) \leq\lambda - 1$.
\end{definition}

As in the directed case, we shall prove that there are no violating cuts of Type 1 in $G^*_i$ for any odd $i\in [2k-1]$.
Essentially the same result also holds when $\lambda$ is an odd number. 
Hence instead of repeating the proof again, we prove the following lemma for any value of $\lambda$.
This will however require that we generalize our notation to accommodate both cases. 
The edge $e^*$ is chosen depending on the value of $\lambda$.
We then have a set $\Z$ which is a subset of $\Z(e^*)$, whose precise definition again depends on the valye of $\lambda$,
however in both cases we have that $e_1 \in Z$.
For any number $i$ such that $e_i \in \Z$, let $\Z_i = Z \cap \{e_1, e_2, \ldots, e_i\}$ and $G_i = G^* \setminus Z_i$.
The violating cuts of Type 1 have the same definition for both cases.

\begin{lemma}\label{lemma:lambda-undir:even type 1 violating cut}
\label{lem:lambda_undir_even type 1}
    For any number $i$ such that $e_i \in \Z$, the graph $G_i^*$ has no violating cuts of Type 1, for any value of $\lambda$.
\end{lemma}

\begin{proof} The proof strategy we employ for this lemma is similar to that used in the proof of Lemma \ref{lemma:lambda-dir:G_i cut properties_type1}.
In the following, whenever we talk of a graph $G_i$ it is implicitly assumed that $e_i \in \Z$.
Now, suppose that for some $i$, the graph $G_i^*$ has a violating cut of Type 1 and let $i$ be the least integer for which this occurs. 
Furthermore, let $\cut{X}$ be a violating cut of Type 1 in $G_i^*$ such that $u^*\in X$ and $X$ is a smallest set with this property.
We first observe that the cut $\cut{X}$ separates the pair $\{u_i,v_i\}$. 
Otherwise, $\cut{X}$ is also a violating cut of Type 1 in $G_{j}^*$ for some $j < i$, which is a contradiction.
Furthermore, recall that $G^*$ is $\lambda$-connected. As a result, we know that $\delta_{G^*}(X)\geq \lambda$.

We now observe that $i>1$.  Suppose to the contrary that $i=1$. Recall that $e^*$ is deletable in $G$. 
This implies that $\delta_{G^*}(X)\geq \lambda$. 
Furthermore $G_1^*=G^*-e_1$, implying that $\delta_{G^*}(X)\geq \lambda-1$, a contradiction to $X$ being a violating cut of Type 1 in $G_1^*$.
By a similar argument, we conclude that there is some $j<i$ such that $e_j=(u_j,v_j) \in \Z$, 
the cut $\cut{X}$ also separates the pair $\{u_j,v_j\}$. 
Going forward, we choose $j$ to be the largest such number.

By Lemma \ref{lem:lambda_special_cuts_1}, we know that there are cuts $C_i,C_j\in {\cal C}(e^*)$ such that $C_i$ separates the endpoints of $e_i$ alone, 
$C_j$ separates the endpoints of $e_j$ alone, and $C_j\subset C_i$.
Since $X,C_i,C_j$ are all cuts of Type 1, it follows that $A\cap B$ and $A\cup B$ are also cuts of Type 1 for every $A,B\in \{X,C_i,C_j\}$. Furthermore, we observe that $\delta_{G_i^*}(C_j)=\lambda-1$. This is because $C_j$ is a $\lambda$-cut in $G^*$ and $e_j$ is the only edge in $\cal Z$ which crosses this cut in $G^*$. We now use the submodularity of the cuts $C_j$ and $X$ to obtain the following inequality.
 $$\delta_{G^*_i}(X\cap C_j)+\delta_{G^*_i}(X\cup C_j)\leq \delta_{G^*_i}(X)+\delta_{G^*_i}(C_j)\leq 2\lambda-3$$
  We infer from this inequality that either $\delta_{G^*_i}(X\cup C_j)\leq \lambda-2$ or $\delta_{G^*_i}(X\cap C_j)\leq \lambda-2$. In the former case, we will demonstrate the presence of a $\lambda$-cut in $G^*$ which is crossed by both $e_i$ and $e_j$ and in the latter case, we will contradict our choice of $X$.
  
  We begin with the first case. That is, $\delta_{G^*_i}(X\cup C_j)\leq \lambda-2$. Note that by Lemma \ref{lem:lambda_special_cuts_1}, for any $r<j$, it must be the case that $u_r,v_r\in C_j$, implying that $u_r,v_r\in C_j\cup X$. On the  other hand, it cannot be the case that for some $r$ such that $j<r<i$, the cut $\cut{X\cup C_j}$ separates the endpoints of $e_r$ because this would contradict our choice of $j$ as the highest integer less than $i$ such that $X$ separates the endpoints of $e_j$. Hence, we conclude that out of the integers $\ell$ from $1$ to $i$ such that $e_\ell \in \Z$, $e_i$ and $e_j$ are the only possible edges crossing the cut $\cut{C_j\cup X}$ in $G^*$. Since $\delta_{G^*}(C_j\cup X)\geq \lambda$ (follows from the fact that $G^*$ is $\lambda$ connected) and $\delta_{G^*_i}(C_j\cup X)\leq \lambda-2$, it follows that $\cut{C_j\cup X}$ separates the endpoints of \emph{both} $e_i$ and $e_j$ in $G_i^*$. However, since $\delta_{G^*}(C_j\cup X)\geq \lambda$, it must be the case that $C_j\cup X$ is a $\lambda$-cut in $G^*$ which is crossed by both $e_i$ and $e_j$, which is a contradiction to the structure guaranteed by Lemma~\ref{lem:lambda_special_edges} and Lemma~\ref{lem:lambda_special_cuts_1}. This concludes the analysis for the first case and we now move on to the second case.
  
      We now take up the case when $\delta_{G^*_i}(X\cap C_j)\leq \lambda-2$. We first observe that  $C_j\neq X$ because $e_i$ crosses $X$ in $G^*$ while $e_i$ cannot cross $C_j$ in $G^*$. We now consider the following two exhaustive subcases.
      
      \begin{description}
      	\item[Subcase 1:] $X\cap C_j\subset X$. In this subcase, we argue that $\cut{X\cap C_j}$ is also a violating cut of Type 1 in $G_i^*$, contradicting our choice of $X$ as a smallest such set. Since $X$ and $C_j$ are both cuts of Type 1 in $G_i^*$, it follows that so is $X\cap C_j$. Since we are in the case when $\delta_{G^*_i}(X\cap C_j)\leq \lambda-2$, it follows that $X\cap C_j$ is in fact a violating cut of Type 1 in $G_i^*$.
      	
      	\item[Subcase 2:] $X\subset C_j$. In this subcase, observe that either $C_j$ separates the endpoints of $e_i$ in $G_i^*$ (implying that $e_i$ crosses $C_j$ in $G^*$) or both endpoints of $e_i$ are contained in $C_j$. In either case, we contradict the properties of $\C(e^*)$ guaranteed by Lemma \ref{lem:lambda_special_cuts_1}. 
      \end{description}

Having obtained a contradiction in each case, we conclude that $G_i^*$ cannot contain a violating cut of Type 1. This completes the proof of the lemma.
\end{proof}

\begin{figure}[t]
\begin{center}
  \includegraphics{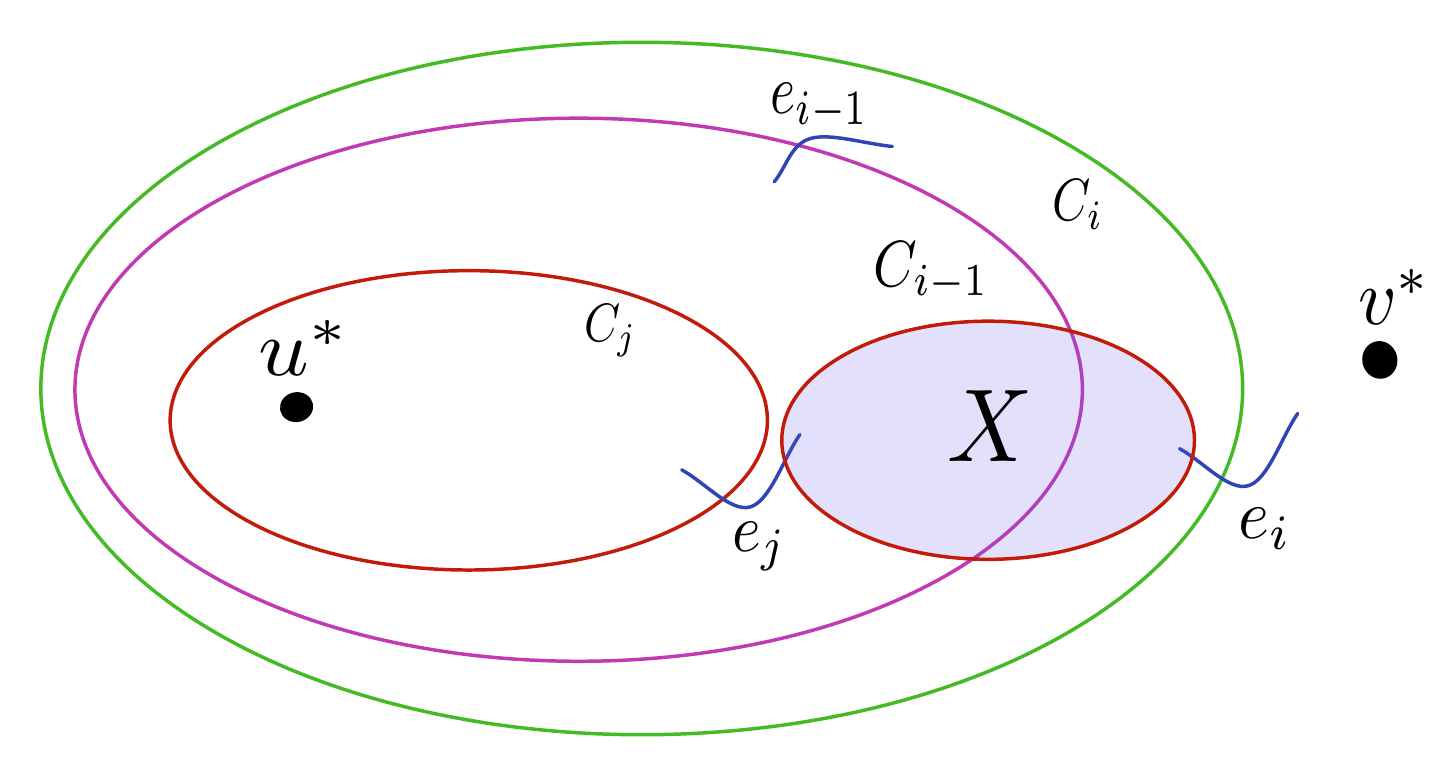}
  \end{center}
  \caption{An illustration of the cuts $C_j\subset C_{i-1}\subset C_i$ , edges $e_j,e_{i-1},e_i$ and the violating cut $X$.}
  \label{fig:undir_triple}
\end{figure}

\begin{lemma}\label{lemma:lambda-undir:even type 2 violating cut}
	For every odd $i\in [2k-1]$, the graph $G_i^*$ has no violating cuts of Type 2.
\end{lemma}

\begin{proof}
	Suppose to the contrary that for some odd $i\in [2k-1]$, the graph $G_i^*$ has a violating cut of Type 2. We choose $i$ to be the least integer for this which happens. Furthermore, we choose $X$ to be a smallest set disjoint from $\{u^*,v^*\}$ such that $\cut{X}$ is a violating cut of Type 2 in $G_i^*$.
	
	We first observe that the cut $\cut{X}$ separates the pair $\{u_i,v_i\}$. Otherwise, $\cut{X}$ is also a violating cut of Type 2 in $G_{i-2}^*$, a contradiction to our choice of $i$. Furthermore, since $e_i$ is deletable in $G$ and $e^*$ does not cross $X$ in $G$, it follows that $\delta_{G^*}(X)\geq \lambda+1$.

We now observe that $i>1$. Suppose to the contrary that $i=1$.  By definition, we know that $G_1^*=G^*-e_1$, implying that $\delta_{G^*}(X)\geq \lambda$, a contradiction to $X$ being a violating cut of Type 2 in $G_1^*$. 
Hence, we conclude that $i>1$, implying that $i\geq 3$. For the same reason, we conclude that for some odd $j<i$, the cut $\cut{X}$ also separates the pair $\{u_j,v_j\}$. Going forward, we choose $j$ to be the largest such integer. Since both $j$ and $i$ are odd, it follows that $j<i-1$. Furthermore, Lemma \ref{lem:lambda_special_cuts_1} guarantees the presence of cuts $C_j\subset C_{i-1}\subset C_i$ in $G^*$ which are crossed by $e_j,e_{i-1},e_i$ respectively (see Figure \ref{fig:undir_triple}). Furthermore, $e_j$ does not cross $C_{i-1}$ or $C_i$, $e_{i-1}$ does not cross $C_j$ or $C_i$ and $e_i$ does not cross $C_j$ or $C_{i-1}$. We now argue that $X$ is disjoint from $C_j$ and contained in $C_i$. This is proved in the following two claims.

\begin{claim}
	$X\subset C_i$.
\end{claim} 

\begin{proof} For this, we consider the cuts $\cut{X}$ and $\cut{C_i}$. Observe that since $e_i$ is the only edge in $\Z_i$ which crosses the cut $\cut{C_i}$ in $G^*$ and $\cut{C_i}$ is a $\lambda$-cut in $G^*$ by definition, 
 we have that $\delta_{G_i^*}(C_i)=\lambda-1$. Using the submodularity of cuts along with the fact that  $\delta_{G_i^*}(C_i)=\lambda-1$ and $\delta_{G_i^*}(X)\leq \lambda-1$, we know that

  $$\delta_{G^*_i}(X\cap C_i)+\delta_{G^*_i}(X\cup C_i)\leq \delta_{G^*_i}(X)+\delta_{G^*_i}(C_i)\leq 2\lambda-2$$

First of all, observe that $X\cup C_i$ is a cut of Type 1 and $X\cap C_i$ is a cut of Type 2 with the property that $u^*,v^*\notin X\cap C_i$. Now, suppose that $\delta_{G^*_i}(X\cup C_i)\leq \lambda-2$. Lemma \ref{lem:lambda_special_cuts_1} guarantees that for every $r<i$, the endpoints of $e_r$ lie inside $C_i$. As a result, $e_i$ is the \emph{only} edge in ${\cal Z}_i$ which may cross $\cut{X\cup C_i}$ in the graph $G^*$. This implies that in the graph $G^*=G_i^*\cup \Z_i$, it holds that $\delta_{G^*}(X\cup C_i)\leq \lambda-1$. But this contradicts the fact that $G^*$ is $\lambda$-connected. Hence, we may assume that $\delta_{G^*_i}(X\cup C_i)\geq \lambda-1$, which in turn implies that $\delta_{G^*_i}(X\cap C_i)\leq \lambda-1$.

 In this case, if $X\setminus C_i\neq \emptyset$, then $X\cap C_i\subset X$. Furthermore, we have already observed that $\cut{X\cap C_i}$ is a cut of Type 2 in $G_i^*$ where $u^*,v^*\notin X\cap C_i$. This contradicts our choice of $X$. Therefore, it must be the case that $X\setminus C_i=\emptyset$, implying that $X\subseteq C_i$. It cannot be the case that $X=C_i$ since $C_i$ is a $\lambda$-cut in $G^*$ and we have already argued that $X$ is a $\lambda+1$-cut in $G^*$. Hence, we conclude that $X\subset C_i$. This completes the proof of the claim.
\end{proof}

\begin{claim}
$X\cap C_j=\emptyset$.
	
\end{claim}

\begin{proof}
Suppose to the contrary that $X\cap C_j\neq \emptyset$.
We consider the cuts $\cut{X}$ and $C_j$. Observe that since $C_j$ is a $\lambda$-cut in $G^*$ and $e_j$ in the only edge of $\Z_i$ which crosses $C_j$ in $G^*$, we have that $\delta_{G_i^*}(C_j)=\lambda-1$. Using the submodularity of cuts along with the fact that  $\delta_{G_i^*}(C_j)=\lambda-1$ and $\delta_{G_i^*}(X)\leq \lambda-1$, we know that

  $$\delta_{G^*_i}(X\cap C_j)+\delta_{G^*_i}(X\cup C_j)\leq \delta_{G^*_i}(X)+\delta_{G^*_i}(C_j)\leq 2\lambda-2$$

First of all, observe that $X\cup C_j$ is a cut of Type 1 and $X\cap C_j$ is a cut of Type 2 with the property that $u^*,v^*\notin X\cap C_j$. We begin with the case when $\delta_{G^*_i}(X\cap C_j)\leq \lambda-1$.

 In this case, if $X\setminus C_j\neq \emptyset$, then $X\cap C_j\subset X$. Furthermore, we have already observed that $\cut{X\cap C_j}$ is a cut of Type 2 in $G_i^*$ where $u^*,v^*\notin X\cap C_j$. This contradicts our choice of $X$. Therefore, it must be the case that $X\setminus C_j=\emptyset$, implying that $X\subseteq C_j$. However, since $e_i$ crosses $X$ in $G^*$, it follows that $e_i$ crosses $C_j$ in $G^*$, a contradiction. Hence, we conclude that $\delta_{G^*_i}(X\cap C_j)\geq \lambda$, implying that $\delta_{G^*_i}(X\cup C_j)\leq \lambda-2$.
 
 In this case, we observe that $e_i$ and $e_j$ are the only edges of $\Z_i$ which cross $X\cup C_j$ in $G^*$. Indeed for any $r<j$, we know that the endpoints of $e_r$ are contained in $C_j$ and for any odd $r$ such that $j<r<i$, we know that $e_j$ does not cross $X$ in $G_i^*$ due to our choice of $j$. Hence, no edge of $\Z_i$ apart from $e_i$ or $e_j$ can cross $X\cup C_j$ in $G^*$. Since $G^*$ is $\lambda$-connected, it must be the case that $\delta_{G^*}(X\cup C_j)\geq \lambda$, implying that both $e_i$ and $e_j$ cross the $\lambda$ cut $\cut{X\cup C_j}$ in $G^*$, a contradiction to the first property of Lemma \ref{lem:lambda_special_cuts_1}.
 This completes the proof of the claim.
\end{proof}

The claims above imply that $X$ lies `between' $C_i$ and $C_j$.
We now study the interaction between $X$ and $C_{i-1}$. Let $X_1=X\cap C_{i-1}$ and $X_2=X\setminus X_1$. We first argue that both $X_1$ and $X_2$ are non-empty. Observe that if $X_1$ is empty, then $X\subseteq C_i\setminus C_{i-1}$. However, we know that $X$ is crossed by the edge $e_j$ in $G^*$, which implies that $C_{i-1}$ is also crossed by the edge $e_j$ in $G^*$, a contradiction to the structure guaranteed by Lemma \ref{lem:lambda_special_cuts_1}. On the other hand, if $X_2$ is empty, then $X\subseteq C_{i-1}$. However we know that $X$ is crossed by the edge $e_i$ in $G^*$, which implies that $C_{i-1}$ is also crossed by the edge $e_i$ in $G^*$, a contradiction to the structure guaranteed by Lemma \ref{lem:lambda_special_cuts_1}. Hence, we conclude that $X_1$ and $X_2$ are both non-empty. In the rest of the proof, we will use the fact that $X_1$ and $X_2$ are non-empty to demonstrate the presence of a violating cut of Type 1 in $G_i^*$, contradicting Lemma \ref{lemma:lambda-undir:even type 1 violating cut}.

  Observe that $X_1$ and $X_2$ are also cuts of Type 2. If $\delta_{G^*_i}(X_1)\leq \lambda-1$, then it contradicts our choice of $X$ as a smallest set with the same properties. Therefore, we conclude that $\delta_{G^*_i}(X_1)\geq \lambda$. By the same argument, we conclude that $\delta_{G^*_i}(X_2)\geq \lambda$. 
  Let $\altdelta_{G^*_i}(X_1,X_2)$ denote the set of edges of $G_i^*$ with one endpoint in $X_1$ and the other in $X_2$ and let  $\delta_{G^*_i}(X_1,X_2)$ denote the size of this set. Observe that $\delta_{G^*_i}(X_1)+\delta_{G^*_i}(X_2)-\delta_{G^*_i}(X)=2\delta_{G^*_i}(X_1,X_2)$. Since $\delta_{G^*_i}(X_1),\delta_{G^*_i}(X_2)\geq \lambda$ it follows that $2\delta_{G^*_i}(X_1,X_2)\geq \lambda+1$. This implies that  $\delta_{G^*_i}(X_1,X_2)\geq \frac{\lambda}{2}+\frac{1}{2}$. However, since $\lambda$ is even, it must be the case that $\delta_{G^*_i}(X_1,X_2)\geq \frac{\lambda}{2}+1$. Finally, since $\altdelta_{G_i^*}(X)=(\altdelta_{G^*_i}(X_1) \cap \altdelta_{G^*_i}(X)) \uplus (\altdelta_{G^*_i}(X_2) \cap \altdelta_{G^*_i}(X))$, it must be the case that either $|\altdelta_{G^*_i}(X_1) \cap \altdelta_{G^*_i}(X)|\leq \frac{\lambda}{2}-\frac{1}{2}$ or $|\altdelta_{G^*_i}(X_2) \cap \altdelta_{G^*_i}(X)|\leq \frac{\lambda}{2}-\frac{1}{2}$. Again, since $\lambda$ is even, it must be the case that either $|\altdelta_{G^*_i}(X_1) \cap \altdelta_{G^*_i}(X)|\leq \frac{\lambda}{2}-1$ or $|\altdelta_{G^*_i}(X_2) \cap \altdelta_{G^*_i}(X)|\leq \frac{\lambda}{2}-1$. We need to consider each case separately.
  
\begin{description}
	\item [Case 1:] $|\altdelta_{G^*_i}(X_1) \cap \altdelta_{G^*_i}(X)|\leq \frac{\lambda}{2}-1$.  
   Let $Y=C_{i-1}\setminus X$. Consider the cut $\cut{Y}$.
   Since $Y$ contains $u^*$ and does not contain $v^*$, $Y$ separates the pair $\{u^*,v^*\}$ in $G_i^*$ and hence $\cut{Y}$ is a cut of Type 1. We now argue that $\delta_{G^*_i}(Y)\leq \lambda-2$. For this, we begin by proving the following.
   \begin{equation}\label{eqn:ineq_1}
\altdelta_{G_i^*}(Y) \subseteq (\altdelta_{G^*_i}(X_1) \cap \altdelta_{G^*_i}(X)) \cup ( \altdelta_{G^*_i}(C_{i-1}) \setminus \altdelta_{G^*_i}(X_1, X_2))
  \end{equation}	
Consider an edge $e\in \altdelta_{G_i^*}(Y)$. Suppose that $e$ crosses $C_{i-1}$. Then, $e\in \altdelta_{G_i^*}(C_{i-1})$. But note that $e\notin \altdelta_{G_i^*}(X_1,X_2)$. This is because $Y$ is disjoint from $X$ and both endpoints of every edge in $\altdelta_{G_i^*}(X_1,X_2)$ are contained within $X$. Therefore, $e\in \altdelta_{G_i^*}(C_{i-1})\setminus \altdelta_{G_i^*}(X_1,X_2)$. On the other hand, suppose that $e$ does not cross $C_{i-1}$. Then, it must be the case that both endpoints of $e$ are in $C_{i-1}$ or disjoint from $C_{i-1}$. Since $e\in \altdelta_{G_i^*}(Y)$, and $Y\subseteq C_{i-1}$, it follows that at least one endpoint (and hence both endpoints) of $e$ are contained in $C_{i-1}$. Furthermore, it must be the case that one endpoint of $e$ is in $C_{i-1}\cap Y=Y$ and the other in $C_{i-1}\setminus Y=X_1$. Since $Y$ is disjoint from $X$ by definition, we conclude that $e\in \altdelta_{G_i^*}(X_1)\cap \altdelta_{G_i^*}(X)$. This completes the proof of (1).

Observe that since $X_1\subseteq C_{i-1}$ and $X_2$ is disjoint from $C_{i-1}$, it follows that $\altdelta_{G_i^*}(X_1,X_2)\subseteq \altdelta_{G_i^*}(C_{i-1})$, implying that $|\altdelta_{G_i^*}(C_{i-1})\setminus \altdelta_{G_i^*}(X_1,X_2)|=\delta_{G_i^*}(C_{i-1})- \delta_{G_i^*}(X_1,X_2)$. Furthermore, we are in the case when 
$|\altdelta_{G^*_i}(X_1) \cap \altdelta_{G^*_i}(X)|\leq \frac{\lambda}{2}-1$. Hence, (1) implies that
$$\delta_{G_i^*}(Y) \leq \frac{\lambda}{2}-1 + \lambda - (\frac{\lambda}{2}+1)  \leq \lambda-2$$
Since $Y$ is a cut of Type 1  in $G_i^*$, we obtain a contradiction to Lemma \ref{lemma:lambda-undir:even type 1 violating cut}. This completes the analysis of Case 1.
		
\item [Case 2:]	$|\altdelta_{G^*_i}(X_2) \cap \altdelta_{G^*_i}(X)|\leq \frac{\lambda}{2}-1$.	The argument for this case is identical with the only difference being the definition of the set $Y$. In this case, we set $Y=\overline{C_{i-1}}\setminus X$. Since $Y$ contains $v^*$ and does not contain $u^*$, the cut $\cut{Y}$ is a cut of Type 1. We now argue that $\delta_{G^*_i}(Y)\leq \lambda-2$. For this, we begin by proving the following.
  \begin{equation}\label{eqn:ineq_1}
\altdelta_{G_i^*}(Y) \subseteq (\altdelta_{G^*_i}(X_2) \cap \altdelta_{G^*_i}(X)) \cup ( \altdelta_{G^*_i}(C_{i-1}) \setminus \altdelta_{G^*_i}(X_1, X_2))
  \end{equation}	
Consider an edge $e\in \altdelta_{G_i^*}(Y)$. Suppose that $e$ crosses $\overline{C_{i-1}}$. Then, $e\in \altdelta_{G_i^*}(\overline{C_{i-1}})$. But note that $e\notin \altdelta_{G_i^*}(X_1,X_2)$. This is because $Y$ is disjoint from $X$ and both endpoints of every edge in $\altdelta_{G_i^*}(X_1,X_2)$ are contained within $X$. Therefore, $e\in \altdelta_{G_i^*}(\overline{C_{i-1}})\setminus \altdelta_{G_i^*}(X_1,X_2)$. On the other hand, suppose that $e$ does not cross $\overline{C_{i-1}}$. Then, it must be the case that both endpoints of $e$ are in $\overline{C_{i-1}}$ or disjoint from $\overline{C_{i-1}}$. Since $e\in \altdelta_{G_i^*}(Y)$, and $Y\subseteq \overline{C_{i-1}}$, it follows that at least one endpoint (and hence both endpoints) of $e$ are contained in $\overline{C_{i-1}}$. Furthermore, it must be the case that one endpoint of $e$ is in $\overline{C_{i-1}}\cap Y=Y$ and the other in $\overline{C_{i-1}}\setminus Y=X_2$. Since $Y$ is disjoint from $X$ by definition, we conclude that $e\in \altdelta_{G_i^*}(X_2)\cap \altdelta_{G_i^*}(X)$. This completes the proof of (2).

Observe that since $X_2\subseteq \overline{C_{i-1}}$ and $X_1$ is disjoint from $\overline{C_{i-1}}$, it follows that $\altdelta_{G_i^*}(X_1,X_2)\subseteq \altdelta_{G_i^*}(\overline{C_{i-1}})$, implying that $|\altdelta_{G_i^*}(\overline{C_{i-1}})\setminus \altdelta_{G_i^*}(X_1,X_2)|=\delta_{G_i^*}(\overline{C_{i-1}})- \delta_{G_i^*}(X_1,X_2)$. Furthermore, since 
$|\altdelta_{G^*_i}(X_2) \cap \altdelta_{G^*_i}(X)|\leq \frac{\lambda}{2}-1$,  (2) implies that
$$\delta_{G_i^*}(Y) \leq \frac{\lambda}{2}-1 + \lambda - (\frac{\lambda}{2}+1)  \leq \lambda-2$$
Since $Y$ is a cut of Type 1  in $G_i^*$, we obtain a contradiction to Lemma \ref{lemma:lambda-undir:even type 1 violating cut}. 
\end{description}

	Having obtained a contradiction in either case, we conclude that $G_i^*$ cannot contain a violating cut of Type 2.
	This completes the proof of the lemma.
\end{proof}

Having proved Lemma \ref{lemma:lambda-undir:even type 1 violating cut} 
and Lemma \ref{lemma:lambda-undir:even type 2 violating cut}, 
we have the following lemma for computing a deletion set from $\Z$.


\begin{lemma}\label{lemma:lambda-undir:alternate solution}
Let $\lambda\in {\mathbb N}$ be an even number and let $G$ be a $\lambda$-connected graph. If there is a deletable edge $e^*\in E(G)$ such that $|\D(e^*)|>2k\lambda$ then there is a set $\Z \subseteq \D(e^*)$ of $k$ edges such that $G- \Z$ is $\lambda$-connected.
\end{lemma}
\begin{proof}
We define the set $\cal Z$ in the statement of the lemma to be the set ${\cal Z}(e^*)={\cal Z}_{2k-1}$. In order to prove that $\cal Z$ satisfies the required properties, we need to argue that $G'=G-{\cal Z}$ remains $\lambda$-connected. If this were not the case then there is a cut $\cut{X}$ in $G'$ such that $\delta_{G'}(X)\leq \lambda-1$. We now consider the following cases. In the first case, $X$ is crossed by the edge $(u^*,v^*)$ in $G$. In this case, it follows that $X$ is a cut of Type 1 in $G_{2k-1}^*$ and furthermore, $\delta_{G_{2k-1}^*}(X)=\delta_{G'}(X)-1\leq \lambda-2$. But this implies the presence of a violating cut of Type 1 in $G_{2k-1}^*$, a contradiction to Lemma \ref{lemma:lambda-undir:even type 1 violating cut}. In the second case, $X$ is not crossed by the edge $(u^*,v^*)$ in $G$. In this case, it follows that $X$ is a cut of Type 2 in $G_k^*$ and $\delta_{G_{2k-1}^*}(X)=\delta_{G'}(X)\leq \lambda-1$. But this implies the presence of a violating cut of Type 2 in $G_{2k-1}^*$, a contradiction to Lemma \ref{lemma:lambda-undir:even type 2 violating cut}. Hence, we conclude that $\cal Z$ indeed satisfies the required properties. This completes the proof of the lemma.
\end{proof}

Based on Lemma \ref{lemma:lambda-undir:alternate solution} 
we obtain a proof of Lemma~\ref{lemma:lambda-undir:upper bound} 
which is similar to that of Lemma \ref{lemma:lambda-dir:upper bound}.
%
%
As a consequence of the above Lemma~\ref{lemma:lambda-undir:upper bound}, we obtain an {\FPT} algorithm for {\lcd} on undirected graphs when $\lambda$ is an even integer.

\begin{lemma}\label{thm:lambda-undir:main} Let $\lambda\in {\mathbb N}$ be an even number. Then, 
	\lcd in undirected graphs can be solved in time $2^{\cO(k \log k)} + n^{\cO(1)}$.
\end{lemma}

This completes the description of our algorithm when $\lambda$ is even and in the rest of the section, we work with odd $\lambda$.

\subsection{Odd Connectivity}

In this subsection we deal with the case when $\lambda$ is odd. This case is significantly more involved when compared to the case when $\lambda$ is even as it is possible that the number of deletable edges is unbounded in $k$ in spite of the presence of a deletion set of size $k$. Indeed, consider the following example. Let $G$ be a cycle on $n$ vertices, $\lambda=1$ and $k=2$. Clearly, every edge in $G$ is deletable, but there is no deletion set of cardinality $2$.
In order to overcome this obstacle, we design a subroutine that either finds a deletion set of cardinality $k$ or detects an edge which is disjoint from some deletion set of cardinality $k$ in the graph. Before we formally state the corresponding lemma, we additionally define a subset, $\R\subseteq E(G)$, of \emph{irrelevant edges}.
From now onward, we denote the input as $(G,k,\R)$, and a deletion set is now defined to be a subset $F$ of $E(G)\setminus \R$ of size $k$ such that $G-F$ is $\lambda$-connected. Finally, we note that the set $\R$ contains all the undeletable edges of $G$. 

\mainodd*

We can then iteratively execute the algorithm of this lemma to either find a deletion set or grow the set of irrelevant edges in the graph. 
%
%
%
%
%
%
We begin by proving  the following lemma which says that if the graph admits a certain kind of decomposition, then certain deletable edges may be safely added to the set $\R$.

\begin{lemma}\label{lemma:lambda-undir:odd irrelevent edge}
    Let $(G,k,\R)$ be the input where $G$ is $\lambda$-connected, and let $X_1, X_2, \ldots X_{2k+2}$ be a partition of $V(G)$ into non-empty subsets  
    such that the following properties hold in the graph $G$.
    \begin{enumerate}
\setlength{\itemsep}{-2pt}
        \item $\delta_G(X_1, X_2) = \delta_G(X_2,X_3) \ldots = \delta_G(X_{2k+2}, X_1) = \frac{\lambda + 1}{2}$.
        
        \item Every edge of the graph either has both endpoints in some $X_i$ for $i\in [2k+2]$,
        or contained in one of the edge sets mentioned above.
        
        \item There are deletable edges $e_1, e_2, \ldots, e_{2k+2}$ in $E(G) \setminus \R$ such that
        $e_i \in \altdelta(X_i, X_{i+1})$ for $i \in [2k+2]$.
        (Here $X_{2k+3}$ denotes the set $X_1$.)
        
    \end{enumerate}
    Then $G$ has a deletion set of cardinality $k$ disjoint from $\R$, if and only if $G$ has a deletion set of cardinality $k$ disjoint from $\R \cup \{ e_1 \}$.

\end{lemma}
\begin{proof}
    The reverse direction is trivially true and hence we consider the forward direction.
    Suppose $S$ is a deletion set of cardinality $k$ for $(G,k,\R)$.
    Let $E_X = \bigcup_{i=1}^{2k+2} \delta(X_i, X_{i+1})$ and observe that the edges $e_1, e_2, \ldots e_{2k+2}$ are all contained in it.
    We call the edges in $E_X$ as \emph{cross edges} and all the other edges as \emph{internal edges}.
    We will first observe that $|S \cap E_X| \leq 1$ i.e. $S$ contains at most one cross edge, or else $G - S$ will not be $\lambda$ connected.
    To see this, let $e$ and $e'$ be two two edges in $S \cap E_X$ such that $e \in \altdelta(X_i, X_{i+1})$, $e' \in \altdelta(X_j, X_{j+1})$ and $1 \leq i \leq j \leq 2k+2$.
    If $i = j$ then let $Y = X_{i}$, else let $Y = X_{i+1} \cup X_{i+2} \cup \ldots \cup X_j$.
    Since $\altdelta_G(Y) = \lambda + 1$ and $e,e' \in \delta_G(Y)$, hence $\delta_{G - S}(Y) \leq \lambda - 2$,
    which contradicts the fact that $G - S$ is $\lambda$-connected.
    Hence $S$ contains at most one cross edge and at most $k-1$ internal edges.
    
    Now, if $e_1 \notin S$, then $S$ is the required deletion set for $(G,k,\R \cup \{ e_1 \})$.
    Otherwise, $S$ has at most $k-1$ internal edges,
    and hence by the pigeonhole principle, there is some $i \in [2k+2] \setminus \{1, 2\}$ such that $(X_i \cup X_{i+1}) \cap V(S) = \emptyset$.
    In other words, $S$ is completely disjoint from all edges that incident on a vertex contained in $X_i \cup X_{i+1}$.
    We will show that $S' = S - e_1 + e_i$ is a  deletion set of cardinality $k$  in $(G,k,\R \cup \{ e_1 \})$.
    Since $|S'| = |S|$ 
    and $S' \cap (R \cup \{e_1\}) = \emptyset$, it only remains to show that $G - S'$ is also $\lambda$ connected.
    
    Now, suppose to the contrary that $G - S'$ is not $\lambda$-connected. 
    This implies that $G -(S \setminus e_1)$ has $\lambda$-cut $\cut{A}$ which is crossed by $e_i$.
    Since $e_1$ cannot cross this cut (as $G - S$ is $\lambda$-connected), it follows that $\cut{A}$ is also a $\lambda$-cut in $G - S$.
    Let $u_1 \in X_1$ and $v_1 \in X_{2}$ be the endpoints of the edge $e_1$.
    Let $Y = X_2 \cup X_3 \cup \ldots \cup X_i \cup X_{i+1}$ and $Z = X_i \cup X_{i+1} \cup \ldots \cup X_{2k+2} \cup X_1$.
    Observe that $e_1$ crosses $\cut{Y}$ and $\cut{Z}$, whereas both the endpoints of $e_i$ are contained in both $Y \cap Z$.
    It follows from the definitions and the properties in the premise of the lemma that,
    $\delta_G(Y) = \delta_G(X) = \lambda + 1$ and $\delta_{G - S}(Y) = \delta_{G\setminus S}(X) = \lambda$.

    Now, $\cut{Y}$, $\cut{Z}$ and $\cut{A}$ are $\lambda$ cuts in $G - S$ and $\lambda$ is odd.
    By switching between $A$ and $\co{A}$ we can ensure that $A \cap Y \neq \emptyset$ and $A \cup Y \neq V(G)$.
    Hence by Proposition~\ref{prop:uncrossing} we have that either $Y \subseteq A$ or $A \subseteq Y$.
    If the first case occurs then both endpoints of $e_i$ are contained in $A$ which contradicts the fact that $e_i \in \altdelta_G(A)$.
    Hence it must be the case that $A \subseteq Y$ and
    furthermore, as $v_1 \notin A$ and $e_1 \notin \altdelta_G(A)$, we have that $u_1 \notin A$ as well.
    
    Now we consider the $A$ and $Z$.
    Suppose that $A \cup Z = V(G)$, which implies that $\co{Z} \subseteq A$.
    But since $e_1$ crosses $\cut{Z}$ and $u_1 \in \co{Z}$, it implies that $u_1 \in A$
    which is a contradiction.
    So it must be the case that $A \cup Z \neq V(G)$, and hence by Proposition~\ref{prop:uncrossing} we have
    that either $Z \subseteq A$ or $A \subseteq Z$.
    As before, the first case again leads to a contradiction and therefore $A \subseteq Z$.
    
    From the above we conclude that $A \subseteq Y \cap Z$, i.e. $A \subseteq X_i \cup X_{i+1}$.
    Now observe that $\cut{A}$ is a $\lambda$-cut in $G - S$ and no edge of $S$ is incident on a vertex in $X_i \cup X_{i+1}$.
    This implies that $\cut{A}$ is a $\lambda$-cut in $G$ as well.
    But this contradicts the fact that $e_i \in \altdelta_G(A)$ is a deletable edge in $G$.
    Having obtained a contradiction in all the cases, we conclude that $G - S'$ is also $\lambda$-connected,
    implying that the set $S'$ is a  deletion set of cardinality $k$  in $(G,k,\R\cup \{e_1\})$.
    This completes the proof of the lemma.
\end{proof}

\noindent {\bf \textsf{Setting up the notation.}}
Before we proceed with the rest of the section, we set up some notation which will be used in subsequent lemmas. We will be dealing with a fixed input $(G,k,\R)$.
Furthermore, we let $S^*$ denote a fixed subset of $E(G)\setminus \R$ of at most $k-1$ edges such that the graph $G_{S^*}=G - S^*$ is $\lambda$-connected. We let $e^*\notin \R$ denote a deletable edge in $G_{S^*}$ such that $\D(e^*)=(\del(G_{S^*})\cap \undel(G_{S^*}- \{e^*\}))\setminus \R$ has at least $\eta \lambda$ edges where $\eta = 3k(2k+3) + 1$.
We denote by $G^*$ the graph $G_{S^*}-\{e^*\}$. Then by Lemma~\ref{lemma:lambda:special edges} and Lemma~\ref{lemma:lambda:special cuts 1}, 
we have a collection $\Z(e^*)=\{e_1,\dots, e_\eta\}$ of edges in $\D(e^*)$,
and a collection $\C(e^*)$ of $\eta$ $\lambda$-cuts in $G^*$ corresponding to $\Z(e^*)$ such that, 
for each $e_i \in \Z(e^*)$ there is a unique cut $C_i \in {\cal C}^*$ which separates the endpoints of $e_i$. 
Furthermore, we may assume that both these collections are known to us. Note that \emph{computing} these collections was not particularly important in the case of digraphs or even $\lambda$ in undirected graphs. This is because the main lemmas we proved were only required to be existential. 
However, in the odd case, it is crucial that we are able to \emph{compute} these collections when given the graph $G_{S^*}$ and the edge $e^*$. 
For every $i\in [\eta]$, we let $(u_i,v_i)$ denote the endpoints of the edge $e_i$.

Let $\widehat{\Z}= \{e_{(2k+3)i + 1} \in \Z(e^*) \mid 0 \leq i \leq 3k \}$ and observe that $|\widehat{\Z}| = 3k + 1$.
Let $\widehat{\C}$ be the subcollection of $\C(e^*)$ corresponding to $\widehat{\Z}$. 
Let $\C$ be defined as the set  $\{C_i \in \widehat{\C} \mid (C_{i} \setminus C_{i-(2k+3)}) \cap V(S^*) = \emptyset\}$ 
where $V(S^*)$ denotes the set of endpoints of edges in $S^*$. 
Since $|S^*|\leq k-1$ at most $2(k-1)$ cuts of $\widehat{\C}$ are excluded from $\C$ and hence, $|\C| \geq k$.
Let $\Z$ be the subcollection of $\widehat{\Z}$ corresponding to $\C$.
For any $i\in [\eta]$ such that $e_i \in \Z$, we define $\Z_i = \{e_j \in \Z \,|\, j \leq i \}$ and $G^*_i = G^* - \Z_i$. 
In the rest of the section, whenever we talk about the set $Z_i$ and graph $G_i$, we assume that the corresponding edge $e_i\in \Z$ and hence these are well-defined.

\begin{definition}
	Let $i\in [\eta]$ such that $e_i \in \Z$.
	A cut $\cut{X}$ in $G^*_i$ {\rm (}for any $i\in [k]${\rm )} is called a cut of {\bf Type 1} if it separates the  pair $\{ u^*, v^* \}$ and a cut of {\bf Type 2} otherwise. We call $\cut{X}$ a {\bf violating cut} if $\cut{X}$ is a cut of Type 1 and $\delta_{G^*_i}(X) \leq\lambda - 2$ or $\cut{X}$ is a cut of Type 2 and $\delta_{G^*_i}(X) \leq\lambda - 1$.
\end{definition}

%
%
As before, we have the following lemma for handling Type 1 cuts,
which follows from Lemma~\ref{lemma:lambda-undir:even type 1 violating cut}
in the previous subsection. 
\begin{lemma}\label{lemma:lambda-undir:odd type 1 violating cut}
    For any $i\in [\eta]$ such that $e_i\in \Z$, the graph $G_i^*$ has no violating cuts of Type 1.
\end{lemma}

To handle the violating cuts of Type 2, we define a violating triple $(X,i,j)$ just like we did in the even case, and we prove several structural lemmas based on this definition.
\begin{definition}
    Let $i\in [\eta]$ such that $e_i \in \Z$. Let $\cut{X}$ be a violating cut of Type 2 in $G_i^*$ such that $u^*,v^*\notin X$, $e_i$ crosses $\cut{X}$ and $X$ is inclusion-wise minimal. Let $j<i$ be such that $e_j\in \Z$, $e_j$ crosses the cut $\cut{X}$ in $G^*$ and there is no $r$ such that $r$ satisfies these properties and $j<r<i$. Then we call the tuple $(X,i,j)$ a {\bf violating triple}.
\end{definition}

Observe that for any violating triple $(X,i,j)$, it holds that $j\leq i-(2k+3)$ and hence, 
there are cuts $C_j \subset C_{i-(2k+2)} \subset C_{i-(2k+1)} \ldots \subset C_{i-1}\subset C_{i}$ 
such that they are all $\lambda$-cuts in $G^*$ and all but $C_{j}$ and $C_{i}$ are $\lambda$-cuts in $G_i^*$ as well.
For the sake of convinience, let us rename these cuts as follows.
Let $C_j \subset C_{2k+2} \subset C_{2k+1} \ldots \subset C_1 \subset C_i$ denote the sets 
$C_j \subset C_{i-(2k+2)} \subset C_{i-(2k+2)} \ldots \subset C_{i-1} \subset C_i$ respectively, 
and let $\C_{ij}$ denote this ordered collection.
Additionally, in some of our arguments we may refer to the cuts $C_0$ and $C_{2k+3}$,
which denote the cuts $C_i$ and $C_j$ respectively.

\begin{lemma}\label{lem:violating_triple_exists} Let $i\in [\eta]$ such that $e_i \in \Z$ and let $\cut{X}$ be a violating cut of Type 2 in $G_i^*$ such that $G_{i-1}^*$ has no such violating cut, $u^*,v^*\notin X$ and $X$ is inclusion-wise minimal. Then, there is a $j<i$ such that $(X,i,j)$ is a violating triple. Furthermore given $G,i,X$, we can compute $j$ in polynomial time. Finally,  the following properties hold with regards to the triple $(X,i,j)$.
    \begin{itemize}
\setlength{\itemsep}{-2pt}
        \item $\delta_{G^*}(X)\geq \lambda+1$.
        \item $X\subseteq C_i \setminus C_j$.
        \item $e_i$ and $e_j$ are the only edges of $\Z$ which cross the cut $\cut{X}$ in $G^*$.
        \item $\delta_{G^*_i}(X)=\lambda-1$.
    \end{itemize}
\end{lemma}
\begin{proof}
    We first argue that the triple $(X,i,j)$ satisfies all the properties stated in the lemma.
    Then we will see how such a triple may be computed in polynomial time.
    
    From the definition of $\cut{X}$ we have that, 
    it separates the edge $e_i$ which is a deletable edge in $G_{S^*}$,
    and there is no such cut in $G^*_j$ for any $j < i$.
    Since $\cut{X}$ is a Type 2 cut, it doesn't separate the pair $u^*,v^*$.
    Hence in $G_{S^*} = G^* + e^*$ we have $\delta_{G_{S^*}}(X) \geq \lambda+1$.
    This implies that $\delta_{G^*}(X) \geq \lambda+1$.
    Furthermore, from the fact that $\delta_{G^*_i}(X) \leq \lambda - 1$ and $G^*_i = G^* - \Z_i$,
    we conclude that there are at least two edges in $\Z_i$ which cross $\cut{X}$.
    Hence, we can also conclude that $i > 1$ and that there is some $j < i$ such that
    $e_j$ is also separated by $X$.
    
    We will show that that $X \subset C_i$.
    Consider the cuts $\cut{X}$ and $\cut{C_i}$ in the graph $G^*_i$.
    By Proposition~\ref{lemma:submod}, we have that 
    $$ \delta(X \cap C_i) + \delta(X \cup C_i) \leq \delta(X) + \delta(C_i) = 2\lambda -2$$
    If $\delta(X \cup C_i) \leq \lambda - 2$, then combined with the fact that $e_i$ and $e^*$
    are the only edges in $G_{S^*}$ which cross the cut $\cut{X \cup C_i}$, we contradict
    the fact that $e^*$ is a deletable edge in $G_{S^*}$.
    Therefore it must be the case that $\delta(X \cap C_i) \leq \lambda -1$ in $G^*_i$.
    But this contradicts the minimality of $X$ if $X \cap C_i$ is a proper subset of $X$.
    Hence $X \subset C_i$.
    We can show that $X \subset \co{C_j}$ in a similar way by considering the cuts $X$ and $\co{C_j}$.
    Together they imply that $X \subseteq C_i \setminus C_j$.
    
    Next, we choose $j$ to be the largest number such that $j < i$, $e_j \in \Z_i$ and $e_j \in \altdelta_{G_{S^*}}(X)$.
    And for $\ell < j$, the endpoints of $e_\ell$ are contained in $C_j$.
    Hence $e_i$ and $e_j$ are the only edges of $\Z_i$ which cross $\cut{X}$.
    This then immediately implies that $\delta_{G^*_i}(X) = \lambda - 1$.
    
    Now we consider the issue of computing the triple $(X,i,j)$.
    Recall that we are given the sets $\C$ and $\Z$.
    We consider each choice of $e_i \in \Z$ in order of increasing value of $i$.
    For each $i$ we consider each choice of $e_j \in Z$ in order of decreasing value of $j$.
    Let $e_i = (u_i,v_i)$ and $e_j = (u_j,v_j)$ where $v_j, u_i \in C_i \setminus C_j$,
    Since $e_i$ and $e_j$ are the only edges of $\Z_i$ which cross $\cut{X}$,
    the cut $\cut{X}$ separates $\{u_i,v_j\}$ from $\{v_i,u_j\}$.
    Further, at most $\lambda + 1$ edges cross this cut in $G_{S^*}$
    and there is no $Y \subset X$ also forms such a cut.
    
    Using standard techniques such an $X$ can be computed in polynomial time, 
    if it exists. (For example, consider each choice of edges $e_i$ in increasing
    order of $i$ and then each choice of $e_j$ in decreasing order of $j$.
    For a fixed $e_i$ and $e_j$, we compute a minimal set of vertices $X$
    such that $\cut{X}$ separates both $e_i$ and $e_j$,
    and  $\delta_{G^*}(X) = \lambda + 1$.
    Such a cut is produced, for example applying by the Ford-Fulkerson algorithm
    if we set $\{u_i, v_j\} \in C_i \setminus C_j$ as the source set
    and $\{u_j, v_j\}$ as the sink set.
    Finally, observe that this process takes polynomial time.)

    Hence, we output $(X,i,j)$ if $e_i$ and $e_j$ are the first pair of edges
    for which the cut $\cut{X}$ exists.
    %
    %
    This completes the proof of this lemma.
\end{proof}

\begin{figure}[t]
    \begin{center}
        \includegraphics[height=200 pt, width=400 pt]{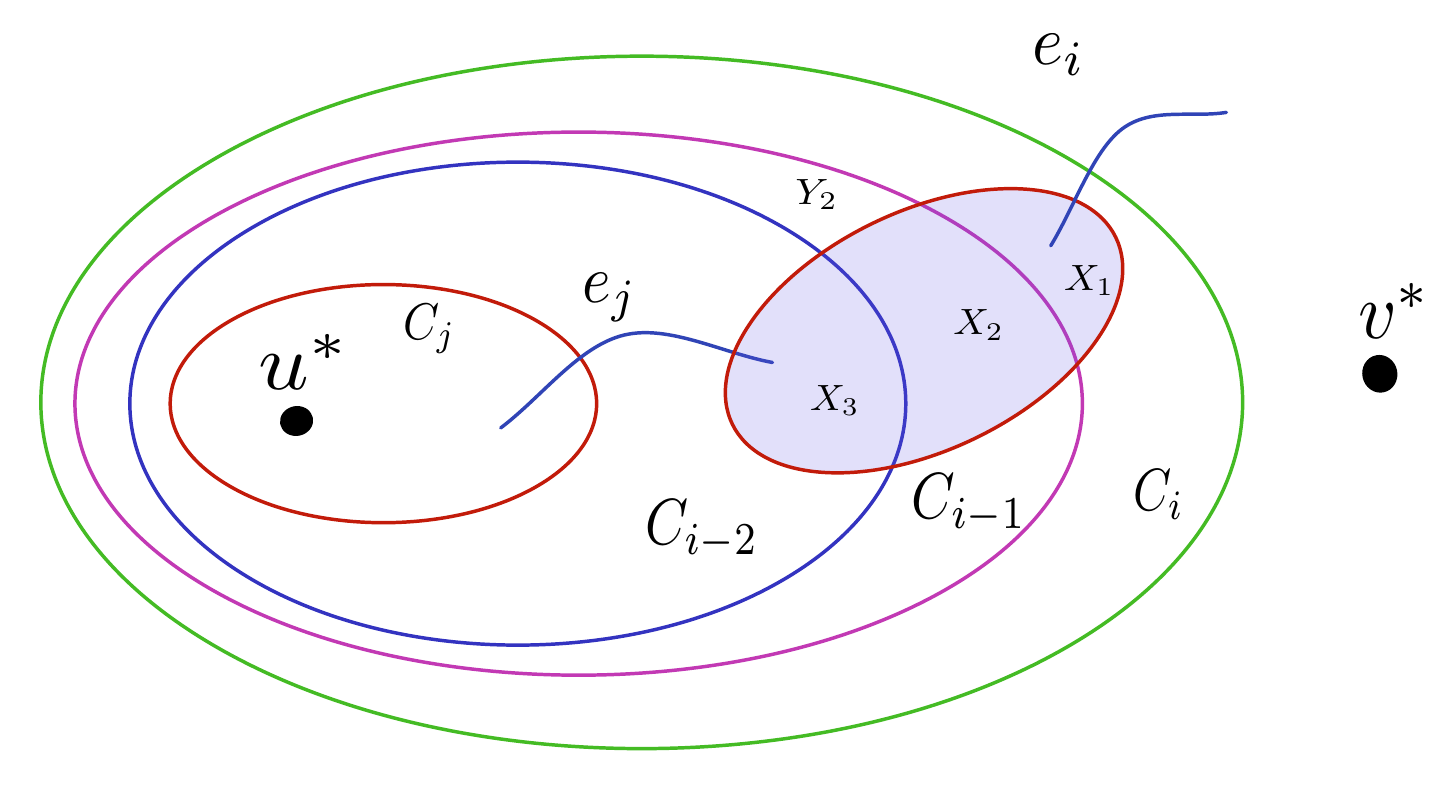}
    \end{center}
    \caption{An illustration of the partition of $X$ based on its intersection with $C_{i-2},C_{i-1},C_i$}
    \label{fig:define_sets}
\end{figure}

Let $C_a$ and $C_b$ be two consecutive cuts in $\C_{ij}$ such that $b = a+1$ and observe that $C_j = C_j \subset C_b \subset C_a \subset C_i = C_i$.
Let $X_1 = X \cap (C_i \setminus C_a)$, $X_2 = X \cap (C_a \setminus C_b)$ and $X_3 = C_b \setminus C_j$.


%

\begin{lemma}\label{lem:define_X_intersection}
    Let $i\in [\eta]$ such that $e_i \in \Z$ and let $(X,i,j)$ be a violating triple. Let $X_1\uplus X_2\uplus X_3$ be the partition of $X$ as defined above.
    The sets $X_1,X_2,X_3$ are all non-empty and furthermore, $X_2=C_a \setminus C_b$.
\end{lemma}

\begin{proof}We first argue that $X_1$ and $X_3$ are non-empty. If $X_3$ is empty, then we infer that $e_j$, which is known to cross the cut $\cut{X}$ in $G^*$ also crosses the cut $\cut{C_b}$ in $G^*$, which is a contradiction. If $X_1$ is empty then we infer that $e_i$, which is known to cross the cut $\cut{X}$ in $G^*$ also crosses the cut $\cut{C_a}$ in $G^*$, which is a contradiction. Before we go ahead, we show that $\delta_{G^*_i}(X_1)=\delta_{G^*_i}(X_3)=\delta_{G^*_i}(X_1\cup X_2)=\delta_{G^*_i}(X_3\cup X_2)=\lambda$. Indeed, since each of these sets is a strict subset of $X$ (due to $X_1$ and $X_3$ being non-empty), the minimality of $X$ implies a lower bound of $\lambda$ on each of these quantities. Hence, we only need to argue that they are upper bounded by $\lambda$. We prove this by invoking the submodularity of cuts on the pairs $(X_1,\overline{C_a})$, $(X,C_b)$, $(X,\overline{C_b})$, and $(X,C_a)$ respectively. Since the arguments are identical for each of the sets, we only describe our argument to show that $\delta_{G^*_i}(X_3)=\lambda$. Consider the application of submodularity on the sets $X$ and $C_b$.
    
    $$\delta_{G^*_i}(X\cap C_b)+\delta_{G^*_i}(X\cup C_b)\leq \delta_{G^*_i}(X)+\delta_{G^*_i}(C_b)\leq \lambda-1 +\lambda= 2\lambda-1$$
    
    This implies that if $\delta_{G^*_i}(X_3)=\delta_{G^*_i}(X\cap C_b)\geq \lambda+1$ then $\delta_{G^*_i}(X\cup C_b) \leq \lambda-2$. However, observe that $e_i$ is the only edge of $\Z_i$ which crosses $X\cup C_b$ in $G^*$. Hence, we infer that $\delta_{G^*}(X\cup C_b) \leq \lambda-1$, contradicting our assumption that $G^*$ is $\lambda$-connected. We now return to the proof of the final statement of the lemma.

    We now argue that the set $X_2$ is also non-empty and even stronger, $X_2=C_a\setminus C_b$.
    Let $Y_2=(C_a\setminus C_b)\setminus X_2$. Suppose to the contrary that $Y_2\neq \emptyset$. Observe that any edge in $\altdelta_{G^*_i}(Y_2)$ is in one of the three sets $\altdelta_{G^*_i}(X_2,Y_2)$ or $\altdelta_{G^*_i}(C_a)\setminus \altdelta_{G^*_i}(X_1,X_2\cup X_3)$ or $\altdelta_{G^*_i}(C_b)\setminus \altdelta_{G^*_i}(X_3,X_2\cup X_1)$. Furthermore, it is straightforward to see that $\altdelta_{G^*_i}(X_1,X_2\cup X_3)\subseteq \altdelta_{G^*_i}(C_a)$ and $\altdelta_{G^*_i}(X_3,X_2\cup X_1)\subseteq \altdelta_{G^*_i}(C_b)$. Hence, we have the following upper bound on $\delta(Y_2)$. All the quantities used below are in the graph $G_i^*$ and hence we avoid explicitly referring to the graph in the subscript.

    \begin{align*}
        \delta(Y_2) \leq\; &  \delta(X_2, Y_2) + |\altdelta(C_a) \setminus \altdelta(X_1, X_2\cup X_3)| 
        + |\altdelta(C_b) \setminus \altdelta(X_3, X_1 \cup X_2)| && 
        \\
        =\; &     \delta(X_2, Y_2)	+ \Big( \delta(C_a) - \delta(X_1, X_2 \cup X_3) \Big) 
        + \Big( \delta(C_b) - \delta(X_3, X_1 \cup X_2) \Big) &&\\
        =\; &     \delta(X_2, Y_2)	+ \Big( \lambda - \delta(X_1, X_2 \cup X_3) \Big) 
        + \Big( \lambda - \delta(X_3, X_1 \cup X_2) \Big) &&\\
        =\; &  \delta(X_2, Y_2) + 2\lambda - \Big( \delta(X_1, X_2 \cup X_3) + \delta(X_3, X_1 \cup X_2) \Big) &&\\
        =\; &  \delta(X_2, Y_2) + 2\lambda - \Big( |\altdelta(X_1)\setminus (\altdelta(X_1)\cap \altdelta(X))| + |\altdelta(X_3)\setminus (\altdelta(X_3)\cap \altdelta(X))| \Big) &&\\
        =\; &  \delta(X_2, Y_2) + 2\lambda - \Big( |\delta(X_1)- |\altdelta(X_1)\cap \altdelta(X))| + \delta(X_3)-|\altdelta(X_3)\cap \altdelta(X))| \Big) &&\\
        =\; &  \delta(X_2, Y_2) + 2\lambda - \Big( \lambda- |\altdelta(X_1)\cap \altdelta(X))| + \lambda-|\altdelta(X_3)\cap \altdelta(X))| \Big) &&\\
        =\; &  \delta(X_2, Y_2) + \Big(  |\altdelta(X_1)\cap \altdelta(X))| +|\altdelta(X_3)\cap \altdelta(X))| \Big) &&\\
        =\; &  \delta(X_2, Y_2) + \Big( \delta(X) -|\altdelta(X_2)\cap \altdelta(X))| \Big) &&\\
        \leq \; &  \delta(X_2, Y_2) + \Big( \lambda-1 -|\altdelta(X_2)\cap \altdelta(X))| \Big) &&\\
        \leq \; &  \lambda-1 + \Big(\delta(X_2, Y_2) -|\altdelta(X_2)\cap \altdelta(X))| \Big) &&\\
        \leq \; &  \lambda-1 &&\\
        %
        %
        %
        %
        %
        %
        %
        %
        %
        %
        %
        %
        %
        %
        %
        %
        %
    \end{align*}
    
    In the above inequalities, we have used the following facts about the graph $G^*_i$ which have been already argued or follow from definition. 
    \begin{enumerate}
\setlength{\itemsep}{-2pt}
        \item $\delta(X)=\lambda-1$, 
        \item $\delta(C_a)=\delta(C_b)=\delta(X_1)=\delta(X_3)=\lambda$, 
        \item $\delta(X_1,X_2\cup X_3)=|\altdelta(X_1,X_2\cup X_3)|=|\altdelta(X_1)\setminus (\altdelta(X_1)\cap \altdelta(X))|$,\\ $\delta(X_3,X_2\cup X_3)=|\altdelta(X_3,X_2\cup X_3)|=|\altdelta(X_3)\setminus (\altdelta(X_3)\cap \altdelta(X))|$
        \item $\delta(X_2,Y_2)=|\altdelta(X_2,Y_2)|$ and $\altdelta(X_2,Y_2)\subseteq \altdelta(X_2)\cap \altdelta(X)$.
    \end{enumerate}
    
    
    We have thus concluded that if $Y_2\neq \emptyset$, then $\cut{Y_2}$ is a $\lambda-1$-cut in $G_i^*$ and since no edge of $\Z$ crosses $\cut{Y_2}$ in $G^*$, it follows that $\cut{Y_2}$ is also a $\lambda-1$ cut in $G^*$, contradicting our assumption that $G^*$ is $\lambda$-connected. This completes the proof of the lemma.
\end{proof}

In the next few lemmas, when dealing with a violating triple $(X,i,j)$, we continue to use the notation defined in the previous lemma. That is, the sets $X_1,X_2, X_3$ are defined to be the intersections of $X$ with the sets $C_i\setminus C_a$, $C_a\setminus C_b$ and $C_b$ respectively with $X_2=C_a\setminus C_b$. 
 Furthermore, we may assume 
(see Proof of Lemma \ref{lem:define_X_intersection}) 
that $\delta_{G^*_i}(X_1)=\delta_{G^*_i}(X_3)=\delta_{G^*_i}(X_1\cup X_2)=\delta_{G^*_i}(X_3\cup X_2)=\lambda$. Finally, $\delta_{G^*_i}(X_2)=\lambda+1$.

Recall that our main objective in the rest of the section is to show that the sets $X_1,X_2,X_3$ satisfy the premises of Lemma \ref{lemma:lambda-undir:odd irrelevent edge}. 
For this, we begin by showing that these sets satisfy similar properties with respect to the graph $G^*$ instead of the graph $G$ (which is what is required for Lemma \ref{lemma:lambda-undir:odd irrelevent edge}). Following this, we show how to `lift' the required properties to the graph $G$.


\begin{lemma}\label{lemma:lambda-undir:odd irrelevent edge_premise1}
    Let $i\in [\eta]$ such that $e_i \in \Z$ and let $(X,i,j)$ be a violating triple. 
    Let $X_1\uplus X_2\uplus X_3$ be the partition of $X$ as defined above. Let $W = V(G) \setminus X$. 
    Then, $\delta_{G_i^*}(W, X_1) = \delta_{G_i^*}(X_3, W)= \frac{\lambda - 1}{2}$, $\delta_{G_i^*}(X_1,X_2) = \delta_{G_i^*}(X_2,X_3)  = \frac{\lambda + 1}{2}$. Furthermore, $\delta_{G_i^*}(X_2,W)=\delta_{G_i^*}(X_1,X_3)=0$.
\end{lemma}

\begin{proof} 
    In the graph $G^*_i$, let us define $\beta_1 = \delta_{G_i^*}(X_1, \co{X}) = |\altdelta_{G_i^*}(X) \cap \altdelta_{G_i^*}(X_1)|$, and we similarly define $\beta_2$ and $\beta_3$.
    Note that $\beta_1 + \beta_2 + \beta_3 = \delta_{G_i^*}(X) = \lambda - 1$.
    Observe that $\beta_1=\delta_{G_i^*}(W, X_1)$ and $\beta_3=\delta_{G_i^*}(W, X_3)$. 
    Recall that $\delta_{G_i^*}(X_2 \cup X_3) = \lambda$, 
    and $\delta_{G^*_i}(X_2 \cup X_3) = |\altdelta_{G_i^*}(X_1,X_2 \cup X_3)| + |\altdelta_{G_i^*}(X)\cap \altdelta_{G_i^*}(X_2)| + |\altdelta_{G_i^*}(X)\cap \altdelta_{G_i^*}(X_3)|=\delta_{G_i^*}(X_1,X_2\cup X_3)+\beta_2+\beta_3$.
    Furthermore, $\delta_{G_i^*}(X_1, X_2 \cup X_3) = |\altdelta_{G_i^*}(X_1)\setminus (\altdelta_{G_i^*}(X_1)\cap \altdelta_{G_i^*}(X))| = \delta_{G_i^*}(X_1) - \beta_1 = \lambda - \beta_1$. Combining the two equations, we infer that  $\beta_1 = \beta_2 + \beta_3$. An analogous argument implies that $\beta_3 = \beta_1 + \beta_2$. Hence, we conclude that $\beta_2=0$, $\beta_1=\beta_3$ and $\beta_1+\beta_3=\delta_{G_i^*}(X)=\lambda-1$. This in turn implies that $\beta_1=\beta_3=\frac{\lambda-1}{2}$ as required by the lemma. We have already argued that $\beta_2=\delta_{G_i^*}(X_2, W)=0$.

    Now, $\delta_{G_i^*}(X_1) = \delta_{G_i^*}(X_1, X_2) + \delta_{G_i^*}(X_1, X_3) + \beta_1$ and $\delta_{G_i^*}(X_3) = \delta_{G_i^*}(X_2, X_3) + \delta_{G_i^*}(X_1, X_3) + \beta_3$ which, along with the fact that $\delta_{G_i^*}(X_1)=\delta_{G_i^*}(X_3)=\lambda$, implies that 
    
    %
    %
    %
    
    \begin{align*}	
        2 \lambda &= 2 \delta_{G_i^*}(X_1, X_3) + \delta_{G_i^*}(X_1, X_2) + \delta_{G_i^*}(X_2, X_3) + \beta_1 + \beta_3 &&\\
        \implies  2\lambda - (\beta_1 + \beta_3)	&= \delta_{G_i^*}(X_2) + 2 \delta_{G_i^*}(X_1, X_3) &&\\
        \implies \qquad \qquad \lambda + 1  &= \delta_{G_i^*}(X_2) + 2 \delta_{G_i^*}(X_1, X_3) &&\\
    \end{align*}
    We now observe that $\delta_{G_i^*}(X_2) \geq \lambda + 1$. Indeed, if $\delta_{G_i^*}(X_2) \leq \lambda$, the fact that $\cut{X_2}$ is not crossed by any edge in $\Z$ in the graph $G^*$ along with the fact that it is crossed by the edges $e_{i-1}$ and $e_{i-2}$ implies that is a $\lambda$-cut in $G$, in turn implying that $e_{i-1}$ and $e_{i-2}$ are undeletable, contradicting our assumption that $\Z\subseteq E(G)\setminus \R$.
    
Since $\delta_{G_i^*}(X_2) \geq \lambda + 1$, the equation above implies that $\delta_{G_i^*}(X_2) = \lambda + 1$ and $\delta_{G_i^*}(X_1, X_3) = 0$.
    Finally, since $\beta_1 = \beta_3 = \frac{\lambda - 1}{2}$ we conclude that  $\delta_{G_i^*}(X_1, X_2) = \delta_{G_i^*}(X) - \beta_1 = \frac{\lambda + 1}{2}$.
    Similarly we conclude that $\delta_{G_i^*}(X_2, X_3) = \frac{\lambda + 1}{2}$. This completes the proof of the lemma.	
\end{proof}

We now extend Lemma \ref{lemma:lambda-undir:odd irrelevent edge_premise1} from the graph $G_i^*$ to the graph 
$G$. 
%

\begin{lemma} \label{lemma:lambda-undir:odd irrelevent edge_premise1_lift1}
    Let $i\in [\eta]$ such that $e_i \in \Z$ and let $(X,i,j)$ be a violating triple. 
    Let $X_1\uplus X_2\uplus X_3$ be the partition of $X$ as defined above. Let $W=V(G)\setminus X$. 
    Then, $\delta_{G}(W, X_1) = \delta_{G}(X_1,X_2) = \delta_{G}(X_2,X_3)  = \frac{\lambda + 1}{2}$. Furthermore, $\delta_{G}(X_2,W)=\delta_{G}(X_1,X_3)=0$.
\end{lemma}

\begin{proof}
    We begin by extending Lemma \ref{lemma:lambda-undir:odd irrelevent edge_premise1} from the graph $G_i^*$ to the graph $G^*$. That is, we show that $\delta_{G^*}(W, X_1) = \delta_{G^*}(X_3, W)= \delta_{G^*}(X_1,X_2) = \delta_{G^*}(X_2,X_3)  = \frac{\lambda + 1}{2}$ and  $\delta_{G^*}(X_2,W)=\delta_{G^*}(X_1,X_3)=0$.
    Recall that $G_i^*=G-\Z_i$ and we have already argued that $e_j$ and $e_i$ are the only edges of $\Z_i$ which cross the cut $\cut{X}$ in $G^*$. Furthermore, since $e_j$ crosses $\cut{C_j}$ and does not cross the cut $\cut{C_b}$, it must be the case that in $G^*$,  $e_j$ has one endpoint in $X_3$ and the other in $W$. Similarly, since $e_i$ crosses $\cut{C_i}$ and does not cross the cut $\cut{C_a}$, it must be the case that in $G^*$,  $e_i$ has one endpoint in $X_1$ and the other in $W$. Finally, every edge in $\Z_i$ has both endpoints in $W$. As a result, Lemma \ref{lemma:lambda-undir:odd irrelevent edge_premise1} implies that $\delta_{G^*}(W, X_1)=\delta_{G_i^*}(W, X_1)+1=\frac{\lambda + 1}{2}$. Similarly, $\delta_{G^*}(W, X_3)=\delta_{G_i^*}(W, X_3)+1=\frac{\lambda + 1}{2}$. Since none of the edges of $\Z_i$ have an endpoint in $X_2$, we conclude that $\delta_{G^*}(X_1,X_2) = \delta_{G^*}(X_2,X_3)=\delta_{G_i^*}(X_1,X_2) = \delta_{G_i^*}(X_2,X_3)= \frac{\lambda + 1}{2}$. For the same reason the sizes of the sets $\delta_{G^*}(X_2,W)$ and $\delta_{G^*}(X_1,X_3)$ are the same as in $G_i^*$, that is, 0.

    We now proceed to the statement of the lemma. Recall that $G^*=G_{S^*}-\{e^*\}$. Observe that no edge of $S^*\cup \{e^*\}$ can have an endpoint in $X_1\cup X_2$. Indeed, $e^*$ clearly has  both endpoints in $W$. Furthermore, if an edge of $S^*$ has an endpoint in $X_1\cup X_2$ then $V(S^*)$ intersects the set $C_i\setminus C_{i-(2k+3)}$, a contradiction to the fact that we added the cut $C_i$ to the collection $\C$. As a result, we conclude that $\delta_{G}(W, X_1) = \delta_{G}(X_1,X_2) = \delta_{G}(X_2,X_3)  = \frac{\lambda + 1}{2}$ and $\delta_{G}(X_2,W)=\delta_{G}(X_1,X_3)=0$. This completes the proof of the lemma.	
\end{proof}


Now we shall apply the above lemmas to construct a partition of $V(G)$ which satisfies the premise of Lemma~\ref{lemma:lambda-undir:odd irrelevent edge}.

\begin{lemma}\label{lemma:lambda-undir:odd type 2 violating cut}
    Let $i\in [\eta]$ such that $e_i \in \Z$ and let $(X,i,j)$ be a violating triple.
    Then there is an deletable edge $e$ such that $(G,k,\R)$ has a deletion set of cardinality $k$ 
    if and only if $(G,k,\R\cup\{e\})$ has a  deletion set of cardinality $k$ .
    
\end{lemma}
\begin{proof}
    Let $(X,i,j)$ be a violating triple in the graph.
    Then by Lemma~\ref{lem:violating_triple_exists} we have that $X \subseteq C_i \setminus C_j$.
    Let us recall that $C_{2k+3} \subset C_{2k+2} \subset C_{2k+1} \ldots \subset C_1 \subset C_0$ denote the sets 
    $C_j \subset C_{i-(2k+2)} \subset C_{i-(2k+2)} \ldots \subset C_{i-1} \subset C_i$ respectively, and they all lie the collection $\C(e^*)$.
%
%
    Let $Y_{2k+3} \uplus Y_{2k+1} \ldots \uplus Y_1$ be a partition of $X$ where $Y_\ell = X \cap (C_{\ell-1} \setminus C_\ell)$ for every $\ell \in [2k+3]$.
    Let $W = V(G) \setminus X$.
    In the following arguments, for any $r \geq 2k+3$ the term $Y_r$ denotes the set $W = V(G) \setminus X$,
    and similarly for any $s \leq 0$, the term $Y_s$ denotes the set $W$ as well.
    
    Now, for any $\ell \in \{2,3, \ldots, 2k+2\}$, let $X^\ell_1 \uplus X^\ell_2 \uplus X^\ell_3$ be a partition of $X$ where
    $X^\ell_1 = Y_1 \cup Y_2 \ldots \cup Y_{\ell-1}$, 
    $X^\ell_2 = Y_\ell$ and 
    $X^\ell_3 = Y_{\ell+1}$ $ \cup Y_{\ell+2} \ldots \cup Y_{2k+3}$.
    Let $C_a = C_\ell$, $C_b = C_{\ell+1}$ and $W = V(G) \setminus X$.
    By Lemma~\ref{lem:define_X_intersection} we have that the sets $X^\ell_1, X^\ell_2, X^\ell_3$ are non-empty,
    and $X^\ell_2 = C_a \setminus C_b = C_\ell \setminus C_{\ell+1}$.
    And by Lemma~\ref{lemma:lambda-undir:odd irrelevent edge_premise1_lift1}, we have in the graph $G$ that,
    $\delta_G(X^\ell_1, X^\ell_2) = \delta_G(X^\ell_2, X^\ell_3) = \frac{\lambda + 1}{2}$ and these are the only edges in $\delta(X^\ell_2)$.
   
    Now consider $2 \leq \ell \leq 2k+1$.
    By applying Lemma~\ref{lemma:lambda-undir:odd irrelevent edge_premise1_lift1} for $\ell + 1$,
    we have that $\delta_G(X^{\ell+1}_1, X^{\ell+1}_3) = \emptyset$.
    Now the fact that $Y_\ell \subset X^{\ell+1}_1$ and every $Y_r \subset X^{\ell+1}_3$ for $r \geq \ell+2$ implies that
    $\delta_G(Y_\ell, Y_r) = \emptyset$.
    And for $\ell = 2k+2$, observe that $X^\ell_3 = Y_{2k+3}$
    and this condition holds by definition.
    Hence for $2 \leq \ell \leq 2k+2$ and for every $r \geq \ell+2$
    we have that $\delta_G(Y_\ell, Y_r) = \emptyset$.
    Similarly we can argue that for every $2 \leq \ell \leq 2k+2$ and $s \leq \ell-2$ we have that $\delta_G(Y_s, Y_\ell) = \emptyset$.
    Together they imply that for any $\ell \in \{2,3,\ldots,2k+2\}$,
    the edges in $\altdelta_G(Y_\ell)$ are divided between $Y_{\ell+1}$ 
    and $Y_{\ell -1}$ as  $\delta_G(Y_{\ell-1}, Y_\ell) = \delta_G(Y_\ell, Y_{\ell+1}) = \frac{\lambda+1}{2}$ .
    
    Now consider the partition $A_1 \uplus A_2 \ldots A_{2k+2}$ of $V(G)$,
    where $A_\ell = Y_{\ell+1}$ for $\ell \in [2k+1]$ and $A_{2k+2} = Y_1 \cup W \cup Y_{2k+3}$.
    Observe that every edge of the graph either has both endpoints within some $A_\ell$,
    or it belongs to $\altdelta_G(A_\ell, A_{\ell+1})$ for $\ell \in [2k+2]$,
    (where $A_{2k+3}$ denotes the set $A_1$).
    Since $A_\ell = C_{\ell+1} \setminus C_{\ell}$ for $\ell \in [2k+1]$ we have that there are 
    deletable edges $e_1, e_2, \ldots e_{2k+2} \in \D(e^*) \setminus \R$ of the graph $G$ 
    such that $e_\ell \in \altdelta_G(A_\ell, A_{\ell - 1})$ for every $\ell \in [2k+2]$ (where $A_0$ denotes the set $A_{2k+2}$).
    This is by the construction of the cuts in $\C(e^*)$.
    
    Finally, we apply Lemma~\ref{lemma:lambda-undir:odd irrelevent edge} to the decomposition $A_1, A_2, \ldots, A_{2k+2}$ of the graph
    and obtain a deletable edge $e$ which has all the required properties.
    This completes the proof of this lemma.	
\end{proof}

Having established Lemma \ref{lemma:lambda-undir:odd type 1 violating cut} and Lemma \ref{lemma:lambda-undir:odd type 2 violating cut}, we complete the proof of Lemma \ref{lem:main_odd}.

\myparagraph{Proof of Lemma \ref{lem:main_odd}.}  Let $F = \{ f_1, f_2, \ldots, f_p\}$ be an arbitrary maximal set of edges disjoint from $\R$ such that $G - F$ is $\lambda$-connected.
If $|F| = p \geq k$, then we already have the required deletion set.
Therefore, we may assume that $p\leq k-1$.

Now, consider the graphs $G_0,\dots, G_p$ with $G_0=G$ and $G_i$ defined as  $G_i = G - \{f_1, \ldots f_i\}$ for all $i\in [p]$.
Note that $G_{i+1} = G_i - f_{i+1}$ and $G_p = G - F$.
Observe that each $G_i$ is $\lambda$-connected by the definition of $F$.
Let $\D_i$ be the set of deletable edges in $G_i$ which are undeletable in $G_{i+1}$.
Observe that $\D_i=\D(f_i)$ (see Definition \ref{def:special_edges}) in the graph $G_i$.

Now consider any deletable edge of $G$.
It is either contained in $F$, 
or there is some $r\in \{0,\dots, p-1\}$ such that it is deletable in $G_i$ but undeletable in $G_{r+1}$.
In other words, the set $F \cup \D_1 \cup \D_2 \ldots \cup \D_p$ covers all the deletable edges of $G$. Since $p \leq k-1$ and the number of deletable edges in $G$ is greater than $\eta\lambda$, it follows that for some $r\in [p]$, the set $\D_r$ has size more than $\eta\lambda$. We fix one such $r\in [p]$ and if $r>1$, then we define $S^*=\{f_1,\dots, f_{r-1}\}$ and $S^*=\emptyset$ otherwise. We define $e^*=e_r$.

We then construct the sets $\Z(e^*), \C(e^*), \widehat \Z, \widehat \C$ (see the paragraph on setting up notation.
Then we construct the  cut-collection $\C$ and the corresponding edge set $\Z$
by using the set $S^*$. Recall that $\Z$ contains at least $k$ edges and is by definition disjoint from $\R$. Consider the graph $G^*=G-S^*\cup \{e^*\}=G-\{f_1,\dots, f_r\}$ and note that $G^*$ is $\lambda$-connected.

We check whether $G^*-\Z$ is $\lambda$-connected. If so, then we are done since $\Z$ is a deletion set of cardinality $k$ in the graph. Otherwise, we know that $G^*-\Z$ contains a violating cut. Lemma \ref{lemma:lambda-undir:odd type 1 violating cut} implies that such a violating cut cannot be of Type 1. Hence, we compute in polynomial time (see Lemma \ref{lem:violating_triple_exists}) a violating triple $(X,i,j)$ in the graph $G_i^*$ for some $i\in [\eta]$. We now invoke Lemma \ref{lemma:lambda-undir:odd type 2 violating cut}  to compute the edge $e\in E(G)\setminus \R$ in polynomial time and return it. The correctness of this step follows from that of Lemma \ref{lemma:lambda-undir:odd type 2 violating cut}. This completes the proof of the lemma. \qed

\medskip
As a consequence of Lemma \ref{lem:main_odd}, we obtain an {\FPT} algorithm for {\lcd} when $\lambda$ is odd, completing the proof of Theorem \ref{thm:lcd-thm}.

\begin{lemma}\label{thm:lambda-undir:main} Let $\lambda\in {\mathbb N}$ be an odd number. Then, 
    \lcd in undirected graphs can be solved in time $2^{\cO(k \log k)} n^{\cO(1)}$.
\end{lemma}

\section{Extension to \wlcd }
\label{sec:weighted}
In this section, we extend our \textsf{FPT} algorithm for \lcd to the weighted version of the problem.
Here the weights are on the edges of the graph and they could be any non-negative real number.
The goal is to delete a set of edges (or arcs) of maximum total weight while satisfying the connectivity constraints.
More formally, let $(G,k,w)$ be the input instance where $w:E(G) \rightarrow {\mathbb R}_{\geq 0}$ be a weight function on the edges. 
A solution $S$ is called \emph{$(k,\alpha)$-solution} if $w(S) \geq \alpha$ and $|S| \leq k$.
We give an \textsf{FPT} algorithm to find such a solution of the input instance parameterized by $k$.
It is clear that we can use this algorithm to find the maximum weight solution with at most $k$ edges.
In the rest of this section, for a given instance, 
we set $\alpha = \max \{ w(S) \mid S \subset E(G), \, |S| \leq k \textit{ and $G \setminus S$ is $\lambda$ connected } \}$.

\subsection{Digraphs and Undirected Even Connectivity}
When the input graph is directed, or when the graph is undirected and $\lambda$ is even,
it turns out that our arguments for the unweighted version can be easily modified to handle weighted instances.
We sort the deletable edges of the input graph by their weight
and let $W$ be the set of the heaviest $2\lambda k^2$ edges.
Now we have the following lemma, which asserts that there is always a $(k,\alpha)$ solution
which intersects $W$.

\begin{lemma}
	There is a $(k,\alpha)$ solution $F$ such that $F \cap W \neq \emptyset$.
\end{lemma}
\begin{proof}
	If $W$ has fewer than $2\lambda k^2$ edges, 
	then it must be the case that $W$ actually contains all the deletable edges of the instance.
	Hence the statement of the lemma is trivially true.
		
	Now let us consider the case $W$ contains exactly $2\lambda k^2$ edges.
	The proof of the lemma is via the following simple observation.
	Suppose that there is a $F_W \subseteq W$ of $k$ or more edges such that $G \setminus F_W$ is $\lambda$-connected.
	Let $F$ be a $(k,\alpha)$ solution of the instance.
	If $F \cap W = \emptyset$ then $w(F_W) \geq w(F)$,
	and hence $F_W$ is the required $(k,\alpha)$ solution.
	Otherwise, $F$ itself satisfies the lemma.
	
	Now it only remains to prove that such a $F_W$ exists.
	For this we consider the set $\del(G)$ restricted to $W$.
	Then we apply Lemma~\ref{lemma:lambda-dir:upper bound} and Lemma~\ref{lemma:lambda-undir:upper bound},
	for the case of digraphs, and undirected graph with an even value of $\lambda$, respectively.
	This gives us the set $F_W$ with the required properties.
\end{proof}

The above lemma implies the following theorem.
\begin{theorem}
	\wlcd can be solved in time $2^{\cO \big( k (\log k + \log \lambda) \big)} n^{\cO(1)}$ on directed graphs for any value of $\lambda$,
	and on undirected graphs when $\lambda$ is an even number.
\end{theorem}

\subsection{Odd Connectivity in undirected graphs}
As in the unweighted case, we define our instance to be of the form $(G,k,\R,w)$ where $\R$ denotes the set of all irrelevant edges in $G$,
and initially $\R$ contains all the undeletable edges of $G$.
We have the following lemma for marking irrelevant edges in weighted instances.
A proof of this lemma follows from a simple modification of 
 Lemma~\ref{lemma:lambda-undir:odd irrelevent edge}
where we return the edge with the least weight
among all the candidate edges.

\begin{lemma}\label{lemma:weighted lambda-undir:odd irrelevent edge}
    Let $(G,k,\R,w)$ be an instance of {\wlcd} and let $X_1, X_2, \ldots X_{2k+2}$ be a partition of $V(G)$ into non-empty subsets  
    such that the following properties hold in the graph $G$.
    \begin{enumerate}
\setlength{\itemsep}{-2pt}
        \item $\delta_G(X_1, X_2) = \delta_G(X_2,X_3) \ldots = \delta_G(X_{2k+2}, X_1) = \frac{\lambda + 1}{2}$.
        
        \item Every edge of the graph either has both endpoints in some $X_i$ for $i\in [2k+2]$,
        or contained in one of the edge sets mentioned above.
        
        \item There are deletable edges $e_1, e_2, \ldots, e_{2k+2}$ in $E(G) \setminus \R$ such that
        $e_i \in \delta(X_i, X_{i+1})$ for $i \in [2k+2]$. (Here $X_{2k+3}$ denotes the set $X_1$.)
    \end{enumerate}
     Let $\ell \in [2k+2]$ such that $w(e_\ell) \leq w(e_i)$ for every  $i \in [2k+2]$.
    Then for any positive real number number $\alpha$, the instance $(G,k,\R,w)$ has a $(k, \alpha)$ solution $S$ if and only if the instance $(G,k,\R \cup \{ e_\ell \}, w)$ has $(k,\alpha)$ solution.

\end{lemma}

The following lemma is proved in the same way as Lemma~\ref{lem:main_odd}.

\begin{lemma}\label{lemma:weighted odd reduction}
	Let $(G,k,\R,w)$ be an instance of $\lcd$ where $\lambda$ is an odd number.
	Let $W$ be the collection of heaviest $7\lambda k^3$ edges in $G$
    disjoint from $\R$.
	Then either $W$ has fewer than $7\lambda k^3$ edges,
	or there is a polynomial time algorithm which given the instance returns
	\begin{itemize}
\setlength{\itemsep}{-2pt}
		\item a subset $F_W$ of $W$ containing $k$ edges such that $G \setminus F_W$ is $\lambda$-connected,
		\item an edge $e \in W$ such that $(G,k,\R,w)$ has a $(k,\alpha)$ solution if and only if $(G,k,\R\cup\{e\},w)$
		      has a $(k,\alpha)$ solution.
	\end{itemize}
\end{lemma}

Since $W$ is disjoint from $\R$, each application of the above lemma either increases the set $\R$, or gives a set $F_W \subseteq W$ of cardinality $k$ which is a solution to the instance.
Therefore we apply the above lemma repeatedly, updating the sets $\R$ and $W$ after each application, until we obtain,
either the solution $F_W \subseteq W$,
or an instance where $W$ has fewer than $7\lambda k^3$ edges.
Note that this process takes polynomial time.
Now, as in the previous subsection, we can show that such instances have a $(k,\alpha)$ solution which intersects $W$. 
This leads to the following theorem.

\begin{theorem}
	\wlcd can be solved in time $2^{\cO \big( k (\log k + \log \lambda) \big)} n^{\cO(1)}$ on undirected graphs for odd values of $\lambda$.
\end{theorem}

\newtheorem{tform}{Transformation}
\newcommand{\set}[1]{\ensuremath{\{ #1 \}}}
\newcommand{\Union}{\ensuremath{\bigcup}}
\newcommand{\union}{\ensuremath{\cup}}

\section{Polynomial Compression for \lcd}
\label{sec:compression}
In this section we design a polynomial compression for the \lcd\ problem. Recall, that a parameterized problem  admits a polynomial kernel, if there is a polynomial time algorithm which given an instance 
$(x,k)\in \Pi$ returns an instance $(x',k')\in \Pi$  such that $(x,k)\in \Pi$  if and only if $(x',k')\in \Pi$  and $|x'|, k'\leq k^{\cO(1)}$.  A {\em polynomial compression} is a relaxation of polynomial kernelization where 
 the output may be an instance of a (fixed) different language than the input language.  That is, polynomial kernelization can be viewed as polynomial time self-reduction while polynomial compression is a polynomial time reduction to a different language. 

We first give a polynomial compression for \lcd, when the input instance is a digraph and then give the required modifications for the undirected case. Let $T$ be the set of end-points of the edges in  $\del(G)$. We will call $T$ as a 
{\em set of terminals}. So from now onwards we will assume that the input consists of $(G,T,k)$.  
To obtain the desired compression, we apply the ideas and methods developed for dynamic graph optimization problems. In particular, we use the results proved in~\cite{AssadiKLT15}. Towards this we first state  the model given in~\cite{AssadiKLT15} verbatim.  For graph problems in the dynamic sketching model,  Assadi et al.~\cite{AssadiKLT15} considered 
the following setup. Given a graph optimization problem $\Pi$, an input graph $G$ on $n$ vertices with $\ell$ vertices identified as \emph{terminals} $T = \{q_1,\ldots,q_\ell \}$, the goal of $\ell$-dynamic sketching for $\Pi$ is to construct a sketch $\Gamma$  such that given any
possible subset of the edges between the terminals (a \emph{query}), we can solve the problem $P$ using only the information contained in the sketch $\Gamma$. Formally,

\begin{definition}[\cite{AssadiKLT15}]
\label{def:k-dynamic}
Given a graph-theoretic problem $\Pi$, a $\ell$-dynamic sketching scheme for
$P$ is a pair of algorithms with the following properties.
\begin{enumerate}[(i)]
\item A \textbf{compression algorithm} that given any input graph $G$ with a set $T$ of $\ell$ terminals,  outputs a data structure $\Gamma$ (i.e, a dynamic sketch).
\item An \textbf{extraction algorithm} that given any subset of the edges between
the terminals, i.e, a \textbf{query} $Q$, and the sketch $\Gamma$, outputs
the answer to the problem $\Pi$ for the graph, denoted by $G^{Q}$, obtained by inserting all edges in $Q$ to $G$  (without further access to $G$).
\end{enumerate}
\end{definition}
Let $\Pi$ be   {\sc Edge-Connectivity} problem. That is, given a digraph $G$ and two designated vertices $s$ and $t$ find the minimum number of edges needed to remove to eliminate all (directed) paths from $s$ to $t$.   The minimum number of edges needed to remove to eliminate all  (directed) paths from $s$ to $t$ in $G$ is denoted by $\mu_G(s,t)$.  For  $\Pi$ being {\sc  Edge-Connectivity} problem,  Assadi et al.~\cite{AssadiKLT15} obtained the following result. 

 \begin{proposition}[\cite{AssadiKLT15}]
\label{prop:stsketch}
 For any $\delta > 0$, there exists a randomized $\ell$-dynamic sketching scheme for the {\sc Edge-Connectivity}  problem with a sketch of size $\cO(\ell^4 \log(1/\delta))$, which answers any query correctly with probability at least $1 - \delta$.
\end{proposition}

Before we prove our main result of this section, we prove an auxiliary lemma which will be crucial to the correctness of our algorithm. 

\begin{lemma}
\label{lem:connectioncompression}
Let $(G,T,k)$ be an instance of \lcd\ and let $Z\subseteq \del(G)$ of size at least $k$. Then $G-Z$ is $\lambda$-connected if and only if for all pairs $\{s,t\}\in T$, $s\neq t$, we have that  
$\mu_{G-Z}(s,t)\geq \lambda$. 
\end{lemma}
\begin{proof}
The forward direction of the proof is straightforward. If $G-Z$ is $\lambda$-connected then for any pair of vertices $x,y\in V(G-Z)$ we have that $\mu_{G-Z}(x,y)\geq \lambda$ and thus it holds for pairs of vertices in $T$.  

For the reverse direction of the proof we show that if  $G-Z$ is \emph{not} $\lambda$-connected then there exists a pair $\{s,t\}\in T$ such that $\mu_{G-Z}(s,t)< \lambda$. Since  $G-Z$ is not 
$\lambda$-connected, there exists a cut $\cut{X}$ in $G-Z$ such that $\delta_{G-Z}(X) <\lambda$. However,  we know that $G$ is $\lambda$-connected and thus $\delta_{G}(X) \geq\lambda$. This  implies that there exists an edge $e=(s,t)\in Z$ such that $s\in X$ and $t\in \overline{X}$.  Furthermore, since $Z\subseteq \del(G)$ and that $T$ is the set of end-points of edges in $\del(G)$ we have that $s,t\in T$. This implies that $X$ separates $\{s,t\}$ in $G-Z$ and thus 
$\mu_{G-Z}(s,t)\leq \delta_{G-Z}(X) < \lambda$. This concludes the proof. 
\end{proof}
Now we give the polynomial compression for \lcd\ on digraphs 
\begin{theorem}
\label{thm:poly-compression-digraphs}
For any $\delta > 0$, there exists a randomized compression for 
\lcd\ of size $ \cO(k^{12} \lambda^6 
(\log k \lambda + \log(1/\delta))$ on digraphs, such that the error probability is upper bounded by $1-\delta$. 
\end{theorem}
\begin{proof}
Let $(G,k)$ be an instance to \lcd\ where $G$ is a digraph. Our starting point is the Lemma~\ref{lemma:lambda-dir:upper bound}. That is, given an input digraph $G$, integers $\lambda$ and $k$, we apply Lemma~\ref{lemma:lambda-dir:upper bound} and either decide that $(G,k)$ is a {\Yes} instance of {\lcd} or we conclude that there are at most $k^2 \lambda$ deletable edges in $G$. That is, $|\del(G)|\leq k^2 \lambda$. This implies that the size of $T$  is upper bounded by 
$2k^2 \lambda$. 

Let $\delta'= \min\{1,\frac{\delta}{8k^4\lambda^2}\}$ , $\ell=|T|$ and $G^\star=G-\del(G)$.  Now for every pair of vertices,   $\{s,t\}\in T$ (an ordered pair), $s\neq t$,  we apply Proposition~\ref{prop:stsketch} with $s,t$ and $G^\star$ and obtain  
a randomized $\ell$-dynamic sketching scheme for the $s$-$t$-edge connectivity with a sketch of size $\cO(\ell^4 \log(1/\delta'))$, which answers any query correctly with probability at least 
$1 - \delta'$. That is, we get a data-structure $\Gamma_{s,t}$  and an extraction algorithm 
${\cal A}_{s,t}$. The extraction algorithm,  given any subset of edges between the terminals  (the query $Q$), and the data-structure $\Gamma_{s,t}$ outputs $\mu_{(G^\star+Q)}(s,t)$, without further access to $G$. 
The family, $\Gamma=\{ (\Gamma_{s,t},{\cal A}_{s,t})~|~s,t \in T \}$ along with the set $\del(G)$ is our compression for \lcd.  This concludes the construction of compression. 

Let $Z\subseteq \del(G)$ of size at least $k$. By Lemma~\ref{lem:connectioncompression} we know that $G-Z$ is $\lambda$-connected if and only if for all pairs $\{s,t\}\in T$, $s\neq t$, we have that  
$\mu_{G-Z}(s,t)\geq \lambda$.  Thus, to check whether $G-Z$ is  $\lambda$-connected  all we need to do is to ask the query $Q=\del(G)-Z$ to every data-structure in the family $\Gamma$. If all the answers return a value at least $\lambda$ then  using Lemma~\ref{lem:connectioncompression} we can conclude that $G-Z$ is  $\lambda$-connected; else we can conclude that $G-Z$ is not  $\lambda$-connected. This completes the proof of correctness of our scheme.

The size of the data-structure is upper bounded by 
\[ \sum_{(s,t)\in T} |\Gamma_{s,t}|\leq  \cO(\ell^6 \log(1/\delta')) \leq \cO(k^{12} \lambda^6  \log(1/\delta'))  \leq \cO(k^{12} \lambda^6 
(\log k \lambda + \log(1/\delta)). \]

Next we bound the error of probability. We can conclude that  $G-Z$ is $\lambda$-connected even though $G-Z$ is not $\lambda$-connected if all the query answers wrongly. Since, all the 
data-structures have been made independently, this happens with probability  at most 
${\delta'}^{|T|^2}\leq \delta'$. On the other hand if $G-Z$ is $\lambda$-connected and we return that $G-Z$ is not $\lambda$ connected if and only if there exists a pair $\{s,t\}\in T$ such that the value returned on the query $Q$ is strictly less than $\lambda$. Using, union bound this probability can be upper bounded by $\delta' \cdot |T|^2 \leq \delta' 4k^4 \lambda^2$. Thus, the error probability of the algorithm (by combining both steps and again using union bound) is upper bounded by 
$\delta' 4k^4 \lambda^2 +\delta' \leq \delta' 8 k^4 \lambda^2$. This is at most $\delta$ by our choice of $\delta'$, this concludes the proof. 
\end{proof}

We now show how the above results can be extended to undirected graphs.
Let $(G,k)$ be an instance of {\lcd} where $G$ is an undirected graph. Again, let $T$ denote the 
end-points of deletable edges in $G$, and we call them the terminal vertices
and rewrite our instance as $(G,T,k)$.
Now, we convert the undirected graph $G$ into a digraph $D_G$, 
by replacing each edge into two anti-parallel directed edges. In other words for each edge $(u,v)$ in 
$E(G)$, we have two directed edges $(u,v)$ and $(v,u)$ in $E(D_G)$. For a set of edges $X$ in $G$, let $D_X$ denote the set of directed edges corresponding to $X$ in $D_G$. Then one can easily 
show that $G-X$ is $\lambda$-connected if and only if $D_G- D_X$ is $\lambda$-connected.
Now, we can construct a compression for $(D_G,T,k)$,  as constructed in 
Theorem~\ref{thm:poly-compression-digraphs} for directed graphs, and show that this is also a polynomial compression for $(G,T,k)$.
Thus we have the following theorem.

\begin{theorem}
\label{thm:poly-compression-undirectedraphs}
	For any $\delta > 0$, there exists a randomized compression for 
	\lcd\ of size $ \cO(k^{18} \lambda^6 
	(\log k \lambda + \log(1/\delta))$ on undirected graphs, such that the error probability is upper bounded by $1-\delta$. 
\end{theorem}



\section{Conclusion}
\label{sec:conclusion}

In this paper, we studied the edge connectivity version of {\sc SNDP with Uniform Demands}. We obtain new structural results on $\lambda$-connected graphs and digraphs, which could be of independent interest.
These results lead to \FPT algorithms for these problems with general weights, and a polynomial compression of the unweighted version.
Our paper opens up several new avenues of research, especially in parameterized complexity. We conclude with a few open problems and future research directions.
\begin{itemize}
\item[{\bf(a)}] Is there an algorithm for  {\sc Survivable Network Design Problem  with Uniform Demands}  running in time $c^k n^{\cO(1)}$ for any fixed value of $\lambda$ ? 



\item[{\bf(b)}] What is the parameterized complexity of the problem
when we are only interested in the connectivity of 
a given subset of terminals (say $T$)? 
This generalizes well studied problems such as {\sc Steiner Tree} and {\sc Strongly Connected Steiner Subgraph}.    

\item[{\bf(c)}] In the context of the above problem, is there a relation between the total number of deletable edges and the cardinality of the largest deletion set ?

\item[{\bf(d)}] The same questions may be asked with respect to vertex connectivity of the graph.

\item[{\bf(e)}] Finally, what is the parameterized complexity of the deletion version of the {\sc Survivable Network Design} problem in its full generality. 
This problem is likely to be quite difficult and resolving its complexity could very well require the development of new algorithmic tools and techniques.

%
%

\end{itemize}

\bibliographystyle{siam}
\bibliography{references}
\end{document}